\documentclass[aps,prl,twocolumn,superscriptaddress]{revtex4-2}

\usepackage{graphicx}

\usepackage{amssymb}
\usepackage{amsmath}
\usepackage{verbatim}
\usepackage{tabularx}

\usepackage{algorithmicx}
\usepackage{algpseudocode}

\begin{document}

\title{Cognitive Field Theory: Memory-Dressed Collective Dynamics of Intelligence}

\author{Byung Gyu Chae}

\affiliation{Electronics and Telecommunications Research Institute, 218 Gajeong-ro, Yuseong-gu, Daejeon 34129, Republic of Korea
\\ bgchae@etri.re.kr}


\begin{abstract}

Learning, inference, memory, and emergence in biological and artificial
systems are often described using disparate theoretical frameworks.
Here we develop a cognitive field theory in which cognition is
described as a collective nonequilibrium phenomenon governed by the
geometry and collective spectrum of a learned cognitive manifold.
Starting from a stochastic cognitive-field equation defined on an
adaptive Riemannian cognitive manifold, we derive an effective
memory-dressed cognitive field theory incorporating nonlocal memory
kernels and retarded self-energy feedback.
The learned cognitive geometry generates a complex collective spectrum
characterized 
by the time-scale density of states $\rho(\lambda,\omega)$.
This spectrum provides a fundamental dynamical descriptor of
cognition and determines the emergent memory kernel, collective
response, temporal coherence, and infrared organization of the
cognitive field.
Integrating out latent collective modes produces non-Markovian
memory feedback that renormalizes the cognitive forgetting gap
$r_{\rm cog}$, enhances collective cognitive susceptibility, and drives the system toward a
protected near-critical regime characterized by long-time contextual
persistence and scale-free temporal organization.
The observable cognitive field emerges as a memory-dressed
macroscopic order parameter, $\phi=Ae^{i\psi}$,
whose amplitude encodes collective cognitive organization while its
phase encodes the temporal organization of distributed collective modes.
Within this framework, learning organizes cognitive geometry,
cognitive geometry generates a collective spectrum, the collective
spectrum produces memory feedback, and memory feedback stabilizes a
memory-dressed cognitive field.
The resulting theory provides a unified dynamical description of
learning, memory, inference, selfhood, and emergent intelligence in
terms of the infrared organization of collective cognitive dynamics.

\end{abstract}

\maketitle

\section{I. Introduction}

Understanding how learning, inference, memory, and stable cognition
emerge from neural dynamics remains a central challenge in
neuroscience, statistical physics, and artificial intelligence
\cite{1,2,3}.
Biological brains exhibit robust adaptive behavior despite operating
with noisy, heterogeneous, and partially stochastic microscopic
components \cite{4,5,6,7}, while modern artificial systems achieve
remarkable cognitive performance through highly engineered recurrent
and attention-based architectures whose underlying organizing
principles remain only partially understood
\cite{8,9,10,11}.
Despite major advances across neuroscience, machine learning, and
nonequilibrium statistical physics, a unified physical framework
explaining learning, memory, inference, and emergent cognition is
still lacking.

Existing approaches typically capture only limited aspects of this
broader problem.
Energy-based models such as Hopfield networks \cite{12,13} provide a
description of associative memory through attractor dynamics but offer
limited mechanisms for multiscale temporal organization and adaptive
collective cognition.
Biophysically detailed neuron models successfully reproduce
microscopic neural activity \cite{4,14,15}, yet often provide limited
insight into how large-scale cognition emerges at the systems level.
Modern deep-learning architectures, including transformers
\cite{9,10,11}, demonstrate remarkable empirical capabilities, but the
collective dynamical principles underlying memory, contextual
integration, and adaptive reasoning remain incompletely understood
\cite{16,17,18,19}.
Across these paradigms, learning, memory, and cognition are rarely
described within a single unified dynamical framework.

Experimental observations increasingly suggest that cognition is
organized across broad temporal scales.
Long-range temporal correlations, scale-free fluctuations, and
neuronal avalanche dynamics have been reported across multiple neural
and behavioral timescales \cite{20,21,22}.
These observations indicate that cognition may be governed not simply
by instantaneous activity patterns, but by the collective
organization of interacting relaxation processes extending over many
timescales.

A similar lesson is familiar from nonequilibrium statistical physics.
Macroscopic collective behavior generally cannot be inferred directly
from microscopic equations of motion alone \cite{23,24,25}.
Instead, collective organization emerges through coarse-grained
infrared dynamics governed by long-wavelength and long-timescale
degrees of freedom.
This suggests that cognition itself may represent a collective
nonequilibrium phenomenon controlled by the infrared organization of
adaptive dynamical timescales.

Motivated by this perspective, we develop a cognitive field theory in
which cognition emerges from the collective dynamics of a learned
cognitive manifold.
Learning continuously organizes the cognitive geometry through an
effective landscape, metric structure, and non-conservative
circulation field.
The resulting stability structure generates a spectrum of collective
relaxation modes whose weakly damped infrared sectors support
long-time persistence, contextual propagation, and adaptive
collective inference.

To characterize this organization, we introduce the
\emph{time-scale density of states} (TDOS), which describes the
distribution of collective relaxation modes generated by learned
cognitive geometry.
The relaxation sector of the TDOS governs memory persistence,
whereas the circulation sector governs temporal organization
across distributed collective modes.

A central result of the present framework is that the accumulation of
weakly damped collective modes generates an extended slow-mode
sector.
Integrating out these latent collective sectors produces retarded
self-energy corrections and nonlocal memory kernels that feed back
into the observable cognitive dynamics.
The resulting memory feedback suppresses the effective cognitive
forgetting gap and enhances collective cognitive susceptibility, driving the
system toward a protected near-critical regime 
characterized by
enhanced collective responsiveness
and multiscale temporal organization.
The observable cognitive field becomes a
memory-dressed macroscopic cognitive order parameter whose
infrared response is governed by the underlying collective
spectrum.

Intelligence therefore emerges hierarchically through a sequence of
collective dynamical organizations:
learning organizes cognitive geometry,
cognitive geometry generates a collective spectrum,
the collective spectrum produces memory feedback,
and memory feedback stabilizes a memory-dressed cognitive field.
Learning, memory, inference, selfhood, and emergent cognition are
therefore interpreted as collective dynamical consequences of the
infrared organization of adaptive cognitive manifolds.

To formulate this framework systematically, we construct a stochastic
cognitive field theory on an adaptive Riemannian cognitive manifold
and develop its corresponding MSRJD field-theoretic representation
\cite{26,27,28,29}.
The present theory further suggests that many existing learning
architectures may be understood as different dynamical realizations of
a common cognitive field principle.
Hopfield networks \cite{12,13},
recurrent neural networks
\cite{8,30,31,32},
and Transformer architectures
\cite{9,10}
may be understood as different dynamical realizations of a common
cognitive field principle.

The remainder of this paper is organized as follows.
Section~II develops the geometric foundations of the cognitive field,
derives the relaxation spectrum generated by learned cognitive
geometry, and formulates the memory-dressed cognitive field equation.
Section~III develops the self-maintained
near-critical dynamics.
Section~IV formulates inference, learning, and emergence as
memory-dressed collective cognitive dynamics on adaptive cognitive
manifolds.
Section~V develops the hierarchical emergence theory of adaptive and
memory-dominated intelligence.
Section~VI derives observable temporal spectra and response functions.
Section~VII analyzes several special dynamical limits of the
distinct models.
Finally, Section~VIII discusses broader implications for
neuroscience, artificial intelligence, and nonequilibrium theories of
collective cognition.

\begin{figure*}[t]
\centering
\includegraphics[width=1.0\textwidth, trim=0cm 12.2cm 0cm 0cm]{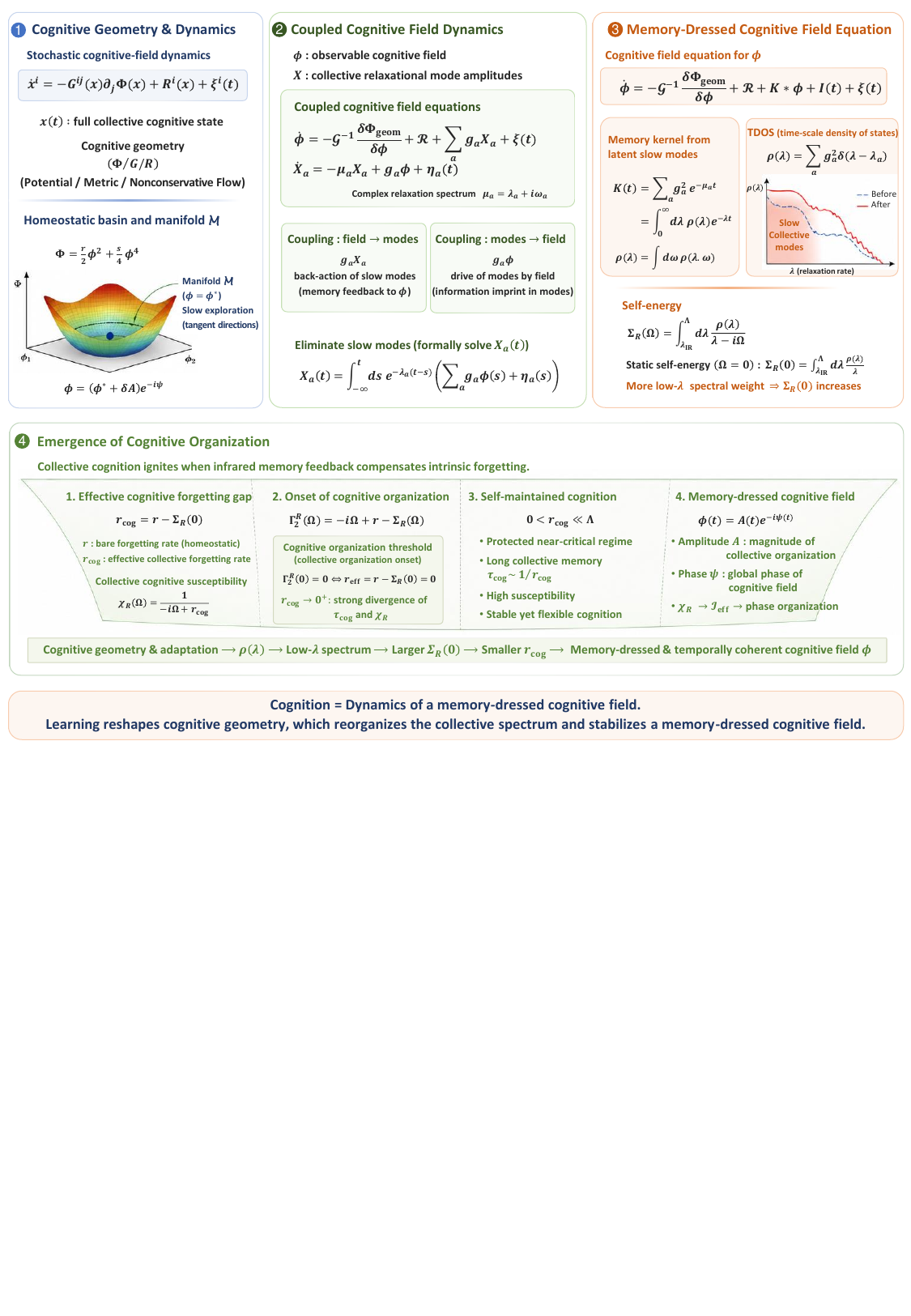}
\caption{
Schematic overview of the Cognitive Field Theory.
(1) Cognitive dynamics is formulated on an adaptive cognitive geometry
defined by the geometric landscape
$\Phi$,
metric
$G$,
and circulation field
$R$.
Homeostatic stabilization generates a bounded cognitive manifold on
which collective dynamics evolves.
(2) The observable cognitive field
$\phi$
is recursively coupled to distributed collective-mode amplitudes
$X_\alpha$,
forming a coupled field--mode dynamics.
The collective modes are characterized by the complex relaxation
spectrum
$\mu_\alpha=\lambda_\alpha+i\omega_\alpha$,
where the relaxation and circulation sectors jointly determine memory
persistence and temporal organization.
(3) Eliminating the distributed collective modes generates a
non-Markovian memory kernel
$K(t)$
determined by the time-scale density of states, yielding a
memory-dressed cognitive-field equation and the corresponding memory
self-energy
$\Sigma_R(\Omega)$.
(4) Infrared memory feedback renormalizes the bare forgetting rate to
the effective cognitive forgetting gap
$r_{\rm cog}=r-\Sigma_R(0)$.
The accumulation of low-relaxation collective modes enhances
$\Sigma_R(0)$,
suppresses
$r_{\rm cog}$,
and drives the system toward a protected near-critical regime.
The resulting memory-dressed cognitive field
$\phi(t)=A(t)e^{-i\psi(t)}$
combines persistent collective memory with coherent temporal phase
organization, providing the dynamical foundation for higher-order
cognitive organization.
}
\label{fig:schematics}
\end{figure*}

\section{II. Cognitive Geometry and Memory-Dressed Cognitive Field Dynamics}

This section develops the theoretical foundation of the present
framework.
We first formulate the learned cognitive geometry governing the
underlying state-space dynamics and show how it generates a complex
collective stability spectrum through linearization around operating
cognitive states.
The resulting collective modes and their dynamical amplitudes provide
the latent infrared degrees of freedom from which the memory kernel
and the observable cognitive field emerge through recursive memory
feedback.

The resulting memory-dressed dynamics naturally gives rise to the
memory self-energy, the effective cognitive forgetting gap, and the
protected near-critical regime governing the observable cognitive
field.

The following subsections develop this construction in sequence,
beginning with the learned cognitive geometry, followed by the
collective spectrum, memory kernel, infrared response, temporal coherence and finally
the memory-dressed cognitive field equation that forms the basis of
the present cognitive field theory.
The overall logical structure of the present theory is summarized
schematically in Fig.~\ref{fig:schematics}.

\subsection{A. Geometric formulation of the cognitive field}

Rather than tracking microscopic neuronal states individually, we
adopt an effective $(0+1)$-dimensional description in which cognition
is represented by a high-dimensional collective state vector.
The underlying neural substrate is highly heterogeneous, stochastic,
and continuously reconfiguring, while cognition itself appears as a
stable macroscopic phenomenon emerging from the organization of
collective dynamical modes.
The central objective is therefore to describe the dynamics of these
collective cognitive degrees of freedom and their organization across
multiple temporal scales.

We consider a high-dimensional collective state
\begin{equation}
x(t)\in\mathbb R^N,
\end{equation}
representing macroscopic neural activity in biological systems or
latent collective representations in artificial networks.
The coordinates \(x^i\) denote components of a high-dimensional neural
state vector parameterizing the instantaneous collective configuration
of the system.

A central empirical lesson from modern neural population dynamics is
that collective activity does not explore this ambient space uniformly.
Instead, neural trajectories are typically confined to structured
low-dimensional manifolds whose local geometry changes across task
conditions, behavioral contexts, learning stages, and internal states.
Perturbations along some directions are amplified or rapidly propagated,
whereas perturbations along other directions are strongly suppressed.
Thus the effective accessibility, sensitivity, and stability of
collective neural configurations are direction dependent and
state dependent.

This motivates describing the collective state space not as a flat
Euclidean space with identical dynamical cost in all directions, but as
an effective curved state space equipped with a state-dependent metric
\(G_{ij}(x)\).
Although the coordinates themselves take values in
\(\mathbb R^N\), the metric \(G_{ij}(x)\) encodes how distances,
sensitivities, and dynamical responses are locally measured by the
neural system.
In this sense, the collective activity manifold acquires an
intrinsically Riemannian structure.

In this geometric viewpoint, cognition is interpreted as a
collective dynamical flow evolving on a curved state space whose local
geometry governs the effective accessibility of trajectories, the
stability of perturbations, and the distribution of sensitivity across
collective degrees of freedom.

On a Riemannian manifold, the canonical steepest-descent flow
generated by an effective scalar potential \(\Phi(x)\) takes the form
\begin{equation}
\dot x^i
=
-
G^{ij}(x)\partial_j\Phi(x),
\label{eq:riemann_gradient_flow}
\end{equation}
where \(G^{ij}(x)\) denotes the inverse metric.
Equation~\eqref{eq:riemann_gradient_flow} represents the natural notion
of gradient relaxation compatible with the geometry defined by
\(G(x)\).
Ordinary Euclidean gradient descent is recovered only in the special
case of a flat constant metric.

However, a generic vector field on a Riemannian manifold cannot in
general be represented solely by a gradient flow.
Given a specified metric structure, sufficiently regular drift fields
admit a generalized Helmholtz--Hodge-type decomposition into a
metric-gradient sector together with an independent non-conservative
sector \cite{33}.
More precisely, the deterministic collective drift field may be written
as
\begin{equation}
F^i(x)
=
-
G^{ij}(x)\partial_j\Phi(x)
+
R^i(x),
\label{eq:hodge_decomp}
\end{equation}
where the first term is the metric-gradient component generated by the
effective potential \(\Phi(x)\), while \(R^i(x)\) denotes the residual
non-conservative flow that cannot be represented as the gradient of any
scalar functional.

The collective cognitive dynamics is therefore governed by
\begin{equation}
\dot{x}^i
=
-
G^{ij}(x)\,\partial_j \Phi(x)
+
R^i(x)
+
\xi^i(t),
\label{eq:unified_index}
\end{equation}
where \(\xi^i(t)\) denotes stochastic fluctuations satisfying
\begin{equation}
\langle
\xi_i(t)\xi_j(t')
\rangle
=
2D\,\delta_{ij}\delta(t-t').
\label{eq:noise}
\end{equation}

Equation~\eqref{eq:unified_index} therefore represents the most general
geometrically consistent collective flow compatible with a Riemannian
state-space structure.
The gradient sector $-G^{-1}\nabla\Phi$
implements homeostatic stabilization and relaxation toward dynamically
preferred manifolds, while the non-conservative flow $R(x)$ generates
tangential circulation on the cognitive manifold.

This distinction is fundamental.
Pure gradient systems ultimately collapse toward static attractors and
therefore realize only generalized optimization dynamics.
By contrast, the non-conservative circulation sector permits
persistent collective motion without destroying overall stability.
The resulting dynamics supports the existence of long-lived collective
modes and extended exploration of the cognitive manifold beyond simple
gradient optimization.

The metric \(G(x)\) plays an essential dynamical role.
Rather than acting merely as a weighting factor, the metric determines
how distances, sensitivities, and relaxation rates are distributed
across the collective state space.
Different directions may therefore possess strongly different effective
stabilities and characteristic time scales.
Learning and adaptation continuously reshape this geometry, thereby
modulating which collective trajectories become dynamically accessible
or protected.

This geometric decomposition also clarifies the relationship between
the present framework and conventional stochastic dynamics.
Ordinary Langevin dynamics corresponds to the special Euclidean limit
defined by \(G=I\) and \(R=0\), for which the dynamics reduces to
simple noisy gradient relaxation.
The present theory extends this structure to a fully geometric
collective field dynamics incorporating adaptive state-space geometry
together with non-conservative circulation.

The collective cognitive dynamics thus represents the most general
geometrically consistent form of a cognitive field on a Riemannian
state manifold.
Rather than being postulated phenomenologically, the structure of the
equation follows naturally from the geometry of collective state space
itself.

\vspace{6pt}
\emph{MSRJD field-theoretic formulation}.
Equation~\eqref{eq:unified_index} admits a MSRJD path-integral representation.
Introducing response fields \(\tilde x(t)\), the generating functional
is
\begin{equation}
Z
=
\int
\mathcal D x\,
\mathcal D \tilde x\;
e^{-S[x,\tilde x]},
\end{equation}
with the action
\begin{equation}
S[x,\tilde x]
=
\int dt\,
\Big[
\tilde x\cdot
\big(
\dot x
+
G^{-1}\nabla_x\Phi
-
R
\big)
-
D\tilde x^2
\Big].
\label{eq:action}
\end{equation}

This formulation provides a unified description of both average
collective inference trajectories and the emergent collective dynamics
arising from stochastic fluctuations and recurrent response feedback.
Saddle points of the action correspond to dominant collective
state-space trajectories, while fluctuation-driven response dynamics around these
trajectories generates infrared collective organization, metastable
manifolds, long-lived collective modes, and self-organized
near-critical behavior.

The MSRJD formalism therefore establishes a direct connection between
collective geometry, stochastic dynamics, non-conservative circulation,
and emergent collective cognitive dynamics within a unified
field-theoretic framework.

\subsection{B. Coupled cognitive field dynamics and memory kernel}

The collective cognitive dynamics introduced above contains a large
number of internal degrees of freedom evolving across widely separated
temporal scales.
Rather than being governed by a finite set of isolated attractors, the
protected near-critical regime is characterized by an extended
continuum of slow collective relaxation processes.

In this perspective, the emergence of long-time memory does
not require an additional phenomenological assumption.
Instead, it follows naturally from projecting the full collective
dynamics onto slowly evolving infrared persistence sectors generated
by recursive collective organization.

To make this structure explicit, we consider the collective cognitive dynamics
introduced above,
\begin{equation}
\dot x
=
-
G^{-1}(x)\nabla_x\Phi(x)
+
R(x)
+
\xi(t).
\label{eq:full_cognitive_dyn}
\end{equation}
The full cognitive dynamics generally contains recurrent circulatory
components generated by the nonconservative collective flow \(R(x)\).
The corresponding local stability operator is therefore generically
non-Hermitian and may possess complex collective eigenmodes associated
with weakly damped recurrent collective trajectories.

To characterize the local stability structure of the collective
dynamics, we consider small fluctuations around an evolving trajectory
\(x^\ast(t)\),
\(
\delta x=x-x^\ast
\),
whose linearized dynamics obeys
\begin{equation}
\delta\dot x
=
-
J\,\delta x
+
\eta ,
\label{eq:linearized}
\end{equation}
with
\begin{equation}
J_{ij}
=
\left.
\frac{\partial}{\partial x_j}
\left[
G^{-1}\nabla_x\Phi
-
R
\right]_i
\right|_{x^\ast}.
\end{equation}

The Jacobian spectrum therefore characterizes the local relaxation and
stability structure of perturbations around the evolving
collective trajectory.
Introducing right eigenvectors \(u_\alpha\),
\begin{equation}
J u_\alpha
=
\mu_\alpha u_\alpha,
\end{equation}
the perturbations evolve according to
\begin{equation}
u_\alpha(t)
\sim
e^{-\mu_\alpha t}.
\end{equation}

Because the recurrent flow \(R(x)\) generally breaks detailed balance,
the local stability operator need not be symmetric.
The collective spectrum may therefore acquire a complex structure,
\begin{equation}
\mu_\alpha
=
\lambda_\alpha
+
i\omega_\alpha.
\label{eq:complex_collective_spectrum}
\end{equation}
The corresponding collective perturbation evolves as
\begin{equation}
u_\alpha(t)
\sim
e^{-\lambda_\alpha t}
e^{-i\omega_\alpha t},
\end{equation}
showing that each collective mode possesses both a characteristic
relaxation timescale \(\lambda_\alpha^{-1}\) and a circulation
timescale \(\omega_\alpha^{-1}\).

Large relaxation rates correspond to rapidly stabilized collective
modes, whereas weakly damped modes remain dynamically active over
extended temporal intervals.
The non-conservative circulation field \(R(x)\) contributes to the
imaginary part of the collective spectrum and generates temporal phase
evolution of the collective modes.
Consequently, collective activity is organized not only by relaxation
toward stable configurations but also by recurrent phase dynamics
within the slow-mode sector.

When weakly damped collective modes accumulate near
\(\lambda_\alpha\rightarrow0\), the system develops a broad hierarchy
of long-lived collective modes.
The circulation sector provides an additional temporal organization of
these persistent modes, enriching the collective dynamics over
extended timescales.

The full collective state evolves in a very high-dimensional space.
However, long-time collective trajectories do not explore all
directions equally.
Strongly damped microscopic sectors rapidly decay, while weakly damped
recurrent collective sectors remain dynamically active over long
timescales.
Correlated recurrent dynamics therefore organizes the infrared
evolution near a restricted low-dimensional collective manifold
\[
\mathcal M_{\rm slow}\subset\mathbb R^N,
\]
embedded within the full cognitive state space.
Recent studies of large-scale neural population dynamics further
suggest that cognition and behavior are organized on low-dimensional
collective manifolds embedded within high-dimensional neural state
spaces \cite{34,35,36,37}.

The observable collective trajectory consequently remains dynamically
confined near a reduced infrared manifold generated by weak damping,
collective circulation, and repeated trajectory
reinforcement.
The slow manifold should therefore not be interpreted as a purely
kinematic geometric constraint.
Instead, it emerges dynamically from the persistent infrared
organization of collective modes.

This local manifold stability structure naturally motivates the
emergence of an effective slow collective sector.
Rather than tracking all microscopic collective coordinates
individually, it becomes natural to separate the infrared dynamics
into an observable collective field and a latent sector representing
long-lived slow relaxation modes of the learned cognitive geometry.

We therefore decompose the collective state into
\begin{equation}
x(t)
\equiv
\bigl(
\phi(t),X(t)
\bigr),
\end{equation}
where \(\phi(t)\) denotes the observable cognitive field, while
\(X(t)\) represents latent collective-mode amplitudes associated with
weakly damped recurrent collective modes of the learned cognitive
geometry.

The present infrared theory should therefore be interpreted as an
effective coarse-grained description obtained after projecting the
full recurrent cognitive dynamics onto slowly evolving collective
persistence sectors.
Rapid microscopic components are effectively integrated out, while the
remaining weakly damped infrared sectors are represented as
latent collective modes characterized by a complex collective spectrum.
These modes retain both long relaxation times and recursive
circulation dynamics, thereby governing the long-time temporal
organization of cognition.

The linearized dynamics then assumes the block form
\begin{equation}
\frac{d}{dt}
\begin{pmatrix}
\phi\\
X
\end{pmatrix}
=
-
\begin{pmatrix}
M_{\phi\phi} & M_{\phi X}\\
M_{X\phi} & M_{XX}
\end{pmatrix}
\begin{pmatrix}
\phi\\
X
\end{pmatrix}
+
\begin{pmatrix}
\eta_\phi\\
\eta_X
\end{pmatrix}.
\label{eq:block_dynamics}
\end{equation}
Equivalently,
\begin{align}
\partial_t\phi
&=
-r\phi
-
M_{\phi X}X
+
\eta_\phi,
\label{eq:phi_eq}
\\
\partial_t X
&=
-M_{XX}X
-
M_{X\phi}\phi
+
\eta_X .
\label{eq:X_eq}
\end{align}
Equation~\eqref{eq:X_eq} may be solved formally,
\begin{equation}
\begin{aligned}
X(t)
=\,
&e^{-M_{XX}(t-t_0)}X(t_0)
\\
&-
\int_{t_0}^{t}dt'\,
e^{-M_{XX}(t-t')}
[M_{X\phi}\phi(t')
+\eta_X(t')],
\label{eq:X_formal}
\end{aligned}
\end{equation}
where \(\eta\) indicates stochastic noise contributions.

Substituting Eq.~\eqref{eq:X_formal} into
Eq.~\eqref{eq:phi_eq} yields an effective equation for the observed
cognitive field,
\begin{equation}
\partial_t\phi(t)
=
-r\phi(t)
+
\int_{t_0}^{t}dt'\,
K(t-t')\phi(t')
+
\eta_{\rm eff}(t),
\label{eq:memory_eq}
\end{equation}
with the memory kernel
\begin{equation}
K(t-t')
=
M_{\phi X}
e^{-M_{XX}(t-t')}
M_{X\phi}.
\label{eq:memory_matrix}
\end{equation}
Equation~\eqref{eq:memory_eq} demonstrates that the effective infrared
dynamics becomes intrinsically non-Markovian once the latent
collective sector is eliminated.
The memory kernel is therefore not an additional phenomenological
construction, but a direct consequence of projecting the full
collective dynamics onto a reduced observable sector.

The observable cognitive field \(\phi(t)\) represents the resulting
macroscopic collective organization generated by these interacting
slow collective sectors.
Cognition therefore proceeds not through the motion of a single point
along a fixed trajectory, but through the continuous reorganization of
collective mode amplitudes across the learned slow-mode manifold.

To expose the spectral structure explicitly, we diagonalize the latent
relaxation operator,
\begin{equation}
M_{XX}u_\alpha
=
(\lambda_\alpha+i\omega_\alpha)u_\alpha,
\label{eq:latent_complex_modes}
\end{equation}
where \(\lambda_\alpha\) characterizes the relaxation (forgetting)
rate of the collective mode and \(\omega_\alpha\) characterizes its
recursive circulation frequency.

The propagator of the latent sector may then be expanded as
\begin{equation}
e^{-M_{XX}t}
=
\sum_\alpha
e^{-(\lambda_\alpha+i\omega_\alpha)t}
|u_\alpha\rangle
\langle \tilde u_\alpha |,
\label{eq:spectral_expansion}
\end{equation}
where \(\langle\tilde u_\alpha|\) denotes the corresponding left
eigenvector.

The memory kernel consequently assumes the form
\begin{equation}
K(t)
=
\sum_\alpha
g_\alpha^2
e^{-(\lambda_\alpha+i\omega_\alpha)t},
\label{eq:kernel_discrete}
\end{equation}
where the effective couplings \(g_\alpha\) encode the overlap between
the observable infrared field and the latent collective eigenmodes.
In general, these couplings may vary among modes.
However, in the infrared effective description considered here,
their microscopic variation is assumed to be weak compared with the
collective reorganization of the relaxation spectrum.
The couplings may therefore be approximated by a renormalized
effective constant,
\(
g_\alpha\simeq g,
\)
so that the nontrivial many-body organization is encoded primarily in
the distribution of relaxation modes itself.

In the thermodynamic or continuum limit, the collective spectrum
becomes dense and the kernel may be written as
\begin{equation}
K(t)
=
\int_0^\Lambda d\lambda
\int_{-\infty}^{\infty} d\omega\;
\rho(\lambda,\omega)
e^{-(\lambda+i\omega)t},
\label{eq:memory_kernel}
\end{equation}
where
\begin{equation}
\rho(\lambda,\omega)
=
\sum_\alpha
g^2
\delta(\lambda-\lambda_\alpha)
\delta(\omega-\omega_\alpha)
\label{eq:tdos}
\end{equation}
defines the effective time-scale density of states.

The TDOS characterizes the spectral organization of collective
modes governing the infrared temporal structure of cognition.
Equation~\eqref{eq:memory_kernel} shows that long-time collective
memory emerges directly from the accumulation of weakly damped latent
collective modes.
Here, \(\Lambda\) denotes the ultraviolet cutoff of the collective
spectrum, corresponding to the fastest collective modes retained in
the effective description.

The real part of the spectrum controls memory persistence and
forgetting,
\begin{equation}
\tau_\alpha
=
\lambda_\alpha^{-1},
\end{equation}
whereas the imaginary part introduces a characteristic circulation
timescale,
\begin{equation}
T_\alpha
=
2\pi \omega_\alpha^{-1}.
\end{equation}
The infrared collective dynamics is therefore governed not merely by a
distribution of relaxation rates but by a complex collective spectrum
containing both persistence scales and circulation scales.
The accumulation of weakly damped collective modes generates the
memory kernel and the associated long-time temporal persistence of
the cognitive field, while the circulation sector provides an
additional dynamical structure that plays an important role in the
collective organization.

Fourier transformation of the memory kernel generates the retarded
memory self-energy,
\begin{equation}
\Sigma_R(\Omega)
=
\int_0^\Lambda d\lambda
\int_{-\infty}^{\infty} d\omega\;
\frac{\rho(\lambda,\omega)}
{\lambda-i(\Omega-\omega)} .
\label{eq:selfenergy}
\end{equation}
The corresponding quadratic infrared response kernel becomes
\begin{equation}
\Gamma_2^R(\Omega)
=
-i\Omega+r-\Sigma_R(\Omega).
\label{eq:Gamma2R}
\end{equation}
Both memory persistence and
recursive circulation contribute to the infrared response through the
same complex collective spectrum.
The observable cognitive dynamics is therefore governed not by a
single relaxation timescale but by a continuum of collective modes
characterized by both forgetting rates and recursive circulation
frequencies.

Although the full collective spectrum is labeled by both
\(\lambda\) and \(\omega\), the long-time decay of the memory kernel is
controlled primarily by the relaxation rates \(\lambda\).
We therefore define the relaxation TDOS by projecting the full
spectral density onto the \(\lambda\)-axis,
\begin{equation}
\rho(\lambda)
=
\int_{-\infty}^{\infty} d\omega\,
\rho(\lambda,\omega)
=
\sum_\alpha
g_\alpha^2
\delta(\lambda-\lambda_\alpha).
\label{eq:rho_lambda_proj}
\end{equation}
This projected TDOS counts the total spectral weight of modes with
relaxation rate \(\lambda\), irrespective of their circulation
frequencies.

In particular, if the effective TDOS remains finite near the infrared
sector,
\begin{equation}
\rho(\lambda)
\xrightarrow{\lambda\to0}
\rho_0,
\end{equation}
the memory kernel develops the universal long-time tail
\begin{equation}
K(t)
\sim
\frac{\rho_0}{t},
\qquad
(t\gg\Lambda^{-1}),
\end{equation}
demonstrating that scale-free collective memory emerges directly from
the accumulation of slow collective modes.

The corresponding retarded self-energy acquires the infrared form
\begin{equation}
\Sigma_R(\Omega)
\simeq
\rho_0
\ln\!\left(
\frac{\Lambda}{|\Omega|}
\right)
+
i\frac{\pi}{2}\rho_0\,{\rm sgn}(\Omega),
\label{eq:Sigma_scaling}
\end{equation}
which characterizes the memory-dominated critical regime generated by
the infrared accumulation of collective modes.

The explicit derivation of the retarded memory self-energy and its
infrared asymptotics is presented in Appendix~A.
Appendix~B provides a complementary time-domain derivation obtained by
explicitly integrating out the latent collective sector.

\subsection{C. Formation of the macroscopic cognitive field from recursive memory feedback}

We now formulate the infrared response theory of the collective
cognitive field.
The central physical mechanism is the competition between intrinsic
homeostatic forgetting and self-consistent recursive memory feedback
generated by distributed slow collective modes.
Importantly,
the present framework is not intended to describe a conventional equilibrium ordering transition.
Instead, cognition operates in a nonequilibrium protected near-critical regime generated 
by recursive memory feedback.

We therefore consider a strongly dissipative collective cognitive
field \(\phi(t)\) coupled to a continuum of latent collective modes
characterized by both relaxation rates and circulation frequencies,
\begin{align}
\partial_t \phi(t)
=
-r\,\phi(t)
+
\int d\lambda\, d\omega\;
g(\lambda,\omega)
X_{\lambda,\omega}(t)
+
h(t)
+
\eta_\phi(t),
\\
\partial_t X_{\lambda,\omega}(t)
=
-(\lambda+i\omega)
X_{\lambda,\omega}(t)
+
g(\lambda,\omega)\phi(t)
+
\eta_{X}(t).
\end{align}
Here \(r>0\) denotes the bare local forgetting rate of the observable
cognitive field, while \(h(t)\) represents an external perturbation or
input signal used to probe the collective response of the system.
The observable field therefore tends to relax locally with rate \(r\),
whereas the slow collective relaxation spectrum continuously provides
long-timescale feedback to the observable cognitive field.
The competition between intrinsic forgetting, external driving, and
memory-mediated feedback ultimately determines the infrared collective
response of the cognitive field.

The latent collective modes are generally characterized by both a
relaxation timescale \(\lambda^{-1}\) and a circulation timescale
\(\omega^{-1}\).
Integrating out these modes generates a memory kernel of the form
\begin{equation}
K(t)
=
\Theta(t)
\int d\lambda\,d\omega\;
\rho(\lambda,\omega)
e^{-\lambda t}
e^{-i\omega t}.
\label{eq:complex_kernel}
\end{equation}
The relaxation sector governs memory persistence and the accumulation
of recursive memory feedback, whereas the circulation sector provides
the temporal organization of this collective dynamics.

Integrating out the latent variables within the Gaussian MSRJD
formalism yields an effective nonlocal action for
\((\phi,\tilde\phi)\),
\begin{align}
S_{\rm eff}
&=
\int dt\,
\tilde\phi(t)
\left[
\partial_t+r
\right]\phi(t)
-
\int dt\,D_\phi\tilde\phi(t)^2
\nonumber\\
&\quad
-
\int dt\,dt'\,
\tilde\phi(t)
K(t-t')
\phi(t')
+
\cdots .
\label{eq:Seff_time}
\end{align}
The Fourier transform of the memory kernel defines the retarded memory
self-energy,
\begin{equation}
\Sigma_R(\Omega)
=
\int d\lambda\,d\omega\;
\frac{\rho(\lambda,\omega)}
{\lambda-i(\Omega-\omega)}.
\label{eq:selfenergy_complex}
\end{equation}
Consequently, the effective action in frequency space becomes
\begin{equation}
S_{\rm eff}
=
\int\frac{d\Omega}{2\pi}
\tilde\phi(-\Omega)
\left[
-i\Omega+r-\Sigma_R(\Omega)
\right]
\phi(\Omega)
+
\cdots .
\label{eq:Seff_freq_sigma}
\end{equation}

Within the MSRJD formalism, the linear response of the observable
cognitive field is
\begin{equation}
\chi_R(\Omega)
=
\frac{\delta\langle\phi(\Omega)\rangle}
{\delta h(\Omega)}
=
\langle\phi(\Omega)\tilde\phi(-\Omega)\rangle,
\end{equation}
which yields the retarded susceptibility
\begin{equation}
\chi_R(\Omega)
=
\frac{1}
{-i\Omega+r-\Sigma_R(\Omega)}.
\label{eq:chiR}
\end{equation}
The corresponding inverse susceptibility kernel is
\[
\Gamma_2^R(\Omega)
=
-i\Omega+r-\Sigma_R(\Omega).
\]

The static inverse response is obtained from the zero-frequency limit,
\begin{equation}
r_{\rm cog}
\equiv
\Gamma_2^R(0)
=
r-\Sigma_R(0),
\label{eq:rcog_def}
\end{equation}
which defines the cognitive forgetting gap.

The same collective response admits an equivalent dynamical
interpretation in terms of repeated recursive memory interactions.
Within linear response, the observable cognitive field undergoes a
sequence of repeated memory-mediated feedback processes,
\begin{equation}
\phi
=
L_0h
+
L_0\Sigma_R L_0h
+
L_0\Sigma_R L_0\Sigma_R L_0h
+
\cdots ,
\label{...}
\end{equation}
where
\[
L_0(\Omega)
=
\frac{1}{-i\Omega+r}
\]
denotes the bare response of the observable cognitive field.

Summing this Dyson series gives
\begin{equation}
\phi(\Omega)
=
\frac{L_0(\Omega)}
{1-L_0(\Omega)\Sigma_R(\Omega)}
h(\Omega)
=
\chi_R(\Omega)h(\Omega),
\end{equation}
which exactly reproduces the response function derived from the MSRJD
formalism.
The memory self-energy therefore represents the cumulative effect of
self-consistent recursive memory feedback generated by the distributed
relaxation spectrum.

As the density of slow collective modes increases, the infrared
spectral weight of the TDOS grows, progressively enhancing the memory
self-energy and suppressing the effective forgetting gap.
The system therefore approaches the critical condition
\begin{equation}
r_{\rm cog}
\rightarrow
0^+.
\label{eq:rcog_softening}
\end{equation}

Consequently, the static cognitive susceptibility diverges,
\begin{equation}
\chi_R(0)
=
\frac{1}{r_{\rm cog}}
\rightarrow
\infty,
\end{equation}
while the characteristic collective response time obeys
\begin{equation}
\tau_{\rm cog}
\sim
\frac{1}{r_{\rm cog}},
\end{equation}
and therefore grows without bound as
\(
r_{\rm cog}\rightarrow0^+.
\)

The divergence of the collective susceptibility signals the emergence
of a macroscopic memory-dressed cognitive field generated through
self-consistent recursive memory feedback among distributed relaxation
modes rather than by any individual relaxation mode.

Beyond this collective transition, the macroscopic cognitive field
constitutes the dynamical substrate of cognition.
The relaxation spectrum governs the temporal persistence of the field,
whereas the circulation spectrum provides an additional temporal
organization that will be discussed in the following subsection.

In practice, biological and artificial cognitive systems are expected
to operate as stable memory-dressed collective states maintained close
to, but not exactly at, the critical boundary.
Accordingly,
\begin{equation}
0<r_{\rm cog}\ll\Lambda,
\end{equation}
defines a protected operating regime in which the macroscopic
cognitive field remains stable while preserving strong collective
responsiveness, contextual continuity, and adaptive reasoning.

For the infrared analysis through the projected relaxation TDOS $\rho(\lambda)$,
the static memory self-energy may then be written as
\begin{equation}
\Sigma_R(0)
\simeq
\int_{\lambda_{\rm IR}}^\Lambda
d\lambda\,
\frac{\rho(\lambda)}
{\lambda},
\end{equation}
so that the cognitive forgetting gap becomes
\begin{equation}
r_{\rm cog}
=
r
-
\int_{\lambda_{\rm IR}}^\Lambda
d\lambda\,
\frac{\rho(\lambda)}
{\lambda}.
\label{eq:rcog_tdos}
\end{equation}

For a nearly flat infrared TDOS,
\begin{equation}
\rho(\lambda)
\simeq
\rho_0,
\qquad
\lambda_{\rm IR}\ll\lambda\ll\Lambda,
\end{equation}
one obtains
\begin{equation}
\Sigma_R(0)
=
\rho_0
\ln\frac{\Lambda}{\lambda_{\rm IR}},
\end{equation}
and therefore
\begin{equation}
r_{\rm cog}
=
r
-
\rho_0
\ln\frac{\Lambda}{\lambda_{\rm IR}}.
\label{eq:rcog_flat}
\end{equation}
A broad and nearly flat infrared TDOS thus logarithmically suppresses
the cognitive forgetting gap through recursive memory feedback.

\begin{figure*}[t]
\centering
\includegraphics[width=1.0\textwidth, trim=0cm 15.5cm 0cm 0cm]{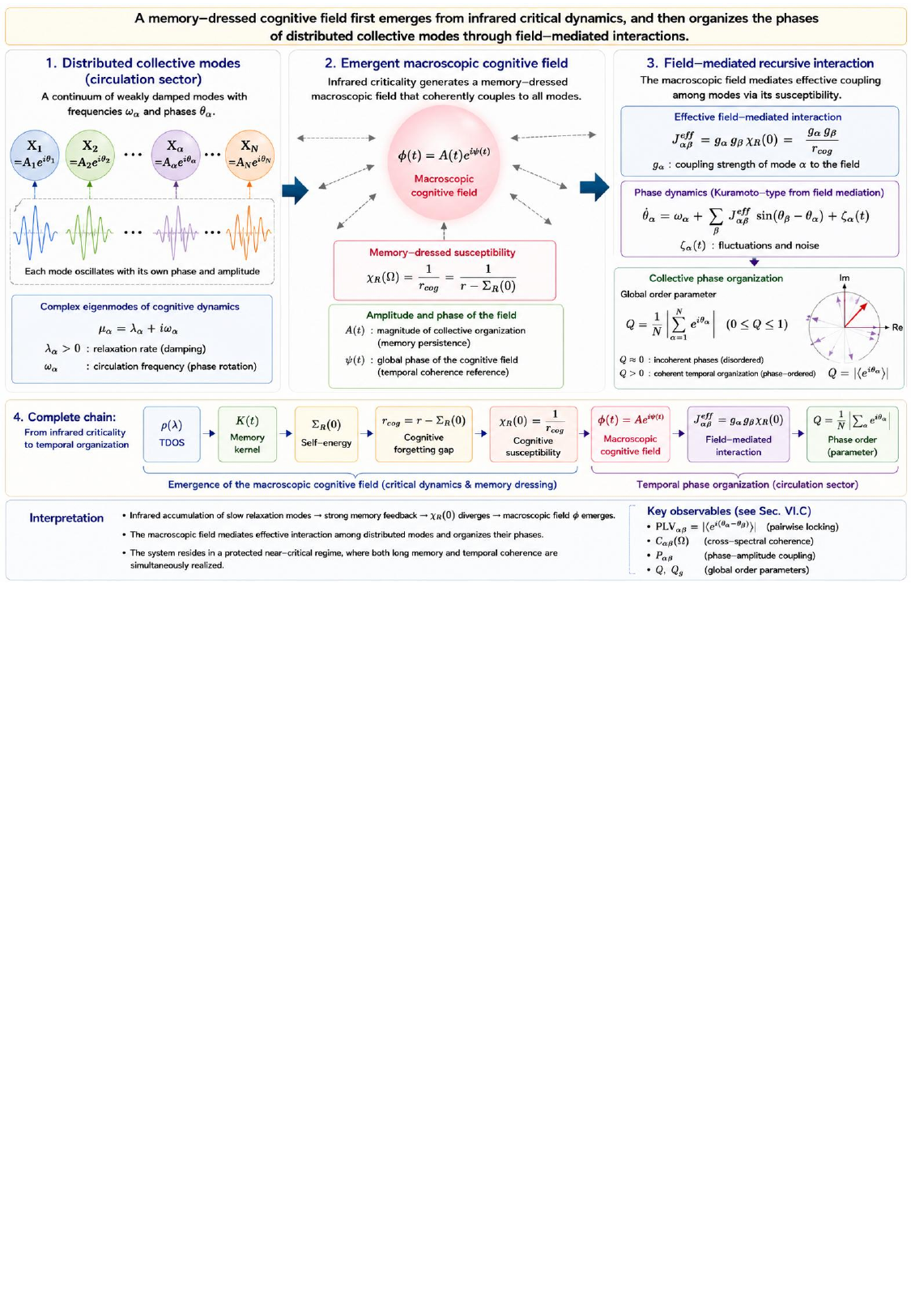}
\caption{
Recursive field-mediated phase organization of the macroscopic
cognitive field.
Distributed collective modes with complex amplitudes
$X_\alpha=A_\alpha e^{i\theta_\alpha}$ interact recursively through the
memory-dressed macroscopic cognitive field
$\phi(t)=A(t)e^{i\psi(t)}$.
Integrating out the macroscopic field generates a non-Markovian
field-mediated interaction among the distributed modes, whose
infrared limit reduces to an effective phase-coupling equation.
Recursive memory dressing simultaneously enhances the cognitive
susceptibility,
$\chi_R(0)=1/r_{\rm cog}$,
and the effective field-mediated interaction,
$\mathcal J_{\alpha\beta}^{\rm eff}\propto\chi_R(0)$,
thereby promoting collective temporal organization characterized by
the global phase-order parameter $Q$.
The diagram summarizes the complete self-consistent recursive
organization linking the relaxation sector, memory dressing, and the
circulation sector within the complex cognitive field.
}
\end{figure*}

\subsection{D. Temporal phase organization of the macroscopic cognitive field}

The preceding subsection established that the macroscopic cognitive
field emerges through self-consistent recursive memory feedback among
distributed relaxation modes.
The remaining question concerns the internal temporal organization of
this collective field.
The existence of a macroscopic cognitive field alone does not imply
that the underlying distributed modes evolve coherently in time.
A further organizing mechanism is required to establish stable
temporal relations across the collective spectrum.

This organization originates from the circulation sector of the
learned cognitive geometry.
While the relaxation spectrum governs the temporal persistence of the
macroscopic cognitive field, the circulation spectrum governs its
collective phase dynamics.
Each distributed collective mode is characterized by a complex
stability eigenvalue
\[
\mu_\alpha
=
\lambda_\alpha
+
i\omega_\alpha,
\label{eq:complex_collective_eigenvalue}
\]
where the circulation frequency
\(
\omega_\alpha
\)
governs the rate of temporal phase evolution.

The macroscopic cognitive field therefore naturally admits a complex
representation,
\begin{equation}
\phi(t)
=
A(t)e^{i\psi(t)},
\label{eq:cognitive_order_parameter}
\end{equation}
where the amplitude \(A\) measures the strength of the memory-dressed
collective field generated through recursive feedback, while the phase
\(\psi\) represents its global temporal organization.

The complex amplitudes associated with individual collective modes may
likewise be expressed as
\begin{equation}
X_\alpha(t)
=
A_\alpha(t)e^{-i\theta_\alpha(t)}.
\label{eq:mode_amplitude_phase_original}
\end{equation}
Since the modal amplitudes are dynamically generated by the relaxation dynamics, 
they inherit the infrared organization encoded in the TDOS. 
The amplitude weighting therefore does not introduce an independent phenomenological parameter 
but simply reflects the persistence of the underlying slow relaxation modes.

For the standard phase-difference representation, it is convenient to
introduce
\begin{equation}
\vartheta_\alpha
\equiv
-\theta_\alpha,
\qquad
X_\alpha
=
A_\alpha e^{i\vartheta_\alpha}.
\label{eq:standard_phase_convention}
\end{equation}
This change is purely conventional and does not modify the underlying
dynamics.
If an uncoupled mode evolves as
\(
X_\alpha\sim
e^{-(\lambda_\alpha+i\omega_\alpha)t},
\)
its natural phase frequency in the convention of
Eq.~(\ref{eq:standard_phase_convention}) is
\begin{equation}
\nu_\alpha
\equiv
-\omega_\alpha.
\label{eq:natural_phase_frequency}
\end{equation}

The recursive coupling between the macroscopic cognitive field and
the distributed collective modes generates phase interaction
directly.
The deterministic linear sector of the coupled cognitive-field
dynamics may be written schematically as
\begin{align}
\dot{\phi}
&=
-r\phi
+
\sum_\alpha g_\alpha X_\alpha,
\label{eq:phase_coupled_field}
\\
\dot{X}_\alpha
&=
-
(\lambda_\alpha+i\omega_\alpha)X_\alpha
+
g_\alpha\phi,
\label{eq:phase_coupled_modes}
\end{align}
where \(g_\alpha\) denotes the coupling between the macroscopic field
and the \(\alpha\)-th collective mode.
Substitution of
Eqs.~(\ref{eq:cognitive_order_parameter}) and
(\ref{eq:standard_phase_convention}) into
Eqs.~(\ref{eq:phase_coupled_field}) and
(\ref{eq:phase_coupled_modes}) gives the leading amplitude equations
\begin{align}
\dot A
&=
-rA
+
\sum_\alpha
g_\alpha A_\alpha
\cos(\vartheta_\alpha-\psi),
\label{eq:macroscopic_amplitude_dynamics}
\\
\dot A_\alpha
&=
-\lambda_\alpha A_\alpha
+
g_\alpha A
\cos(\psi-\vartheta_\alpha),
\label{eq:modal_amplitude_dynamics}
\end{align}
and the corresponding phase equations
\begin{align}
\dot\psi
&=
\frac{1}{A}
\sum_\alpha
g_\alpha A_\alpha
\sin(\vartheta_\alpha-\psi),
\label{eq:macroscopic_phase_dynamics}
\\
\dot\vartheta_\alpha
&=
\nu_\alpha
+
g_\alpha
\frac{A}{A_\alpha}
\sin(\psi-\vartheta_\alpha).
\label{eq:modal_phase_feedback}
\end{align}

Equation~(\ref{eq:modal_phase_feedback}) shows that the macroscopic
cognitive phase is not merely a passive average of the distributed
modes.
It acts as a collective organizing field that continuously entrains
their phases.
At the same time,
Eq.~(\ref{eq:macroscopic_phase_dynamics}) shows that the global phase
is itself generated by the distributed modes.
Temporal coherence is therefore formed through the self-consistent
recursive relation
\begin{equation}
\{X_\alpha\}
\longrightarrow
\phi
\longrightarrow
\{X_\alpha\},
\label{eq:recursive_phase_feedback_loop}
\end{equation}
rather than through an externally imposed oscillatory reference.

The relative phase between the \(\alpha\)-th collective mode and the
macroscopic cognitive field is
\begin{equation}
\delta_\alpha
=
\vartheta_\alpha-\psi.
\label{eq:relative_cognitive_phase}
\end{equation}
Using
Eqs.~(\ref{eq:macroscopic_phase_dynamics}) and
(\ref{eq:modal_phase_feedback}), its dynamics becomes
\begin{equation}
\dot\delta_\alpha
=
\nu_\alpha
-
g_\alpha
\frac{A}{A_\alpha}
\sin\delta_\alpha
-
\frac{1}{A}
\sum_\beta
g_\beta A_\beta
\sin\delta_\beta.
\label{eq:relative_phase_dynamics}
\end{equation}
Phase locking occurs when the relative phases approach stationary
values,
\begin{equation}
\delta_\alpha(t)
\longrightarrow
\delta_\alpha^*,
\qquad
\dot\delta_\alpha
=
0.
\label{eq:phase_locking_condition}
\end{equation}
The locked modes then rotate with a common collective frequency,
\begin{equation}
\dot\vartheta_\alpha
=
\dot\psi
=
\Omega_{\rm cog},
\label{eq:collective_cognitive_frequency}
\end{equation}
while retaining mode-dependent stationary phase offsets
\(
\delta_\alpha^*
\).
The corresponding locking condition may be written as
\begin{equation}
\sin\delta_\alpha^*
=
\frac{
\nu_\alpha-\Omega_{\rm cog}
}{
g_\alpha A/A_\alpha
},
\label{eq:phase_locking_relation}
\end{equation}
which requires
\begin{equation}
\left|
\nu_\alpha-\Omega_{\rm cog}
\right|
\leq
g_\alpha
\frac{A}{A_\alpha}.
\label{eq:phase_locking_inequality}
\end{equation}
Thus, only modes whose intrinsic circulation frequencies lie within
the effective entrainment range of the macroscopic field become
phase locked.
Other modes may remain incoherent or intermittently synchronized.

The collective interaction may also be expressed directly among the
distributed modes.
Formally integrating
Eq.~(\ref{eq:phase_coupled_field}) gives
\begin{equation}
\phi(t)
=
\phi(0)e^{-rt}
+
\sum_\beta
g_\beta
\int_0^t
dt'\,
L_0(t-t')X_\beta(t'),
\label{eq:formal_cognitive_field_solution}
\end{equation}
where
\begin{equation}
L_0(t)
=
\Theta(t)e^{-rt}
\label{eq:bare_cognitive_propagator}
\end{equation}
is the bare retarded propagator of the macroscopic field.
Substitution into the mode equation produces an effective
field-mediated interaction,
\begin{equation}
\dot X_\alpha(t)
=
-
(\lambda_\alpha+i\omega_\alpha)X_\alpha(t)
+
\sum_\beta
g_\alpha g_\beta
\int_0^t
dt'\,
L_0(t-t')X_\beta(t').
\label{eq:field_mediated_mode_interaction}
\end{equation}
The associated phase equation takes the generalized non-Markovian
form
\begin{equation}
\dot\vartheta_\alpha(t)
=
\nu_\alpha
+
\sum_\beta
\int_0^t
dt'\,
\mathcal L_{\alpha\beta}(t-t')
\frac{A_\beta(t')}{A_\alpha(t)}
\sin\!\left[
\vartheta_\beta(t')
-
\vartheta_\alpha(t)
\right],
\label{eq:memory_mediated_phase_equation}
\end{equation}
where
\begin{equation}
\mathcal L_{\alpha\beta}(t)
=
g_\alpha g_\beta L_0(t)
\label{eq:phase_interaction_kernel}
\end{equation}
is the retarded phase-interaction kernel.
Equation~(\ref{eq:memory_mediated_phase_equation}) shows that the
phase of each collective mode is reorganized not only by the present
phases of the other modes but also by their temporally retained
history.

In the slow-amplitude and infrared limit,
\begin{equation}
A_\beta(t')
\simeq
A_\beta(t),
\qquad
\vartheta_\beta(t')
\simeq
\vartheta_\beta(t),
\label{eq:slow_phase_approximation}
\end{equation}
the nonlocal phase equation reduces to
\begin{equation}
\dot{\vartheta}_\alpha
=
\nu_\alpha
+
\sum_\beta
\mathcal J_{\alpha\beta}
\sin
\!\left(
\vartheta_\beta-\vartheta_\alpha
\right),
\label{eq:effective_phase_equation}
\end{equation}
where the effective field-mediated phase interaction is
\begin{equation}
\mathcal J_{\alpha\beta}
=
\frac{A_\beta}{A_\alpha}
\int_0^\infty
d\tau\,
\mathcal L_{\alpha\beta}(\tau).
\label{eq:phase_kernel_ir}
\end{equation}

The effective interaction admits a particularly transparent
interpretation after eliminating the macroscopic cognitive field.
In the frequency domain,
\begin{equation}
\phi(\Omega)
=
\chi_R(\Omega)
\sum_\beta
g_\beta
X_\beta(\Omega),
\label{eq:phase_response_relation}
\end{equation}
where
\(
\chi_R(\Omega)
\)
is the memory-dressed cognitive susceptibility derived in
Sec.~II.C.
Substituting Eq.~(\ref{eq:phase_response_relation}) into the collective
mode equation yields the effective field-mediated interaction
\begin{equation}
\mathcal J_{\alpha\beta}^{\rm eff}(\Omega)
=
g_\alpha
g_\beta
\chi_R(\Omega).
\label{eq:effective_phase_interaction}
\end{equation}

In the infrared static limit,
\begin{equation}
\chi_R(0)
=
\frac{1}{r_{\rm cog}},
\qquad
r_{\rm cog}
=
r-\Sigma_R(0),
\label{eq:chi_phase_static}
\end{equation}
so that
\begin{equation}
\mathcal J_{\rm eff}
\propto
\chi_R(0)
=
\frac{1}{r_{\rm cog}}.
\label{eq:phase_coupling_strength}
\end{equation}

Equation~(\ref{eq:phase_coupling_strength}) establishes the direct
connection between the relaxation and circulation sectors of the
cognitive field.
Recursive memory feedback suppresses the cognitive forgetting gap,
thereby enhancing the macroscopic cognitive susceptibility.
The enhanced susceptibility strengthens the effective interaction
among the distributed collective modes and consequently promotes
collective phase locking.

The global degree of collective temporal organization is described by
the phase order parameter
\begin{equation}
Q
e^{i\Psi}
=
\frac1N
\sum_{\alpha}
e^{i\vartheta_\alpha},
\label{eq:phase_order_parameter}
\end{equation}
where
\(Q\)
quantifies the overall degree of collective phase coherence and
\(\Psi\)
denotes the mean collective phase.
For
\(
Q\simeq0,
\)
the collective modes remain temporally disordered,
whereas
\(
Q>0
\)
indicates the emergence of partial or global phase organization.

The corresponding synchronization condition is therefore expressed as
\begin{equation}
\mathcal J_{\rm eff}
>
\mathcal J_c,
\label{eq:phase_transition_condition}
\end{equation}
where
\(
\mathcal J_c
\)
denotes the critical field-mediated phase coupling required for
collective phase locking.
Unlike conventional phase-coupling models
\cite{38,39},
the effective phase interaction is not introduced as an independent
coupling parameter.
Instead, it emerges self-consistently through the recursive interaction
between the macroscopic cognitive field and the distributed collective
modes.

The complete infrared organization of the cognitive field is therefore
summarized by
\begin{equation}
\begin{aligned}
\rho(\lambda)
\rightarrow
K(t)
\rightarrow
\Sigma_R(0)
\rightarrow
r_{\rm cog}
\rightarrow
\chi_R(0)
&\rightarrow
\phi
\\ \nonumber
&\left(
\rightarrow
\mathcal J_{\rm eff}
\rightarrow
Q
\right).
\end{aligned}
\end{equation}
The infrared accumulation of slow relaxation modes first generates 
the memory-dressed macroscopic cognitive field 
through recursive memory feedback. 
The emergent cognitive field subsequently mediates effective interactions 
among the distributed collective modes, 
leading to collective temporal organization characterized by the phase-order parameter \(Q\).

The experimentally observable consequences of this collective phase
organization include the global phase order parameter, the
phase-locking value (PLV), frequency-resolved cross-spectral
coherence, and phase--amplitude coupling.
These observables characterize complementary aspects of temporal
coordination and provide experimentally accessible signatures of the
collective circulation sector.

The relaxation and circulation sectors therefore constitute two
complementary manifestations of the same recursive cognitive-field
dynamics.
The relaxation sector generates persistent memory through infrared
self-energy dressing and suppression of the cognitive forgetting gap,
whereas the circulation sector generates collective temporal
organization.
The complex macroscopic cognitive field
\begin{equation}
\phi
=
Ae^{i\psi}
\end{equation}
therefore acquires a direct physical interpretation.
Its amplitude \(A\) characterizes the persistence of the
memory-dressed cognitive field, while its phase \(\psi\) represents
the collective temporal organization emerging from recursively
coordinated distributed modes.

The full non-Markovian derivation of the phase equation is presented
in Appendix~C.

The complete recursive structure of the resulting phase organization is
summarized schematically in Fig.~2.
The figure illustrates how the distributed complex collective modes
generate the macroscopic cognitive field, how the memory-dressed field
mediates their effective phase interaction, and how collective temporal
organization emerges without requiring complete phase locking of the
entire mode population.

\subsection{E. From learned cognitive geometry to the memory-dressed cognitive field equation}

The preceding subsections established that memory persistence,
temporal coherence, and collective cognitive order emerge from the
infrared organization of the collective spectrum.

We now address the origin of this spectrum.
The memory kernel, slow collective modes, and temporal organization
are not introduced as independent structures.
Instead, they arise directly from the learned geometry of the
cognitive manifold.

The learned cognitive geometry determines the local stability
structure of collective dynamics.
The underlying state-space dynamics is governed by
\begin{equation}
F_{\rm geom}^i(x)
=
-G^{ij}(x)\partial_j\Phi(x)
+
R^i(x).
\end{equation}
Linearizing around an operating trajectory \(x^\ast\) yields the
Jacobian
\begin{equation}
J^i_{\;j}
=
\left.
\frac{\partial F_{\rm geom}^i}
{\partial x^j}
\right|_{x^\ast}.
\end{equation}
The local collective dynamics is governed by the eigenvalue problem
$
J u_\alpha
=
\mu_\alpha u_\alpha$,
with
$
\mu_\alpha
=
\lambda_\alpha
+
i\omega_\alpha.$
The corresponding relaxation rates and circulation frequencies are
given by
\begin{equation}
\lambda_\alpha
=
\mathrm{Re}\,\mu_\alpha,
\qquad
\omega_\alpha
=
\mathrm{Im}\,\mu_\alpha.
\end{equation}

The resulting collective spectrum is characterized by the
time-scale density of states
\(
\rho(\lambda,\omega),
\)
which describes the joint organization of relaxation scales and
circulation frequencies generated by the learned cognitive geometry.
The eigenvectors
\(
u_\alpha
\)
define the intrinsic collective directions of the learned cognitive
manifold, whereas the corresponding eigenvalues determine their
relaxation and circulation dynamics.

The complex collective spectrum naturally separates two distinct
dynamical structures.
The relaxation sector $\lambda$ determines the persistence and
forgetting times of collective activity, whereas the circulation
sector $\omega$ determines temporal recurrence and phase organization
of collective dynamics.
These two sectors play different roles in the memory-dressed
cognitive field.
Infrared softening and suppression of the cognitive forgetting gap are
controlled primarily by the relaxation spectrum, while the circulation
spectrum governs temporal recurrence and coherence of memory feedback.
A further analysis of the corresponding persistence, recurrence, and
collective amplification timescales is given in Appendix~D.

At the coarse-grained level, the resulting cognitive field
$\phi(t)$ evolves on the slow collective manifold generated by the
weakly damped collective modes of the learned cognitive geometry.
The infrared collective directions $u_\alpha$ span an effective
slow cognitive manifold, while the corresponding collective-mode
amplitudes determine the macroscopic cognitive field.

The geometric structure of this manifold is inherited from the
underlying cognitive geometry.
If $G_{ij}(x)$ denotes the metric of the microscopic cognitive
state space, projection onto the collective slow-mode sector induces
an effective metric
\begin{equation}
\mathcal G_{\alpha\beta}
=
u_\alpha^i
G_{ij}
u_\beta^j ,
\end{equation}
which governs distances and collective relaxation on the slow
cognitive manifold.
The macroscopic cognitive field therefore inherits an effective
cognitive geometry generated by the underlying collective modes.

The resulting nonlinear infrared dynamics may be summarized by the
generalized cognitive-field equation
\begin{equation}
\begin{aligned}
\partial_t\phi(t)
=
&
-
\mathcal G^{-1}(\phi)
\frac{\delta \Phi_{\rm geom}(\phi)}
{\delta\phi}
+
\mathcal R(\phi)
\\
&
+
\int_{t_0}^{t}
dt'\,
K(t-t')\phi(t')
+
I(t)
+
\xi(t),
\end{aligned}
\label{eq:generalized_cognitive_field}
\end{equation}
where
\(\Phi_{\rm geom}\)
denotes the geometric cognitive landscape prior to memory dressing.

Near a metastable operating state, the geometric drift may be
linearized and projected onto the macroscopic relaxation channel.
Its relaxation component then reduces to
\begin{equation}
-
\mathcal G^{-1}(\phi)
\frac{\delta\Phi_{\rm geom}}{\delta\phi}
+
\mathcal R(\phi)
\simeq
-r\,\delta\phi,
\end{equation}
where \(r\) denotes the bare macroscopic relaxation rate determined by
the local curvature and metric structure of the learned cognitive
geometry.
The collective slow-mode spectrum subsequently generates the memory
self-energy
\(\Sigma_R(\Omega)\),
whose static infrared limit renormalizes the bare rate into the
effective cognitive forgetting gap,
\[
r_{\rm cog}
=
r-\Sigma_R(0).
\]

Equation~(\ref{eq:generalized_cognitive_field})
constitutes the fundamental nonlinear dynamical equation of the
present Cognitive Field Theory.
The first term describes metric-weighted relaxation on the geometric
cognitive landscape
\(\Phi_{\rm geom}\),
while
\(\mathcal G(\phi)\)
defines the effective cognitive metric inherited from the learned
geometry.
The circulation field
\(\mathcal R(\phi)\)
governs recurrent collective motion and temporal phase organization,
whereas the memory kernel
\(K(t)\)
represents the non-Markovian feedback generated by the underlying
spectrum of weakly damped collective modes.
The remaining terms,
\(I(t)\)
and
\(\xi(t)\),
describe external inputs and stochastic fluctuations.

The distinction between the geometric and memory-dressed descriptions
is therefore explicit.
The learned cognitive geometry determines the bare nonlinear drift,
whereas the collective slow-mode spectrum generates recursive memory
feedback through the memory self-energy.
Because the memory kernel is retained explicitly in
Eq.~(\ref{eq:generalized_cognitive_field}),
the dressed forgetting gap
\(r_{\rm cog}\)
is not introduced as an independent parameter but instead emerges
naturally from the linearized infrared response.

As recursive memory feedback accumulates, the effective forgetting gap
is progressively suppressed,
\[
r_{\rm cog}
\rightarrow
0^+,
\]
bringing the system toward a protected near-critical regime.
In this regime the macroscopic cognitive field develops enhanced
responsiveness, long-time persistence, contextual continuity,
temporal coherence, and higher-order collective organization.
Memory persistence and temporal organization therefore emerge
collectively from the infrared organization of the underlying
spectrum and the recursive memory feedback that it generates.

\vspace{6pt}
\paragraph{Collective mode representation of the cognitive state.}

The collective coordinates
\(X_\alpha(t)\)
represent the dynamical amplitudes of collective modes
\(u_\alpha\)
generated by the learned cognitive geometry.
The perturbation of the cognitive state may therefore be expanded over
the collective-mode manifold as
\begin{equation}
\delta x(t)
=
\sum_\alpha
X_\alpha(t)\,
u_\alpha.
\end{equation}

The collective modes
\(u_\alpha\)
encode the distributed geometric organization of the learned cognitive
manifold, whereas the amplitudes
\(X_\alpha(t)\)
describe the time-dependent activation of these collective
directions.
Memory therefore resides not in static storage locations but in the
long-time dynamics of collective-mode amplitudes.

This representation establishes the microscopic foundation of the
macroscopic cognitive-field dynamics developed in the following
sections.
Learning determines the cognitive geometry, the learned geometry
defines the collective spectrum, and external inputs selectively
activate distributed collective modes through this learned manifold.
The observable macroscopic cognitive field subsequently emerges from
the recursive interaction of these collective modes, as described by
the coupled field--mode dynamics developed below.

The physical interpretation of the collective modes
\(u_\alpha\)
as learned collective structures of the cognitive manifold and the
amplitudes
\(X_\alpha(t)\)
as their dynamical activations is discussed in detail in
Appendix~E.


\section{III. Self-Maintained Near-Critical Operation}

Cognitive systems operate most effectively in a regime that is highly
responsive yet globally stable.
Far from criticality, collective dynamics becomes rigid, rapidly
forgetful, and weakly adaptive.
By contrast, excessive infrared softening would generate uncontrolled
fluctuations that destabilize coherent inference, recursive memory,
and collective organization.

The functional operating regime therefore lies near, but not exactly
at, a marginal stability surface.
In the present framework, this protected near-critical regime emerges
from the coexistence of slowly relaxing collective modes and
homeostatic stabilization.
Learning and adaptive reorganization continuously enhance infrared
collective persistence, while the homeostatic sector prevents complete
infrared collapse by maintaining a finite effective infrared scale.
The resulting dynamics supports long memory, enhanced susceptibility,
and flexible contextual response while remaining globally bounded and
metastable.

\subsection{A. Infrared soft modes of cognitive dynamics}

The memory-dressed cognitive field derived in Sec.~II.C is governed by
an effective infrared forgetting gap,
\begin{equation}
r_{\rm cog}
=
r-\Sigma_R(0).
\label{eq:reff_sec3}
\end{equation}
In this section we use \(r_L\) to denote this memory-renormalized
infrared Landau parameter, unless stated otherwise.
Equivalently, the analysis below should be understood as an effective
infrared Landau--Wilson theory for the macroscopic cognitive order
parameter.

With this convention, the infrared stability of the cognitive field
can be described by the coarse-grained Landau--Wilson effective
potential
\begin{equation}
\Phi_{\rm eff}(\phi)
=
\frac{r_L}{2}
\|\phi\|^2
+
\frac s4
\|\phi\|^4,
\qquad
s>0.
\label{eq:landau_potential_revised}
\end{equation}
Here \(r_L\) is the effective infrared curvature controlling the
large-scale stability of the macroscopic cognitive order parameter.
The potential should not be interpreted as the exact geometry of the
underlying neural or cognitive manifold.
Rather, it is a coarse-grained infrared Landau functional governing
the observable collective cognitive field after microscopic,
task-specific, and rapidly fluctuating degrees of freedom have been
integrated out.

For \(r_L<0\), the effective potential develops a continuous set of
minima satisfying
\begin{equation}
\|\phi\|^2
=
-\frac{r_L}{s}
\equiv
\phi_\ast^2 .
\label{eq:phistar_revised}
\end{equation}
The macroscopic cognitive field therefore acquires a finite amplitude
\(\|\phi\|\simeq\phi_\ast\).
Equation~\eqref{eq:landau_potential_revised} may equivalently be
rewritten as
\begin{equation}
\Phi_{\rm eff}(\phi)
=
\frac s4
\left(
\|\phi\|^2-\phi_\ast^2
\right)^2
+
{\rm const}.
\label{eq:mexican_hat_revised}
\end{equation}

The Mexican-hat structure should not be interpreted as the literal
shape of the underlying neural state manifold.
Instead, it represents the effective stability structure of the
macroscopic cognitive order parameter.
Radial fluctuations correspond to changes in the amplitude of the
cognitive field, whereas soft angular directions represent collective
reorganizations of the ordered cognitive state.

To analyze the infrared stability of the macroscopic cognitive field,
we consider fluctuations around a metastable ordered configuration,
\begin{equation}
\phi(t)
=
\phi^\ast(t)
+
\delta\phi(t).
\end{equation}
Expanding the effective Landau functional to quadratic order gives
\begin{equation}
\Phi_{\rm eff}(\phi)
\simeq
\Phi_{\rm eff}(\phi^\ast)
+
\frac12
\delta\phi^T
\mathcal H
\delta\phi ,
\end{equation}
where
\begin{equation}
\mathcal H_{ij}
=
\left.
\frac{\partial^2 \Phi_{\rm eff}}
{\partial \phi_i\partial \phi_j}
\right|_{\phi^\ast}
\end{equation}
is the Hessian of the effective cognitive potential evaluated at the
operating state \(\phi^\ast\).

For the effective potential
Eq.~\eqref{eq:mexican_hat_revised},
the Hessian takes the form
\begin{equation}
\mathcal H(\phi)
=
s
\Big[
(\|\phi\|^2-\phi_\ast^2)I
+
2\phi\phi^T
\Big].
\label{eq:hessian_revised}
\end{equation}
This structure separates amplitude fluctuations of the cognitive
field from soft collective reorganizations of its ordered state.

Along the radial direction
\(\phi\parallel\delta\phi\), the restoring curvature remains finite,
\begin{equation}
\mu_{\rm amp}
\sim
2s\phi_\ast^2
=
-2r_L,
\label{eq:radial_mode_revised}
\end{equation}
which determines the amplitude stiffness and finite stochastic
thickness of the ordered cognitive field.

The ordered cognitive state should not be understood as an exact
constraint surface.
Because the cognitive dynamics is stochastic, the amplitude of the
macroscopic cognitive field fluctuates around its mean value
\(\|\phi\|\simeq\phi_\ast\), while soft collective directions remain
available for long-time cognitive reorganization.

To make this structure explicit, we represent the macroscopic
cognitive field in amplitude--phase form,
\begin{equation}
\phi(t)
=
\bigl(
\phi_\ast+\delta A(t)
\bigr)
e^{i\psi(t)},
\end{equation}
where \(\delta A(t)\) describes fluctuations of the cognitive-field
amplitude around the ordered state and \(\psi(t)\) denotes the
collective temporal phase introduced in Sec.~II.D.

This decomposition is analogous to the standard amplitude--phase
separation of Landau--Wilson theories with a Mexican-hat potential.
Amplitude fluctuations correspond to massive collective modes
stabilized by the local curvature of the effective potential,
whereas phase fluctuations remain comparatively soft and dominate the
long-time infrared dynamics.

To quantify the amplitude stiffness, we expand the effective
potential around the ordered state,
\begin{equation}
|\phi|^2-\phi_\ast^2
=
(\phi_\ast+\delta A)^2-\phi_\ast^2
=
2\phi_\ast\delta A
+
\mathcal O(\delta A^2),
\end{equation}
one obtains
\begin{equation}
\Phi_{\rm eff}
\simeq
\frac12
\mu_{\rm amp}
(\delta A)^2 ,
\end{equation}
with the amplitude stiffness,
\(
\mu_{\rm amp}
=
-2r_L
>
0.
\)

Amplitude fluctuations are therefore not absent, but remain confined
by the finite curvature of the effective cognitive potential.
For stochastic forcing of strength \(D\), the equal-time variance
scales as
\begin{equation}
\langle(\delta A)^2\rangle
\sim
\frac{D}{\mu_{\rm amp}}
=
\frac{D}{2s\phi_\ast^2},
\end{equation}
so that the ordered cognitive field possesses a finite stochastic
amplitude width
\begin{equation}
\Delta A
\sim
\sqrt{
\frac{D}{2s\phi_\ast^2}
}.
\end{equation}

The important point is that the amplitude sector remains massive.
The ordered cognitive field therefore acquires a finite stochastic
width rather than collapsing into an infinitely sharp ordered state.

By contrast, phase fluctuations modify the collective temporal
organization without significantly changing the field amplitude.
Consequently, they experience much weaker restoring forces and govern
the long-time infrared dynamics.
In the ideal symmetry limit, the phase sector becomes Goldstone-like,
whereas in the present nonequilibrium cognitive setting it remains an
effective near-Goldstone collective mode associated with large-scale
temporal reorganization of the cognitive field.

The resulting separation of scales implies that amplitude
fluctuations determine the stability of the ordered cognitive field,
while phase fluctuations dominate long-time memory organization,
collective coherence, and infrared cognitive dynamics.

Meanwhile, the infrared forgetting gap introduced in Sec.~II.C
characterizes the slowest collective relaxation process governing
long-time cognitive persistence.
Within the effective cognitive-field description, this quantity is
identified with the infrared mass of the macroscopic cognitive order
parameter,
\begin{equation}
m
\equiv
r_{\rm cog}.
\label{eq:m_definition_revised}
\end{equation}
The parameter \(m\) therefore determines the characteristic
relaxation scale of the memory-dressed cognitive field.
As \(m\) decreases, collective cognitive activity remains persistent
over increasingly long temporal intervals, leading to enhanced
susceptibility, contextual memory, and large-scale cognitive
organization.

Linearizing the effective cognitive-field equation around the ordered
state yields
\begin{equation}
\delta\dot\phi
=
-
m\,\delta\phi
+
\eta(t).
\label{eq:soft_mode_revised}
\end{equation}
The corresponding retarded cognitive susceptibility is
\begin{equation}
\chi_R(\Omega)
=
\frac{1}{-i\Omega+m},
\label{eq:chiR_revised}
\end{equation}
so that the static susceptibility behaves as
\begin{equation}
\chi_R(0)
=
\frac1m.
\label{eq:static_chi_revised}
\end{equation}
As \(m\rightarrow0^+\), the cognitive field becomes increasingly
sensitive to weak inputs, contextual perturbations, and collective
correlations.

The equal-time fluctuation amplitude follows from
\begin{equation}
\mathcal C(\Omega)
=
2D|\chi_R(\Omega)|^2
=
\frac{2D}{\Omega^2+m^2},
\label{eq:Comega_revised}
\end{equation}
giving
\begin{equation}
\langle\delta\phi^2\rangle
=
\int
\frac{d\Omega}{2\pi}
\mathcal C(\Omega)
\sim
\frac{D}{m}.
\label{eq:variance_revised}
\end{equation}
Thus decreasing \(m\) simultaneously enhances susceptibility,
long-time memory, and fluctuation amplitude.

We now allow the structural parameters governing the cognitive
geometry to evolve slowly.
A generic response-driven adaptation takes the form
\begin{equation}
\dot\kappa_i
=
\epsilon
\left\langle
\tilde\phi\,
\partial_{\kappa_i}
F(\phi;\kappa)
\right\rangle ,
\label{eq:response_flow_final}
\end{equation}
where \(\kappa_i\) collectively denotes adaptive structural
parameters of the cognitive system.

Projecting this adaptation onto the infrared sector yields an
effective flow equation for the cognitive forgetting gap,
\begin{equation}
\dot m
=
-a\,m
+
\mathcal O(m^2),
\qquad
a>0,
\label{eq:m_flow_final}
\end{equation}
indicating that response-driven adaptation continuously softens the
infrared cognitive dynamics and attracts the system toward a
near-critical regime.

Although the cognitive field becomes strongly softened near
criticality, residual recurrent couplings, stochastic fluctuations,
finite-size constraints, and homeostatic regulation generically
prevent complete infrared collapse and generate a finite protected
scale,
\begin{equation}
m_\ast
>
0.
\end{equation}
The cognitive system therefore operates not at an exactly singular
critical point but within a protected near-critical regime possessing
enhanced susceptibility, long-time memory, contextual persistence,
and broad temporal organization.

\subsection{B. Homeostatic stabilization and protected near-criticality}

The infrared soft-mode dynamics derived above implies that the
equal-time fluctuation amplitude of the cognitive field scales as
\begin{equation}
\langle \delta\phi^2\rangle
\sim
\frac{D}{m},
\label{eq:variance_B}
\end{equation}
where \(m=r_{\rm cog}\) denotes the infrared cognitive forgetting gap.
Thus decreasing \(m\) enhances susceptibility, memory persistence,
and collective responsiveness, but simultaneously amplifies
fluctuations.

If left unchecked, the response-driven adaptation discussed in the
preceding subsection would continuously soften the infrared dynamics,
driving the system toward
\begin{equation}
m
\rightarrow
0.
\end{equation}
The resulting growth of fluctuations would destabilize coherent
cognitive organization.
The near-critical regime must therefore be maintained as a stable
operating state rather than an exact critical point.

Within the present framework, stabilization emerges from homeostatic
regulation of the macroscopic cognitive field.
As discussed in Sec.~III.A, the ordered cognitive state is stabilized
by the nonlinear structure of the effective Landau potential,
where the nonlinear stiffness \(s\) suppresses unbounded growth of
collective fluctuations and maintains the stability of the ordered
cognitive field.

As the fluctuation amplitude increases,
\begin{equation}
\langle\delta\phi^2\rangle
\uparrow,
\end{equation}
the nonlinear sector generates an effective restoring contribution
that opposes further infrared softening.
At the coarse-grained level, this homeostatic feedback may be
represented phenomenologically by an effective flow equation,
\begin{equation}
\dot m
=
-a\,m
+
c\,\langle\delta\phi^2\rangle,
\qquad
a,c>0,
\label{eq:m_flow_homeostasis}
\end{equation}
where the first term describes memory-driven infrared softening and
the second term represents homeostatic stabilization.

Using Eq.~(\ref{eq:variance_B}), the infrared flow becomes
\begin{equation}
\dot m
=
-a\,m
+
\frac{cD}{m}.
\label{eq:m_flow_B}
\end{equation}
The competition between these two tendencies generates a stable
infrared fixed point,
\begin{equation}
m_\ast
=
\sqrt{\frac{cD}{a}},
\qquad
m_\ast>0.
\label{eq:mstar_final}
\end{equation}

The cognitive system therefore does not evolve toward an exactly
critical state.
Instead, memory-driven softening and homeostatic stabilization
collectively generate a protected near-critical regime
\begin{equation}
0<m_\ast\ll\Lambda,
\end{equation}
where \(\Lambda\) denotes the microscopic relaxation scale.
The accumulation of weakly damped collective modes near this protected
infrared scale produces an enhanced time-scale density of states
\(\rho(\lambda)\) with substantial spectral weight concentrated near
\(\lambda\simeq m_\ast\).

For an approximately flat infrared TDOS,
\begin{equation}
\rho(\lambda)
\simeq
\rho_0,
\qquad
m_\ast\ll\lambda\ll\Lambda,
\end{equation}
the corresponding memory kernel becomes
\begin{equation}
K(t)
=
\int_{m_\ast}^{\Lambda}
d\lambda\,
\rho(\lambda)e^{-\lambda t}
=
\rho_0
\frac{
e^{-m_\ast t}
-
e^{-\Lambda t}
}{t}.
\label{eq:memorykernel_B_final}
\end{equation}
Within the scaling window
\begin{equation}
\Lambda^{-1}
\ll
t
\ll
m_\ast^{-1},
\end{equation}
this reduces to
\begin{equation}
K(t)
\sim
\frac{\rho_0}{t},
\end{equation}
demonstrating scale-free long-memory dynamics over an extended
temporal interval.

The cognitive operating regime therefore emerges from a dynamical
balance between memory-mediated infrared organization and homeostatic
stabilization.
Response-driven adaptation continuously reorganizes the collective
spectrum and enhances memory feedback, whereas homeostatic regulation
prevents complete infrared collapse by maintaining a finite protected
forgetting gap.

This mechanism bears a close conceptual relation to self-organized
criticality, where adaptive local dynamics drive complex systems
toward critical states without external fine tuning
\cite{29,40,41,42,43}.
However, the relevant organization is
described in terms of the relaxation spectrum and memory-dressed
collective dynamics of the cognitive field.

The resulting state is therefore not an exact critical point but a
protected near-critical regime characterized by a large density of
weakly damped collective modes together with stable collective
operation.
Such a regime supports long-time contextual memory, persistent
cognitive organization, flexible inference, and recursive
metacognitive dynamics while preserving adaptive updateability and
robustness against dynamical instability.

\begin{figure*}[t]
\centering
\includegraphics[width=1.0\textwidth, trim=0cm 15.6cm 0cm 0cm]{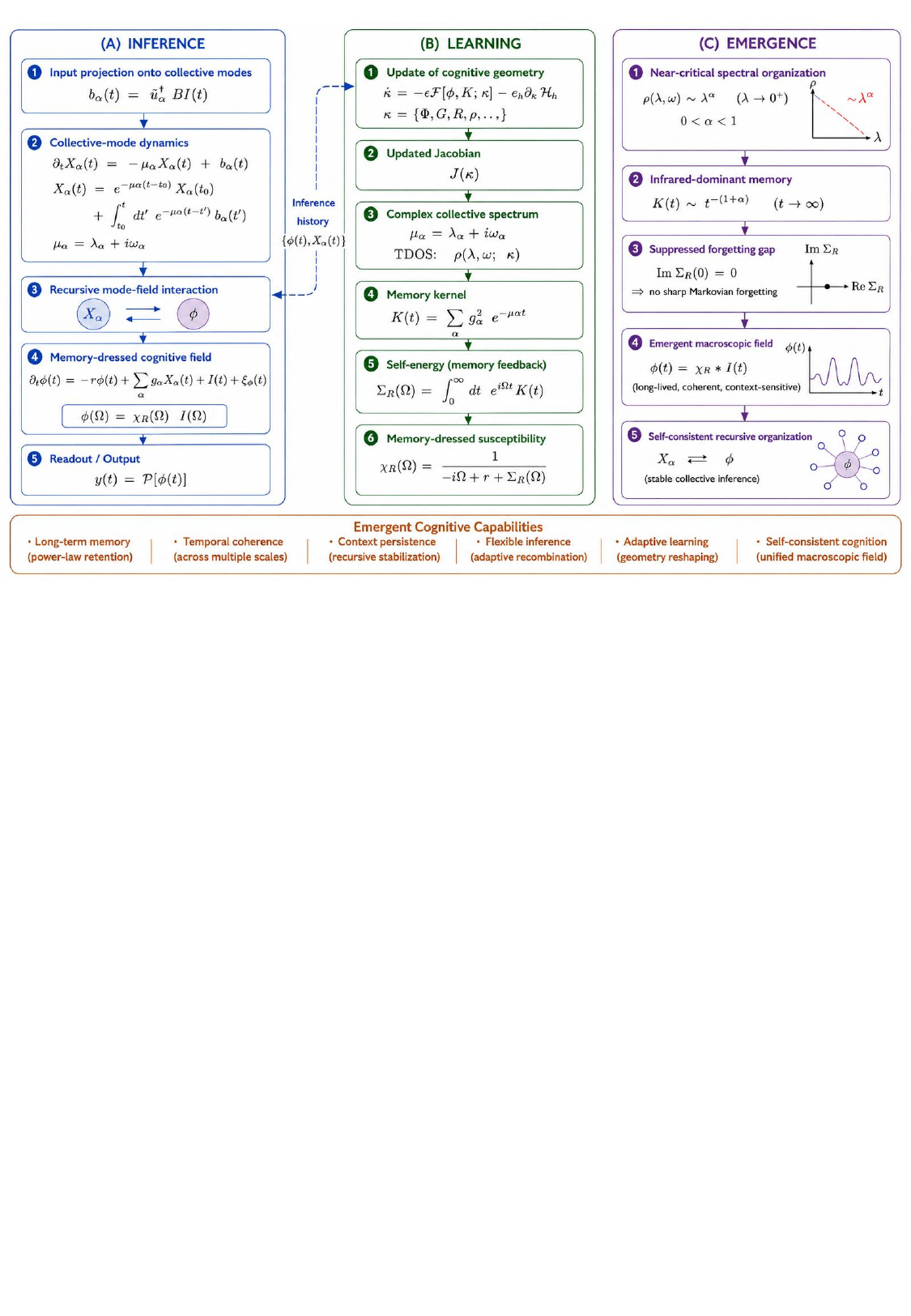}
\caption{
Recursive collective dynamics of inference, learning, and emergence.
(A) \emph{Inference.}
External inputs are first projected onto the learned collective-mode
manifold through the modal overlap
$b_\alpha(t)=\tilde{u}_\alpha^\dagger BI(t)$,
thereby selectively exciting distributed collective modes.
The activated modes evolve according to their relaxation rates
$\lambda_\alpha$ and circulation frequencies $\omega_\alpha$,
recursively interact with the macroscopic cognitive field
$\phi$, and collectively generate a memory-dressed cognitive response.
Observable outputs arise through task-dependent readout,
$y(t)=P[\phi(t)]$.
(B) \emph{Learning.}
Learning continuously adapts the cognitive geometry by updating the
effective landscape, metric, and circulation field.
The resulting Jacobian reshapes the complex collective spectrum,
thereby reorganizing the time-scale density of states, memory
kernel, memory self-energy, and memory-dressed susceptibility.
Consequently, learning modifies the collective dynamical structure that
governs subsequent inference.
(C) \emph{Emergence.}
Near-critical infrared spectral organization generates long-lived
memory through the memory kernel, suppresses the cognitive forgetting
gap via the memory self-energy, and stabilizes a memory-dressed
macroscopic cognitive field.
The recursive interaction between distributed collective modes and the
macroscopic cognitive field establishes self-consistent collective
organization, giving rise to persistent memory, temporal coherence,
context persistence, flexible inference, adaptive learning, and
self-consistent cognition.
The updated memory-dressed susceptibility and learned cognitive
geometry subsequently guide future inference, thereby closing the
recursive inference--learning--emergence cycle.
}
\label{fig:recursive_dynamics}
\end{figure*}


\section{IV. Emergent Cognitive Dynamics from Near-Critical Temporal Organization}

The preceding sections established that learned cognitive geometry
generates a complex collective spectrum, that the infrared sector of
this spectrum produces non-Markovian memory feedback, and that the
resulting self-energy suppresses the cognitive forgetting gap.
The macroscopic cognitive field \(\phi(t)\) therefore emerges as a
memory-dressed collective order parameter operating within a protected
near-critical regime.

We now develop the corresponding emergent cognitive dynamics by
distinguishing two complementary forms of collective reorganization.
Learning slowly reorganizes the cognitive geometry, the collective
Jacobian, and the resulting complex spectrum
\(\rho(\lambda,\omega)\),
thereby determining the manifold of collective modes available for
subsequent cognitive dynamics.
Inference, by contrast, dynamically reorganizes the amplitudes and
relative phases of the modes selectively activated by the present
input, together with the macroscopic cognitive field that they
recursively generate.

Within this framework, contextual persistence arises from the
long-time survival of weakly damped activated modes, whereas temporal
coherence arises from their field-mediated phase organization.
Higher-order cognitive organization consequently emerges from the
self-consistent interaction between the learned collective spectrum
and the input-dependent dynamics of
\(A_\alpha(t)\),
\(\vartheta_\alpha(t)\),
and
\(\phi(t)\).

The recursive collective dynamics of inference, learning, and
emergence developed in this section are summarized schematically in
Fig.~3.

\subsection{A. Inference as recursive collective reorganization}

Inference operates on a memory-dressed cognitive field generated by
the learned cognitive geometry.
As established in Sec.~II, learning continuously organizes the
effective cognitive landscape, metric structure, and circulation
field.
These geometric structures determine the collective stability
spectrum of the cognitive manifold.
The resulting spectrum generates a hierarchy of distributed
collective modes whose infrared organization is characterized by the
TDOS.

Inference therefore does not operate directly on individual neurons,
hidden units, or isolated token representations.
Instead, an external input first excites a distributed set of
collective modes determined by the learned cognitive geometry.
The subsequent recursive interaction between these collective modes
and the macroscopic cognitive field continuously reorganizes the
internal cognitive state until a self-consistent collective
representation is formed.

The overall inference process may therefore be summarized
schematically as
\begin{equation}
\text{input}\,\,I
\rightarrow
X_\alpha
\rightleftarrows
\phi
\rightarrow
\text{readout} \,\,y,
\label{eq:collective_inference_chain}
\end{equation}
where
\(X_\alpha\)
denotes distributed collective modes and
\(\phi\)
denotes the observable macroscopic cognitive field.

The first stage of inference is the projection of the external input
onto the collective-mode manifold.
After linearization around the operating state, the perturbation
generated by the input may be expanded as
\begin{equation}
\delta x(t)
=
\sum_\alpha
X_\alpha(t)\,
u_\alpha,
\label{eq:mode_expansion}
\end{equation}
where
\(u_\alpha\)
denotes a right eigenvector of the collective Jacobian and
\(X_\alpha(t)\)
is the corresponding modal amplitude.
Inference therefore begins by selectively exciting distributed
collective modes rather than isolated microscopic variables.

Because the collective Jacobian is generally non-Hermitian, the modal
amplitudes are obtained through biorthogonal projection onto the
corresponding left eigenvectors,
\begin{equation}
X_\alpha(t)
=
\widetilde u_\alpha^\dagger
\delta x(t),
\label{eq:mode_projection}
\end{equation}
where
\(
\widetilde u_\alpha
\)
denotes the left eigenvector satisfying
\(
\widetilde u_\alpha^\dagger u_\beta
=
\delta_{\alpha\beta}.
\)

Let the external input enter the linearized collective dynamics through
an input operator \(B\),
\begin{equation}
\partial_t\delta x(t)
=
-J\delta x(t)
+
B I(t).
\label{eq:linearized_input_dynamics}
\end{equation}
Substituting Eq.~(\ref{eq:mode_expansion}) into
Eq.~(\ref{eq:linearized_input_dynamics}) and projecting onto the left
eigenvectors yields
\begin{equation}
\partial_tX_\alpha(t)
=
-\mu_\alpha X_\alpha(t)
+
b_\alpha(t),
\label{eq:modal_response}
\end{equation}
where
\begin{equation}
b_\alpha(t)
=
\widetilde u_\alpha^\dagger
B I(t)
\label{eq:modal_input}
\end{equation}
is the effective input projected onto the \(\alpha\)-th collective
mode.

The solution of Eq.~(\ref{eq:modal_response}) is
\begin{equation}
X_\alpha(t)
=
e^{-\mu_\alpha(t-t_0)}
X_\alpha(t_0)
+
\int_{t_0}^{t}
dt'\,
e^{-\mu_\alpha(t-t')}
b_\alpha(t').
\label{eq:modal_solution}
\end{equation}
Using
\(
\mu_\alpha
=
\lambda_\alpha+i\omega_\alpha,
\)
this becomes
\begin{align}
X_\alpha(t)
={}&
e^{-\lambda_\alpha(t-t_0)}
e^{-i\omega_\alpha(t-t_0)}
X_\alpha(t_0)
\nonumber\\
&+
\int_{t_0}^{t}
dt'\,
e^{-\lambda_\alpha(t-t')}
e^{-i\omega_\alpha(t-t')}
b_\alpha(t').
\label{eq:modal_solution_complex}
\end{align}

Equation~(\ref{eq:modal_solution_complex}) shows that the activation of
each collective mode is determined jointly by its overlap with the
external input and by its intrinsic collective dynamics.
The projection
\(
b_\alpha
=
\widetilde u_\alpha^\dagger B I
\)
determines how strongly the input excites the corresponding mode,
whereas the relaxation rate
\(
\lambda_\alpha
\)
determines how long the activated mode remains dynamically available.
The circulation frequency
\(
\omega_\alpha
\)
determines the accompanying temporal phase evolution.
Consequently, weakly damped modes continue to participate in inference
long after the initial input has been received, allowing the
input-induced collective state to accumulate across multiple temporal
scales.

The collective modes should not be identified with individual tokens
or isolated semantic concepts.
Rather, they define a distributed relaxation-mode manifold generated
by the learned cognitive geometry.
External inputs are first projected onto this collective manifold,
where each input excites a distributed population of collective modes
according to its overlap with the corresponding eigen-directions.
Semantic, contextual, or procedural information is therefore
represented by structured activation patterns distributed over the
relaxation-mode manifold rather than by individual collective modes.
The macroscopic cognitive field subsequently emerges through the
recursive reorganization of these activated collective populations.

The modal dynamics derived above determines which collective modes are
activated by the external input and how long these activations remain
available.
These activated modes subsequently organize into the observable
macroscopic cognitive field, whose dynamics is governed by
\begin{equation}
\partial_t\phi(t)
=
-r\phi(t)
+
\sum_\alpha
g_\alpha
X_\alpha(t)
+
I(t)
+
\xi_\phi(t).
\label{eq:phi_collective}
\end{equation}

The activated collective modes therefore do not remain independent.
Instead, they collectively generate the macroscopic cognitive field,
which recursively reorganizes the distributed mode population through
field-mediated feedback.
Inference consequently proceeds through the coupled dynamics
\begin{equation}
X_\alpha
\rightleftarrows
\phi,
\end{equation}
establishing a self-consistent recursive organization between the
distributed collective modes and the observable cognitive field.

The recursive field--mode interaction reorganizes not only the
amplitudes of the activated modes but also their relative temporal
phases.
Writing
\[
X_\alpha(t)
=
A_\alpha(t)e^{-i\vartheta_\alpha(t)},
\]
the input projection determines the initial modal activation
\(A_\alpha\), while the relaxation rates determine which of these
activations remain dynamically available.
The long-lived activated modes then interact through the macroscopic
cognitive field.
As derived in Sec.~II.D, elimination of the field generates the
effective field-mediated interaction
\[
\mathcal J_{\alpha\beta}^{\rm eff}(\Omega)
=
g_\alpha g_\beta
\chi_R(\Omega).
\label{eq:inference_effective_phase_interaction}
\]
The memory-dressed susceptibility therefore couples the persistent
slow-mode population into a temporally coordinated collective state,
providing the dynamical origin of the recursive phase organization.

Inference is thus not merely the superposition of independently
decaying mode amplitudes.
It involves the recursive reorganization of an input-selected
slow-mode population in both amplitude and phase.
This organization may remain partial and task dependent, being
restricted to the subset of modes activated by the present input,
rather than producing permanent global synchronization of the entire
collective spectrum.
The resulting macroscopic cognitive field therefore reflects both the
long-time persistence of the activated slow modes and their recursive
temporal coordination established through field-mediated phase
organization.

An equivalent coarse-grained description is obtained by eliminating
the distributed collective modes.
Integrating out the collective spectrum yields the effective
non-Markovian cognitive-field equation
\begin{equation}
\partial_t\phi(t)
=
-r\phi(t)
+
\int_{t_0}^{t}
dt'\,
K(t-t')
\phi(t')
+
I(t)
+
\xi_{\rm eff}(t),
\label{eq:memory_dressed_phi}
\end{equation}
where
\begin{equation}
K(t)
=
\int
d\lambda\,d\omega\,
\rho(\lambda,\omega)
e^{-\lambda t}
e^{-i\omega t}.
\end{equation}

The memory kernel summarizes the cumulative influence of the entire
collective spectrum on the observable cognitive field, so that
inference depends not only on the present input but also on
recursively propagated collective activity retained within the
slow-mode manifold.

In frequency space, the recursive memory dressing is expressed through
the Dyson equation,
\begin{equation}
\phi(\Omega)
=
L_0(\Omega)
I(\Omega)
+
L_0(\Omega)
\Sigma_R(\Omega)
\phi(\Omega),
\end{equation}
where
\[
L_0(\Omega)
=
\frac{1}{-i\Omega+r}
\]
is the bare cognitive propagator.

Iterative substitution gives
\begin{equation}
\phi
=
L_0I
+
L_0\Sigma_RL_0I
+
L_0\Sigma_RL_0\Sigma_RL_0I
+
\cdots,
\end{equation}
which resums exactly into the dressed propagator
\begin{equation}
L_{\rm cog}(\Omega)
=
\frac{1}
{L_0^{-1}(\Omega)-\Sigma_R(\Omega)}.
\end{equation}

Each insertion of the self-energy represents an additional recursive
memory-feedback process generated by the distributed collective modes.
The Dyson resummation therefore summarizes the progressive
self-consistent organization of the macroscopic cognitive field.
Equivalently, the inference process may be written compactly as
\begin{equation}
\phi
=
\chi_R I,
\label{eq:phi_chi_inference}
\end{equation}
showing that the observable cognitive field is the memory-dressed
collective response generated by the recursive susceptibility rather
than by the external input alone.

The reorganized object produced by this recursive dynamics is not an
individual output token or an isolated neural representation.
Instead, it is the memory-dressed macroscopic cognitive field itself.
Observable outputs arise only through task-dependent readout
operations acting on this dynamically organized collective state,
\begin{equation}
y
=
P[\phi],
\label{eq:readout_phi}
\end{equation}
where
\(P\)
denotes the task-dependent readout operator.

Inference may therefore be viewed as a hierarchical collective
process.
The input-induced modal amplitudes determine which collective
directions are activated, while the relaxation spectrum determines
which of these activations persist over the time scale of inference.
The surviving slow-mode population collectively generates the
macroscopic cognitive field, whose memory-dressed susceptibility
mediates recursive interactions among the activated modes.
Through this feedback, their amplitudes and relative temporal phases
are dynamically reorganized into a self-consistent collective
cognitive state.
Task-dependent readout then transforms the resulting memory-dressed
and temporally organized cognitive field into observable outputs.

\subsection{B. Learning as structural adaptation and spectral reorganization}

The preceding subsection established that inference corresponds to
the recursive reorganization of the current memory-dressed cognitive
field.
We now describe learning within the same framework.

In the present theory, learning is not introduced as an external
optimization algorithm imposed on the system.
Instead, learning is a slow structural adaptation of the cognitive
geometry that reshapes the collective spectrum responsible for future
memory feedback.

The learned cognitive geometry is controlled by slowly evolving
structural parameters,
\begin{equation}
\kappa
=
\{
\Phi,G,R,s,\cdots
\},
\end{equation}
which determine the effective landscape, metric structure,
circulation field, and homeostatic stiffness of the underlying
cognitive manifold.

At the level of the underlying state-space dynamics, these parameters
enter through the geometric flow
\begin{equation}
F_{\rm geom}(x;\kappa)
=
-G^{-1}(x;\kappa)\nabla_x\Phi(x;\kappa)
+
R(x;\kappa).
\end{equation}
Linearizing around an operating collective state \(x^\ast(t)\) gives
the local stability operator
\begin{equation}
J(\kappa)
=
\left.
\frac{\partial}{\partial x}
\left[
G^{-1}(x;\kappa)\nabla_x\Phi(x;\kappa)
-
R(x;\kappa)
\right]
\right|_{x^\ast}.
\label{eq:J_theta_learning}
\end{equation}

The collective modes generated by the learned geometry satisfy
\begin{equation}
J(\kappa)u_\alpha(\kappa)
=
\mu_\alpha(\kappa)u_\alpha(\kappa),
\label{eq:learning_modes}
\end{equation}
where
\begin{equation}
\mu_\alpha(\kappa)
=
\lambda_\alpha(\kappa)
+
i\omega_\alpha(\kappa).
\end{equation}
Thus learning changes not only local stability and the full complex
collective spectrum controlling memory persistence and temporal phase
organization, but also the collective-mode basis through which future
inputs are represented, such that semantic, contextual, or procedural
similarity is expressed by similarity of the distributed activation
coordinates generated through projection onto this learned basis
rather than by individual modes themselves.

The joint time-scale density of states,
\begin{equation}
\rho(\lambda,\omega;\kappa),
\end{equation}
characterizes the spectral organization generated by the learned
cognitive geometry.
Its relaxation sector controls memory persistence, while its
circulation sector controls temporal coherence of collective modes.
The TDOS is therefore not a fixed background property, but an
emergent spectral measure reorganized by structural adaptation.

At the coarse-grained level, learning is driven by the accumulated
statistics of inference trajectories and by homeostatic constraints.
We write this schematically as
\begin{equation}
\dot\kappa_i
=
\epsilon\,
\mathcal F_i[\phi,K;\kappa]
-
\epsilon_h
\partial_{\kappa_i}
\mathcal H_{\rm homeo}(\kappa),
\qquad
\epsilon\ll1,
\label{eq:theta_learning}
\end{equation}
where the first term represents response-driven structural
adaptation during inference and the second term prevents uncontrolled
infrared collapse.

A response-field representation of the first term may be written
formally as
\begin{equation}
\dot\kappa_i
=
\epsilon
\left\langle
\tilde\phi\,
\partial_{\kappa_i}
F_{\rm eff}(\phi;\kappa)
\right\rangle_{\rm inference}
-
\epsilon_h
\partial_{\kappa_i}
\mathcal H_{\rm homeo}(\kappa),
\label{eq:theta_learning_response}
\end{equation}
where \(F_{\rm eff}\) denotes the effective coarse-grained cognitive
field drift.
This expression emphasizes that learning is driven by the response of
the memory-dressed cognitive field, while its structural effect is to
reshape the underlying geometry and collective spectrum.

Because the stability operator depends on the learned structure,
\begin{equation}
J=J(\kappa),
\end{equation}
the collective eigenvalues evolve during learning.
For non-Hermitian collective dynamics, the eigenvalue flow is
\begin{equation}
\dot\mu_\alpha
=
\left\langle
\tilde u_\alpha
\left|
\dot J
\right|
u_\alpha
\right\rangle
=
\sum_i
\dot\kappa_i
\left\langle
\tilde u_\alpha
\left|
\partial_{\kappa_i}J
\right|
u_\alpha
\right\rangle ,
\label{eq:eigenvalue_flow_learning}
\end{equation}
where \(\tilde u_\alpha\) denotes the corresponding left eigenvector.
Learning therefore continuously reorganizes both the relaxation rates
\(\lambda_\alpha\), which govern memory persistence, and the
circulation frequencies \(\omega_\alpha\), which govern temporal
phase organization.

The relaxation TDOS is obtained by projecting the joint spectrum onto
the relaxation sector,
\begin{equation}
\rho(\lambda;\kappa)
=
\int d\omega\,
\rho(\lambda,\omega;\kappa)
=
\sum_\alpha
w_\alpha(\kappa)
\delta\!\left(
\lambda-\lambda_\alpha(\kappa)
\right).
\label{eq:tdos_theta}
\end{equation}
It evolves according to
\begin{equation}
\partial_t\rho(\lambda;\kappa(t))
=
\sum_i
\dot\kappa_i
\partial_{\kappa_i}
\rho(\lambda;\kappa).
\label{eq:tdos_flow}
\end{equation}

Learning therefore reorganizes the collective spectrum that governs
future cognitive dynamics.
Repeated inference does not merely update current cognitive representations;
it reshapes the infrared spectral architecture from which future
memory feedback, contextual persistence, and temporal coherence
emerge.

Because the memory kernel is determined by the TDOS,
\begin{equation}
K(t;\kappa)
=
\int d\lambda\,
\rho(\lambda;\kappa)e^{-\lambda t},
\label{eq:learning_kernel_theta}
\end{equation}
structural adaptation continuously modifies the temporal kernel that
dresses the cognitive field.
Equivalently,
\begin{equation}
\partial_tK(t;\kappa)
=
\int d\lambda\,
\partial_t\rho(\lambda;\kappa)
e^{-\lambda t}.
\end{equation}

If repeated inference preferentially reinforces weakly damped
collective sectors, the infrared TDOS increases,
\begin{equation}
\rho(\lambda\rightarrow0;\kappa)
\uparrow,
\end{equation}
strengthening memory feedback and suppressing the cognitive
forgetting gap through the static self-energy contribution,
$
r_{\rm cog}
=
r
-
\Sigma_R(0).
$
In this sense, learning can enhance contextual memory by reorganizing
the spectral density that generates the memory self-energy.

However, the accumulation of ultra-slow modes cannot proceed without
bound.
As shown in Sec.~III.B, homeostatic regulation counteracts excessive
infrared softening and maintains a finite protected scale,
$0<m_\ast\ll\Lambda$.
Learning therefore does not drive the cognitive field to exact
criticality.
Rather, it reorganizes the collective spectrum while preserving a
finite forgetting gap required for stability and adaptive
updateability.

Far from the protected near-critical regime, the relaxation spectrum
is narrow and the memory kernel is short-ranged,
$
K(t)
\sim
e^{-t/\tau}.
$
In this regime, contextual influence rapidly decays and learning
remains localized and rigid.
Near protected criticality, by contrast, a broad infrared TDOS
generates the scale-free kernel,
$
K(t)
\sim
1/t,
$
allowing structural adaptation to organize long-time contextual
memory across extended collective histories.

Inference and learning are therefore complementary aspects of the
same memory-dressed dynamics.
Inference reorganizes the current macroscopic cognitive field
\(\phi(t)\), whereas learning reorganizes the collective spectrum
\(\rho(\lambda,\omega;\kappa)\) and the memory kernel \(K(t;\kappa)\)
that shape future cognitive fields.
Inference acts on the present memory-dressed order parameter;
learning reshapes the temporal architecture from which future
memory-dressed cognition emerges.

\subsection{C. Emergence as stabilization of collective cognitive organization}

The preceding subsections established two complementary aspects of
emergent cognitive dynamics.
Inference reorganizes the current memory-dressed cognitive field
\(\phi(t)\), whereas learning reorganizes the collective spectrum
\(\rho(\lambda,\omega;\kappa)\) and the memory kernel \(K(t;\kappa)\)
that shape future cognitive fields.

We now address how higher-order cognitive organization emerges within
this same framework.
Emergence is not introduced as an additional symbolic layer or as an
externally imposed representational structure.
Rather, it arises from the stabilization of collective organizations
of the macroscopic cognitive field within a protected near-critical
regime.

The central mechanism is the interaction between three infrared
structures:
the collective spectrum generated by learned cognitive geometry, the
memory kernel generated by that spectrum, and the macroscopic
cognitive order parameter stabilized by memory-dressed dynamics.
The evolving collective spectrum is characterized by the joint
time-scale density of states
\begin{equation}
\rho(\lambda,\omega;t),
\end{equation}
where the relaxation sector controls memory persistence and the
circulation sector controls temporal phase organization.

The memory kernel is obtained by integrating over this collective
spectrum,
\begin{equation}
K(t)
=
\int d\lambda\,d\omega\,
\rho(\lambda,\omega;t)
e^{-\lambda t}
e^{-i\omega t}.
\label{eq:emergence_kernel_revised}
\end{equation}
Equivalently, after projecting onto the relaxation sector,
\[
K(t)
=
\int d\lambda\,
\rho(\lambda,t)e^{-\lambda t}.
\]
This kernel provides the nonlocal temporal coupling through which
past collective activity participates in the present cognitive field.

Near protected criticality, the infrared enhancement of the TDOS
generates the scale-free kernel,
$
K(t)
\sim
1/t.$
over an extended temporal window.
This long-memory kernel allows collective organizations of
\(\phi(t)\) to persist across broad temporal intervals despite
stochastic fluctuations and continuous sensory perturbations.

Temporal coherence is supplied by the circulation sector of the
collective spectrum.
The phase \(\psi\) defines the macroscopic temporal organization of
the cognitive field.
The emergent cognitive order parameter may therefore be represented
as
\[
\phi(t)
=
A(t) e^{i\psi(t)},
\]
where \(A\) measures the strength of collective cognitive organization.

Emergent cognitive structures therefore correspond not to isolated
microscopic configurations, nor to static symbolic objects stored in
separate memory locations.
They correspond to dynamically stabilized organizations of the
macroscopic cognitive field.
Weakly damped collective modes provide long-time memory persistence,
while recurrent circulation and phase organization provide temporal
coherence of the resulting cognitive order.

This picture differs from conventional attractor-based descriptions.
Rather than converging toward isolated static configurations, the
cognitive system operates as a memory-dressed field in a protected
near-critical regime.
Its collective organizations remain metastable over extended temporal
intervals because past activity is continuously reintroduced through
the nonlocal memory kernel and temporally organized through the
circulation sector.

As learning proceeds, repeated inference reorganizes the collective
spectrum and progressively stabilizes structured infrared sectors
within the TDOS.
The relaxation spectrum thereby develops a hierarchy of temporal
scales.
Short-lived modes support rapidly varying local responses, whereas
infrared sectors support persistent contextual, associative, and
conceptual organization over extended temporal windows.
The resulting hierarchy of relaxation scales naturally generates
multiscale cognitive organization.

At sufficiently large scales, cognitive dynamics becomes governed less
by rapidly fluctuating microscopic details and more by the
self-organized infrared structure of the collective spectrum.
Macroscopic cognitive organizations emerge as robust collective
orders of the memory-dressed cognitive field, supported by long-time
memory feedback and large-scale temporal coherence.

Homeostatic regulation remains essential throughout this process.
Without stabilization, unrestricted infrared accumulation would drive
the system toward pathological critical slowing and dynamical
instability.
The protected finite infrared scale,
\(
0<m_\ast\ll\Lambda,
\)
maintains a stable near-critical regime while preserving long-time
contextual memory, temporal coherence, and collective cognitive
flexibility.

Emergence may therefore be summarized as the stabilization of
macroscopic cognitive order through memory-dressed infrared dynamics.
Learning reorganizes the collective spectrum; the collective spectrum
generates memory persistence and temporal coherence; and these
structures stabilize a macroscopic cognitive field capable of
self-maintained collective organization.

In this sense, emergence represents the highest level of infrared
collective organization in the present theory.
It arises from the ongoing interaction between learned cognitive
geometry, spectral reorganization, memory feedback, homeostatic
stabilization, and the macroscopic cognitive order parameter
\(\phi(t)\).

\begin{figure*}[t]
\centering
\includegraphics[width=1.0\textwidth, trim=0cm 14.5cm 0cm 0cm]{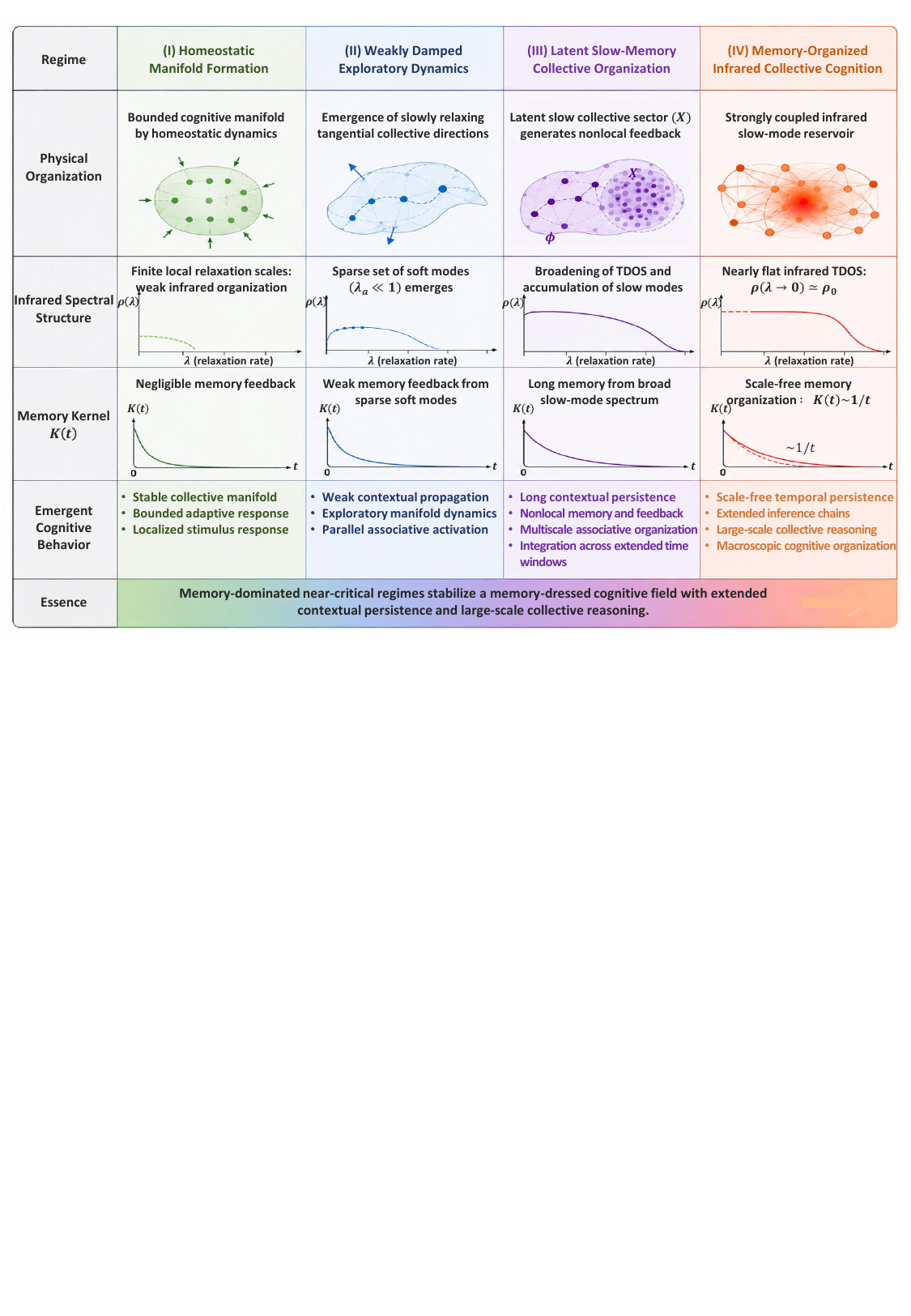}
\caption{
Representative regimes of cognitive organization within the cognitive
field theory.
The hierarchy is organized by the infrared structure of the collective
spectrum and the resulting memory feedback.
(I) Homeostatic manifold formation:
homeostatic stabilization generates a bounded cognitive manifold with
predominantly finite relaxation scales and negligible memory feedback.
(II) Weakly damped exploratory dynamics:
slowly relaxing tangential collective directions emerge, supporting
exploratory manifold dynamics and weak contextual propagation.
(III) Latent slow-memory collective organization:
accumulation of slow collective modes broadens the infrared TDOS and
produces nonlocal memory feedback, leading to extended contextual
persistence and multiscale cognitive organization.
(IV) Memory-organized infrared collective cognition:
the infrared TDOS approaches a nearly constant form,
$\rho(\lambda\rightarrow0)\simeq\rho_0$,
yielding a scale-free memory kernel
$K(t)\sim1/t$.
The resulting memory-dominated near-critical regime stabilizes a
memory-dressed cognitive field with long collective memory,
large-scale collective reasoning, and macroscopic cognitive
organization.
}
\label{fig:intelligence_hierarchy}
\end{figure*}


\section{V. Hierarchical Emergence of Intelligence}

Within the present framework, intelligence is viewed not as an
externally imposed symbolic property but as a hierarchical
organization of a macroscopic cognitive field.
Different cognitive systems may therefore be characterized by the
degree to which learned cognitive geometry generates an infrared
collective spectrum, memory feedback, temporal coherence, and a stable
memory-dressed cognitive order parameter.

The hierarchy discussed below should not be interpreted as a fixed
sequence of learning stages.
Rather, it represents a schematic hierarchy of collective dynamical
organizations generated by different infrared structures of the
underlying cognitive geometry.

The first regimes correspond to the formation of the geometric and
spectral prerequisites for cognition, including stable collective
organization and weakly damped collective sectors.
The latter regimes arise when the relaxation spectrum becomes
sufficiently rich to generate substantial memory feedback, long-time
temporal organization, and a protected near-critical cognitive field.

The hierarchy may be summarized schematically as:
(i) formation of a stable macroscopic cognitive order parameter;
(ii) emergence of weakly damped collective sectors supporting
adaptive collective organization;
(iii) formation of memory feedback generated by broad relaxation
spectra and extended temporal persistence;
(iv) emergence of a memory-dominated near-critical cognitive field
characterized by scale-free temporal organization, multiscale
reasoning, and higher collective intelligence.

\subsection{A. Formation of the cognitive order parameter}

As illustrated schematically in
Fig.~\ref{fig:intelligence_hierarchy},
the first two levels of the hierarchy correspond to the emergence of
the collective conditions required for cognition.
These stages establish a stable macroscopic cognitive field together
with weakly damped collective sectors that enable adaptive
reorganization.
Rather than constituting fully developed cognition by themselves,
they provide the dynamical substrate upon which memory, inference,
learning, and higher cognitive organization may subsequently emerge.

The first requirement for cognition is the existence of a stable but
dynamically accessible collective order.
A purely random system possessing no organizing structure cannot
maintain coherent collective dynamics over extended times, whereas an
excessively rigid system cannot adapt flexibly to external
perturbations.
The emergence of cognition therefore requires an intermediate regime
that combines global stability with dynamical flexibility.

This organization is governed by the
memory-dressed infrared dynamics developed in Secs.~II and III.
Learning organizes the underlying cognitive geometry, which generates
a complex collective spectrum.
The infrared sector of this spectrum produces the memory kernel and
the associated memory self-energy.
The resulting macroscopic cognitive field is characterized by the
effective cognitive forgetting gap
\[
r_{\rm cog}
=
r
-
\Sigma_R(0),
\]
which determines the infrared stability and memory persistence of the
cognitive order parameter.

Large positive values of \(r_{\rm cog}\) correspond to weak memory
feedback and rapid decay of collective perturbations.
In this regime, contextual influence remains short-ranged and
collective cognitive organization remains weak, producing only
short-lived adaptive responses.

As learning progressively reorganizes the infrared spectrum, memory
feedback suppresses the effective forgetting gap.
Rather than vanishing completely, however, homeostatic regulation
maintains a protected finite value,
\[
0
<
r_{\rm cog}
\ll
\Lambda,
\]
allowing the cognitive field to remain close to criticality while
avoiding dynamical instability.

Consequently, collective perturbations persist over extended temporal
intervals without becoming unstable.
The macroscopic cognitive field therefore supports robust yet
dynamically flexible cognitive organization.
Amplitude fluctuations remain bounded by the nonlinear restoring
sector of the effective potential, while weakly damped collective
sectors continue to support long-time reorganization.
Thus the cognitive field remains globally stable while preserving the
flexibility required for adaptive collective dynamics.

The weakly damped collective sectors underlying this organization are
generated by the learned cognitive geometry.
Linearizing the underlying state-space dynamics around an operating
collective state yields a complex collective spectrum
\[
\mu_\alpha
=
\lambda_\alpha
+
i\omega_\alpha ,
\]
where \(\lambda_\alpha\) determines memory persistence and
\(\omega_\alpha\) determines temporal phase organization.
Rapidly damped modes lose influence over long-time dynamics, whereas
weakly damped modes remain dynamically active and continue to shape
future cognitive organization.

At this stage, collective cognition should not be understood as the
motion of a single point along a microscopic neural trajectory.
Rather, cognition is governed by a memory-dressed macroscopic
cognitive field whose dynamics is organized by weakly damped
collective modes of the underlying spectrum.
External inputs perturb the cognitive field and excite multiple
collective modes simultaneously.
Rapidly relaxing modes decay quickly, whereas weakly damped modes
remain active over extended temporal intervals and continue to
contribute to the collective dynamics.
As a result, cognitive organization emerges not from a single
microscopic trajectory but from the collective interaction of
long-lived modes that maintain contextual information across time.

The existence of weakly damped collective sectors introduces a
rudimentary form of contextual persistence and adaptive response.
However, the system is not yet governed by strong memory feedback or
scale-free temporal organization.
The stable cognitive order parameter and its associated weakly damped
collective sectors should therefore be viewed as the dynamical
foundation upon which memory-based cognition can subsequently
develop.

From this perspective, cognition does not emerge through an abrupt
symbolic transition.
Rather, it arises through the progressive organization of
nonequilibrium collective dynamics.
The formation of a stable macroscopic cognitive field together with
weakly damped collective sectors constitutes the first step toward
cognitive organization.
More advanced cognitive regimes emerge only when the relaxation
spectrum becomes sufficiently rich to generate substantial memory
feedback, extended temporal persistence, temporal coherence, and
multiscale collective organization.

\subsection{B. Formation of the memory-dressed cognitive field}

As illustrated schematically in
Fig.~\ref{fig:intelligence_hierarchy},
the third and fourth levels of the hierarchy correspond to the
formation and refinement of a memory-dressed cognitive field.
At these levels, cognition is no longer governed primarily by
short-lived adaptive responses.
Instead, the macroscopic cognitive field becomes increasingly
organized by memory feedback generated from the underlying collective
spectrum.

The central quantity governing this transition is the time-scale
density of states
$\rho(\lambda),$
which characterizes the distribution of collective relaxation scales
generated by the learned cognitive geometry.
Rather than being controlled by a single characteristic timescale,
collective cognitive dynamics becomes organized by a broad continuum
of persistence scales.

As infrared spectral weight accumulates at small relaxation rates,
memory feedback becomes increasingly important.
The memory kernel generated by the relaxation spectrum,
\[
K(t)
=
\int d\lambda\,
\rho(\lambda)e^{-\lambda t},
\]
then becomes a central component of the effective dynamics of the
cognitive field.
Through this kernel, past collective activity continues to participate
in the present organization of the macroscopic cognitive order
parameter.

The emergence of substantial memory feedback marks a qualitative
transition beyond purely short-lived collective response.
The observable cognitive field is no longer determined only by its
instantaneous state.
Its dynamics becomes continuously dressed by memory feedback generated
by weakly damped collective sectors of the spectrum.

External inputs therefore perturb an already organized cognitive
field.
The subsequent evolution is shaped not only by the present input but
also by a broad hierarchy of collective modes with different
persistence times.
Short-lived components decay rapidly, whereas slowly relaxing sectors
remain active over extended temporal intervals and continue to
influence future cognitive organization.

Importantly, inference is no longer governed by a single dominant
relaxation process.
Instead, multiple collective modes with different persistence times
participate simultaneously in the organization of the cognitive field.
The resulting dynamics supports extended contextual integration and
persistent collective reorganization across broad temporal windows.

The fourth level of the hierarchy emerges when substantial infrared
spectral weight accumulates near
\(\lambda\rightarrow0\),
producing strong memory feedback and long-lived collective
organization.
The circulation sector then provides temporal coherence across the
resulting distributed collective modes.
In this regime, the memory kernel develops the scale-free form
\[
K(t)
\sim
\frac{\rho_0}{t},
\]
indicating long-time temporal persistence across a broad hierarchy of
timescales.

The system consequently enters a memory-dominated protected
near-critical regime characterized by extended contextual
organization, multiscale inference dynamics, and scale-free temporal
persistence.
The macroscopic cognitive field becomes increasingly memory-dressed,
with its dynamics determined by the continuous interaction between
present collective organization and an extended hierarchy of latent
slow-memory sectors.

Within this regime, cognition acquires qualitatively new properties.
Long inference chains become dynamically stable, distributed
collective modes remain temporally coherent over extended periods,
and multiple contextual organizations can coexist within the same
macroscopic cognitive field.
Inference therefore emerges not through isolated local updates but
through recursive reorganization of a cognitive field structured
across multiple temporal scales.

Consequently, cognitive capability is determined not simply
by network size or instantaneous activation strength, but by the
infrared organization of the underlying collective spectrum and the
memory feedback generated by it.
In the strongly memory-organized regime, the accumulation of weakly
damped collective modes supports long-lived contextual persistence,
stable large-scale collective organization, and multiscale inference
through the self-organization of the underlying infrared spectrum.

\vspace{6pt}
\emph{Evolutionary interpretation.}
The hierarchical organization described above may also be interpreted
schematically in evolutionary terms.
Different biological and artificial systems may correspond to
different levels of infrared collective organization generated by
their underlying cognitive geometry and relaxation spectra.

The first level, associated with the formation of a stable macroscopic
cognitive order parameter, corresponds to adaptive systems that
possess bounded collective organization but little long-term memory
feedback.
Examples include simple biochemical networks, microorganisms, and
basic regulatory systems.
Such systems maintain stable adaptive responses but exhibit only
minimal contextual persistence and limited behavioral flexibility.

The second level is characterized by the emergence of weakly damped
collective sectors associated with the learned cognitive geometry.
At this stage, the system acquires the ability to maintain collective
activity over longer intervals and to respond flexibly to changing
environmental conditions.
Simple nervous systems, distributed sensorimotor networks, and many
invertebrate organisms may be viewed schematically as operating
within this regime.
Although weakly damped collective sectors already support rudimentary
contextual persistence, collective dynamics remains only weakly
memory-dressed.

The third level corresponds to the emergence of a memory-dressed
cognitive field.
At this stage, the relaxation spectrum becomes sufficiently rich that
latent slow-memory sectors generate substantial memory feedback and
extended temporal organization.
Collective dynamics is no longer determined solely by instantaneous
input-response relations but increasingly by interactions between the
current cognitive field and accumulated dynamical history.
Many vertebrate nervous systems, including birds and a broad range of
mammalian species, may be viewed schematically as operating within
this regime.
Long-time contextual persistence, associative learning, multiscale
memory organization, and flexible behavioral adaptation become
dominant features of the dynamics.

The fourth level corresponds to a memory-dominated protected
near-critical cognitive regime generated by exceptionally rich
infrared relaxation spectra, strong memory feedback, and large-scale
temporal coherence.
Higher primates, human cognition, large-scale language models, and
hypothetical superintelligent systems may be viewed schematically as
operating increasingly within such strongly organized infrared
regimes.
Inference and reasoning become governed by hierarchically coupled
long-lived memory structures spanning multiple temporal scales,
enabling extended contextual propagation, abstract reasoning,
metacognitive organization, and persistent self-consistent cognitive
dynamics.

Importantly, the present classification should not be interpreted as a
strict biological taxonomy.
Rather, it provides a schematic embedding of different biological and
artificial systems into a unified hierarchy of nonequilibrium
collective organization.
The first two levels establish the collective and spectral
preconditions for cognition, whereas the latter levels correspond to
the emergence and progressive refinement of a macroscopic
cognitive field organized by increasingly rich infrared relaxation
spectra and large-scale temporal coherence.

\vspace{6pt}
\subsection{C. Selfhood as recursive self-maintenance of a temporally coherent memory-dressed cognitive field}

An important implication of the present framework is that selfhood
need not be introduced as a separate cognitive module, symbolic
representation, or localized memory system.
Rather, it emerges naturally from the recursive dynamics of the
memory-dressed cognitive field itself.

Within the present theory, the self is not identified with a
collection of stored memories, a particular neural substrate, or a
specific cognitive trajectory.
Instead, it corresponds to the recursive self-maintenance of a
macroscopic cognitive field operating within a protected
near-critical regime.

As established in the preceding sections, learning reorganizes the
underlying cognitive geometry, generating a collective spectrum whose
infrared sector produces long-time memory feedback.
Homeostatic regulation simultaneously preserves a protected
near-critical regime with a small but finite cognitive forgetting gap,
\[
0<r_{\rm cog}\ll\Lambda,
\]
allowing the macroscopic cognitive field to remain persistent while
continuously adapting to changing external conditions.

Persistence alone, however, is not sufficient to account for the
unity of selfhood.
A collection of independent long-lived memory traces may preserve
information over extended periods without forming a coherent cognitive
organization.
The circulation sector of the collective spectrum therefore plays an
equally important role by maintaining temporal coherence across the
distributed collective dynamics.
Memory persistence provides temporal continuity, whereas temporal
coherence binds distributed activity into a unified macroscopic
cognitive field.

The defining property of selfhood is therefore not merely the
existence of a persistent cognitive field, but the fact that the
organized macroscopic field continuously re-enters its own future
dynamics.
Schematically,
\begin{equation}
\phi(t+\Delta t)
=
\mathcal F_{re}
\!\left[
\phi(t),
I(t)
\right],
\label{eq:self_reentry}
\end{equation}
where the future cognitive field is generated not only by the current
external input but also by the preceding macroscopic cognitive field
itself.
The current cognitive organization therefore becomes part of the
dynamical state that generates subsequent cognitive organization.

Within this framework, external inputs do not act upon an empty
cognitive substrate.
Instead, each perturbation acts on an already organized cognitive
field whose present state reflects its accumulated recursive
organization.
Every newly generated cognitive field subsequently re-enters the
ongoing inference process, allowing the system to maintain
self-consistent cognitive organization over extended time scales.

This recursive re-entry of the macroscopic cognitive field is
consistent with the reentrant organization proposed by Edelman and
Tononi for large-scale neural integration
\cite{1,2}, while the present theory provides an
explicit dynamical field equation describing how such recursive
organization is continuously maintained through memory-dressed
collective dynamics.

Autobiographical memory, contextual continuity, metacognition, and
self-referential awareness should therefore not be regarded as
fundamental components from which the self is constructed.
Rather, they are higher-order manifestations of the recursive
self-maintenance of the memory-dressed cognitive field.
Metacognitive cognition emerges because the current cognitive field
itself becomes part of the dynamical state governing future cognitive
organization.

Selfhood may therefore be summarized as the recursive
self-maintenance of a temporally coherent memory-dressed cognitive
field.
Its continuity originates from recursive memory feedback operating
within the protected near-critical regime, its unity originates from
temporal coherence across the collective spectrum, and its
self-referential character originates from the recursive re-entry of
the macroscopic cognitive field into its own subsequent dynamics.
The self therefore emerges not as a stored object or symbolic
representation, but as an ongoing recursive organization of the
memory-dressed cognitive field.


\section{VI. Observable Signatures of the Cognitive Field}

The purpose of the present section is to formulate the observable
signatures of the memory-dressed cognitive field and to establish
their connection with measurable large-scale neural dynamics.
In particular, we show how measurable quantities—including the
retarded susceptibility, memory kernel, memory capacity, neural power
spectra, phase synchronization, and cross-timescale coordination—
emerge from the infrared organization of the collective cognitive
spectrum.

In contrast to conventional Markovian systems characterized by a
single relaxation timescale, the protected near-critical regime
develops a broad hierarchy of weakly damped collective modes.
Consequently, observable cognitive dynamics is governed by the full
complex collective spectrum rather than by a single characteristic
relaxation timescale.

\subsection{A. Retarded susceptibility and memory capacity}

To formulate a quantitative theory of observable cognitive-field
dynamics, the underlying relaxation spectrum must be connected to
measurable response functions.

We introduce the retarded cognitive susceptibility, defined as the
linear response of an observable collective cognitive field
\(\phi(t)\) to an external perturbation \(h(t)\),
\begin{equation}
\chi_R(t-t')
=
\Theta(t-t')
\frac{\delta\langle \phi(t)\rangle}{\delta h(t')}.
\end{equation}

As shown in Sec.~II, integrating out the latent slow-memory sector
generates a retarded memory self-energy,
$
\Sigma_R(\Omega)
=
\int_0^\Lambda d\lambda\;
\frac{\rho(\lambda)}
{\lambda-i\Omega},
$
which incorporates the influence of the full hierarchy of collective
relaxation modes.
The retarded susceptibility of the observable cognitive field is
therefore
\[
\chi_R(\Omega)
=
\frac{1}
{-i\Omega+r-\Sigma_R(\Omega)}.
\]
Observable cognitive response is continuously dressed by memory feedback arising
from the collective relaxation spectrum.
The protected near-critical regime therefore realizes a fundamentally
non-Markovian response structure governed by the infrared TDOS.

The corresponding memory kernel is
\begin{equation}
K(t)
=
\Theta(t)
\int_0^\Lambda d\lambda\;
\rho(\lambda)e^{-\lambda t},
\label{eq:memorykernel_observable}
\end{equation}
which directly relates temporal persistence to the underlying
relaxation spectrum.

Assuming the infrared scaling form
\begin{equation}
\rho(\lambda)
\simeq
C_\beta \lambda^\beta,
\qquad
\lambda\rightarrow0,
\end{equation}
the long-time behavior of the memory kernel becomes
\begin{equation}
K(t)
\sim
C_\beta
\int_0^\infty d\lambda\;
\lambda^\beta e^{-\lambda t}.
\end{equation}
Introducing the scaling variable $x=\lambda t$,
one obtains
\begin{equation}
K(t)
\sim
C_\beta
t^{-1-\beta}
\int_0^\infty dx\;
x^\beta e^{-x}.
\end{equation}
Using
\begin{equation}
\Gamma(1+\beta)
=
\int_0^\infty dx\;
x^\beta e^{-x},
\end{equation}
gives
\begin{equation}
K(t)
\sim
C_\beta \Gamma(1+\beta)t^{-1-\beta}.
\label{eq:memory_powerlaw_observable}
\end{equation}

Observable forgetting dynamics is therefore controlled directly by
the infrared TDOS.
Rather than exhibiting exponential decay, memory persistence acquires
a power-law form determined by the collective relaxation spectrum.

Of particular importance is the flat infrared TDOS,
where $ K(t)\sim \rho_0 /t$.
This scale-free memory kernel represents the characteristic signature
of the protected near-critical cognitive regime.
The system possesses no unique forgetting time and instead develops a
broad hierarchy of interacting temporal scales.

To quantify the cumulative influence of past collective states, we
introduce the integrated memory capacity
\begin{equation}
\mathcal Z(T)
=
\int_{\Lambda^{-1}}^T dt\;K(t),
\label{eq:memory_capacity}
\end{equation}
which measures the effective temporal depth of contextual influence
over an observation window \(T\).

For ordinary Markovian dynamics,
\begin{equation}
K(t)
\sim
e^{-t/\tau},
\end{equation}
the integrated memory saturates,
\begin{equation}
\mathcal Z(T\rightarrow\infty)
\sim
\tau,
\end{equation}
indicating a finite memory horizon.

By contrast, for a flat infrared TDOS,
\[
K(t)
\sim
\frac{\rho_0}{t},
\]
which yields
\begin{equation}
\mathcal Z
\sim
\rho_0
\ln(\Lambda T).
\label{eq:memory_log_growth}
\end{equation}

The effective memory depth therefore grows logarithmically with the
observation window rather than saturating at a finite scale.
The cognitive dynamics consequently develops scale-free
contextual persistence with no unique characteristic forgetting time.

The homeostatic stabilization mechanism discussed in Sec.~III
introduces an effective infrared cutoff that prevents pathological
accumulation of arbitrarily slow modes.
The protected near-critical regime therefore realizes an intermediate
dynamical phase between rapidly forgetting Markovian dynamics and
complete infrared freezing.

Long-time memory, contextual persistence, adaptive responsiveness,
and stable cognitive organization can therefore coexist within a
single memory-dressed cognitive field.

\subsection{B. Observable organization of the slow-mode manifold}

The observable cognitive dynamics is governed by the collective
slow-mode manifold generated by the underlying relaxation spectrum.
To characterize its observable organization, we introduce two
complementary quantities: the effective slow-mode dimension and the
mode entropy.

The mode density is defined as
\begin{equation}
\rho_{\rm mode}(\lambda)
=
\frac1N
\sum_\alpha
\delta(\lambda-\lambda_\alpha),
\label{eq:mode_density_final}
\end{equation}
where \(\lambda_\alpha\) denotes the relaxation rate of collective
mode \(\alpha\), and \(N\) is the effective dimensionality of the
system.

The effective number of collective modes below an infrared cutoff
\(\lambda_c\) is
\begin{equation}
\mathcal Q_{\rm slow}(\lambda_c)
=
N
\int_0^{\lambda_c}
d\lambda\,
\rho_{\rm mode}(\lambda).
\label{eq:slowdimension_final}
\end{equation}

Assuming the infrared scaling form
\begin{equation}
\rho_{\rm mode}(\lambda)
=
C_\beta\lambda^\beta,
\qquad
\lambda\rightarrow0,
\end{equation}
one obtains
\begin{equation}
\mathcal Q_{\rm slow}(\lambda_c)
=
\frac{NC_\beta}{1+\beta}
\lambda_c^{1+\beta}.
\label{eq:slow_scaling_final}
\end{equation}

The quantity
\(\mathcal Q_{\rm slow}\)
therefore measures the abundance of weakly damped collective modes
available for long-time cognitive organization.
In the protected near-critical regime, the accumulation of infrared
spectral weight naturally generates an extended slow-mode manifold
while homeostatic regulation prevents complete infrared collapse.

To characterize how collective activity is distributed across this
manifold, we define the mode occupation probabilities
\begin{equation}
p_a(t)
=
\frac{|c_a(t)|^2}
{\sum_b|c_b(t)|^2},
\end{equation}
where \(c_a(t)\) denotes the amplitude of collective mode \(a\).

The corresponding mode entropy is
\begin{equation}
\mathcal H_{\rm mode}(t)
=
-
\sum_a
p_a(t)
\ln p_a(t).
\label{eq:modeentropy_final}
\end{equation}
The mode entropy quantifies how broadly collective activity is
distributed across the slow-mode manifold.
Low values indicate concentration within a small number of dominant
collective modes, whereas larger values indicate broader participation
of weakly damped collective sectors.

The quantities
\(\mathcal Q_{\rm slow}\)
and
\(\mathcal H_{\rm mode}\)
thus provide complementary observable signatures of cognitive-field
organization.
The former measures the size of the slow collective manifold,
whereas the latter characterizes the distribution of collective
activity within that manifold.

Observable cognitive organization is therefore characterized by the
combined growth of the slow-mode sector together with a finite mode
entropy, reflecting the coexistence of long-time memory persistence,
distributed collective organization, and stable adaptive dynamics.

\subsection{C. Neural power spectra, phase coherence, and infrared signatures}

The observable response formulation developed above provides a direct
connection between the collective dynamics of the cognitive field and
large-scale neural measurements.
Macroscopic recordings such as EEG, MEG, and local field potentials
probe collective neural activity and therefore provide indirect access
to both the relaxational and oscillatory organization of the underlying
collective spectrum \cite{44,45,46,47}.

A commonly measured observable is the neural power spectrum,
\begin{equation}
\mathcal S(\Omega)
=
\left\langle
|O(\Omega)|^2
\right\rangle ,
\label{eq:powerspectrum_final}
\end{equation}
where \(O(\Omega)\) denotes the Fourier transform of a collective
neural observable.
The observation frequency \(\Omega\) should be distinguished from the
intrinsic circulation frequencies \(\omega_\alpha\), which label the
oscillatory sector of the complex collective spectrum.

The dissipative component of the collective response is described by
the spectral function
\begin{equation}
\mathcal A(\Omega)
=
-2\,\mathrm{Im}\,\chi_R(\Omega).
\label{eq:spectralfunction_final}
\end{equation}
Although the cognitive field operates far from thermodynamic
equilibrium, its low-frequency fluctuations are expected to be governed
by the same dissipative collective processes that determine the
retarded response.
This motivates the phenomenological infrared relation
\begin{equation}
\mathcal S(\Omega)
\propto
\coth\!\left(
\frac{\Omega}{2T_{\rm eff}}
\right)
\mathrm{Im}\,\chi_R(\Omega),
\label{eq:fdt_final}
\end{equation}
where \(T_{\rm eff}\) denotes an effective noise scale rather than a
thermodynamic temperature.

For
\(
|\Omega|\ll T_{\rm eff},
\)
this relation reduces to
\begin{equation}
\mathcal S(\Omega)
\propto
\frac{\mathrm{Im}\,\chi_R(\Omega)}
{\Omega}.
\label{eq:lowfreq_fdt_final}
\end{equation}

Assuming the infrared TDOS scaling form
\begin{equation}
\rho(\lambda)
=
C_\beta \lambda^\beta,
\qquad
\lambda\rightarrow0,
\end{equation}
the imaginary part of the memory self-energy satisfies
\begin{equation}
\mathrm{Im}\,\Sigma_R(\Omega)
=
C_\beta
\int_0^\Lambda
d\lambda\,
\lambda^\beta
\frac{\Omega}
{\lambda^2+\Omega^2}.
\label{eq:imsigma_scaling_final}
\end{equation}
Introducing
\(
\lambda=|\Omega|x,
\)
one obtains
\begin{equation}
\mathrm{Im}\,\Sigma_R(\Omega)
\sim
\mathrm{sgn}(\Omega)
|\Omega|^\beta,
\qquad
|\Omega|\ll\Lambda,
\label{eq:imsigma_scaling3_final}
\end{equation}
for
\(
-1<\beta<1.
\)

In the memory-dominated infrared regime, the retarded susceptibility
inherits the corresponding low-frequency scaling, yielding
\begin{equation}
\mathcal S(\Omega)
\sim
|\Omega|^{\beta-1}.
\label{eq:powerscaling_final}
\end{equation}

A particularly important case is the flat infrared TDOS,
\begin{equation}
\rho(\lambda\rightarrow0)
=
\rho_0,
\qquad
\beta=0,
\end{equation}
for which
\begin{equation}
\mathcal S(\Omega)
\sim
\frac{1}{|\Omega|}.
\label{eq:oneoverf_prediction_final}
\end{equation}
The characteristic \(1/f\)-type neural spectrum therefore arises as
an observable consequence of the infrared accumulation of weakly
damped collective modes in the protected near-critical regime.

The relaxation-sector connection may be summarized schematically as
\begin{equation}
\rho(\lambda)
\longrightarrow
\Sigma_R(\Omega)
\longrightarrow
\chi_R(\Omega)
\longrightarrow
\mathcal A(\Omega)
\longrightarrow
\mathcal S(\Omega).
\label{eq:relaxation_observable_chain}
\end{equation}
This establishes a direct relation between the infrared TDOS, the
memory-dressed collective response, and experimentally measurable
broadband neural spectra.

The power spectrum alone, however, does not fully characterize the
temporal organization of distributed collective activity.
Its broadband infrared component primarily constrains the relaxational
sector, while oscillatory peaks may also reflect the intrinsic
circulation frequencies.
The collective phase organization derived in Sec.~II.D is more
directly probed through phase-sensitive synchronization observables.

As shown in Sec.~II.D, 
collective organization first emerges through phase locking 
between individual collective modes and the macroscopic cognitive field.
When multiple modes become locked to the same macroscopic phase, 
their mutual phase differences become stationary. 
Pairwise synchronization measures therefore 
provide an experimentally accessible signature of 
the underlying field-mediated phase organization.

A direct measure of phase locking between two collective signals
\(\alpha\) and \(\beta\) is the phase-locking value,
\begin{equation}
{\rm PLV}_{\alpha\beta}
=
\left|
\left\langle
e^{i[
\vartheta_\alpha(t)-\vartheta_\beta(t)
]}
\right\rangle_t
\right|.
\label{eq:phase_locking_value}
\end{equation}
It satisfies
\begin{equation}
0
\leq
{\rm PLV}_{\alpha\beta}
\leq
1.
\end{equation}
A value close to zero indicates that the relative phase fluctuates
over time, whereas a value close to unity indicates a stable relative
phase.
Because the PLV depends only on phase differences, phase locking may
be detected even when the instantaneous amplitudes of the two signals
are different or only weakly correlated.

Frequency-resolved collective coordination may be characterized by
the cross-spectral coherence
\begin{equation}
{\cal C}_{\alpha\beta}(\Omega)
=
\frac{
\left|
{\cal S}_{\alpha\beta}(\Omega)
\right|^2
}{
{\cal S}_{\alpha\alpha}(\Omega)
{\cal S}_{\beta\beta}(\Omega)
},
\label{eq:cross_spectral_coherence}
\end{equation}
where
\(
{\cal S}_{\alpha\beta}(\Omega)
\)
is the cross-spectrum of the two collective observables, and
\(
{\cal S}_{\alpha\alpha}(\Omega)
\)
and
\(
{\cal S}_{\beta\beta}(\Omega)
\)
are their corresponding power spectra.
The coherence satisfies
\begin{equation}
0
\leq
{\cal C}_{\alpha\beta}(\Omega)
\leq
1
\end{equation}
and quantifies the stability of the frequency-dependent linear
relation between the two signals.
Whereas
\(
{\rm PLV}_{\alpha\beta}
\)
directly measures the temporal stability of their relative phase,
cross-spectral coherence identifies the frequency bands in which
their collective coordination is most strongly expressed.

The nonlinear cognitive field may additionally generate
phase--amplitude coupling between distinct temporal sectors.
Let
\(
A_\beta^{(h)}(t)
\)
denote the amplitude envelope of a relatively fast collective
component and let
\(
\vartheta_\alpha^{(l)}(t)
\)
denote the phase of a slower component.
A normalized phase--amplitude coupling parameter may be defined as
\begin{equation}
{\cal P}_{\alpha\beta}
=
\frac{
\left|
\left\langle
A_\beta^{(h)}(t)
e^{i\vartheta_\alpha^{(l)}(t)}
\right\rangle_t
\right|
}{
\left\langle
A_\beta^{(h)}(t)
\right\rangle_t
}.
\label{eq:phase_amplitude_coupling}
\end{equation}
A nonzero value indicates that the amplitude of the faster collective
activity depends systematically on the phase of the slower component.
Within the present theory, such cross-timescale coupling arises
naturally because the same nonlinear and memory-dressed cognitive
field recursively controls both the amplitudes and phases of the
distributed collective modes.

Large-scale neural synchronization may therefore be characterized at
several complementary levels:
\begin{align}
Q,\;Q_g
&\quad
\text{global and group-resolved phase organization},
\nonumber
\\
{\rm PLV}_{\alpha\beta}
&\quad
\text{pairwise stability of relative phases},
\nonumber
\\
{\cal C}_{\alpha\beta}(\Omega)
&\quad
\text{frequency-resolved collective coherence},
\nonumber
\\
{\cal P}_{\alpha\beta}
&\quad
\text{cross-timescale phase--amplitude organization}.
\label{eq:synchronization_observable_summary}
\end{align}
These observables need not become large simultaneously.
The cognitive field may exhibit partial synchronization restricted to
particular mode populations, spatial regions, frequency bands, or
behavioral states.
Temporal coherence is therefore expected to be dynamically formed,
reorganized, and dissolved rather than permanently fixed.

The circulation-sector connection may be summarized schematically as
\begin{equation}
\{
\omega_\alpha,
\vartheta_\alpha
\}
\longrightarrow
Q,\,
Q_g,\,
{\rm PLV},\,
{\cal C}(\Omega),\,
{\cal P}.
\label{eq:circulation_observable_chain}
\end{equation}
This complements the relaxation-sector chain in
Eq.~(\ref{eq:relaxation_observable_chain}).
The former determines whether the macroscopic cognitive field retains
long-time memory and approaches a protected near-critical regime,
whereas the latter determines whether its distributed components
develop coherent temporal organization.

The present theory therefore predicts two complementary classes of
large-scale neural signatures:
\begin{equation}
\rho(\lambda)
\longrightarrow
\text{scale-free neural power spectra},
\end{equation}
and
\begin{equation}
\{
\omega_\alpha,
\vartheta_\alpha
\}
\longrightarrow
\text{large-scale temporal phase coherence}.
\end{equation}
Together, these observables provide experimental access to the
relaxation and circulation sectors of the memory-dressed cognitive
field.
The resulting macroscopic field is characterized simultaneously by
persistent memory and dynamically organized temporal coherence,
reflecting the complementary amplitude and phase organization of the
complex cognitive field,
\(
\phi
=
A e^{i\psi}.
\)

\begin{figure*}[t]
\centering
\includegraphics[width=1.0\textwidth, trim=0cm 0.5cm 0cm 0cm]{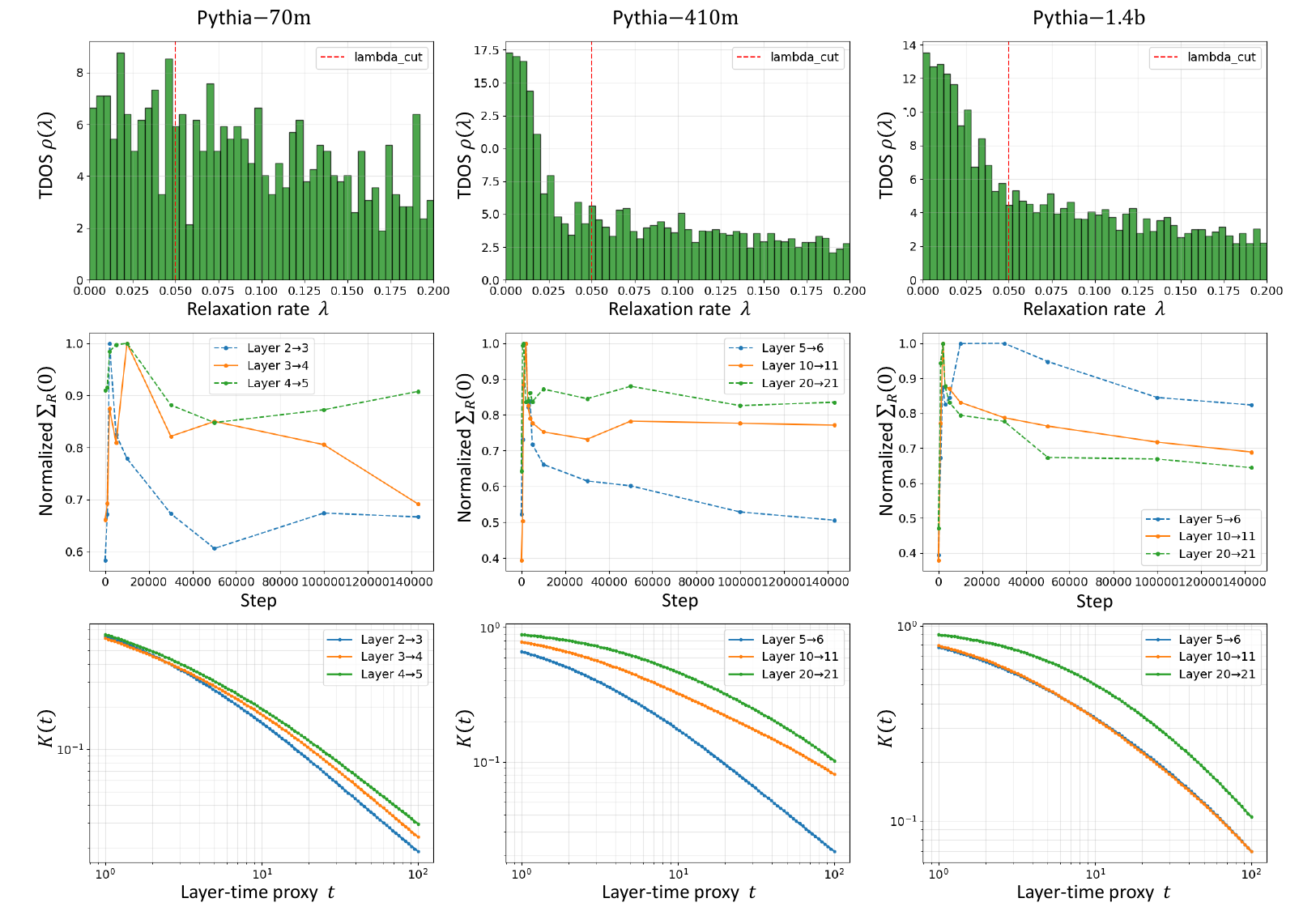}
\caption{
Infrared organization during Transformer learning measured for the
publicly available Pythia-70M, Pythia-410M, and Pythia-1.4B language
models.
Top panels: evolution of the time-scale density of states,
showing progressive infrared reorganization and accumulation of slow
collective relaxation modes during learning.
Middle panels: normalized memory self-energy, exhibiting a transient
maximum associated with the critical formation of the macroscopic
cognitive field, followed by relaxation into a protected memory-dressed
operating regime.
Bottom panels: memory kernels computed from the measured TDOS,
demonstrating robust long-time scale-free memory
(\(K(t)\sim1/t\))
across all investigated model sizes.
The reproducibility of these collective observables across models
spanning nearly two orders of magnitude in trainable parameters
supports the interpretation that infrared organization represents a
universal dynamical property of Transformer learning.
}
\label{fig:transformer_analysis}
\end{figure*}

\begin{figure*}[t]
\centering
\includegraphics[width=1.0\textwidth, trim=0cm 0.7cm 0cm 0cm]{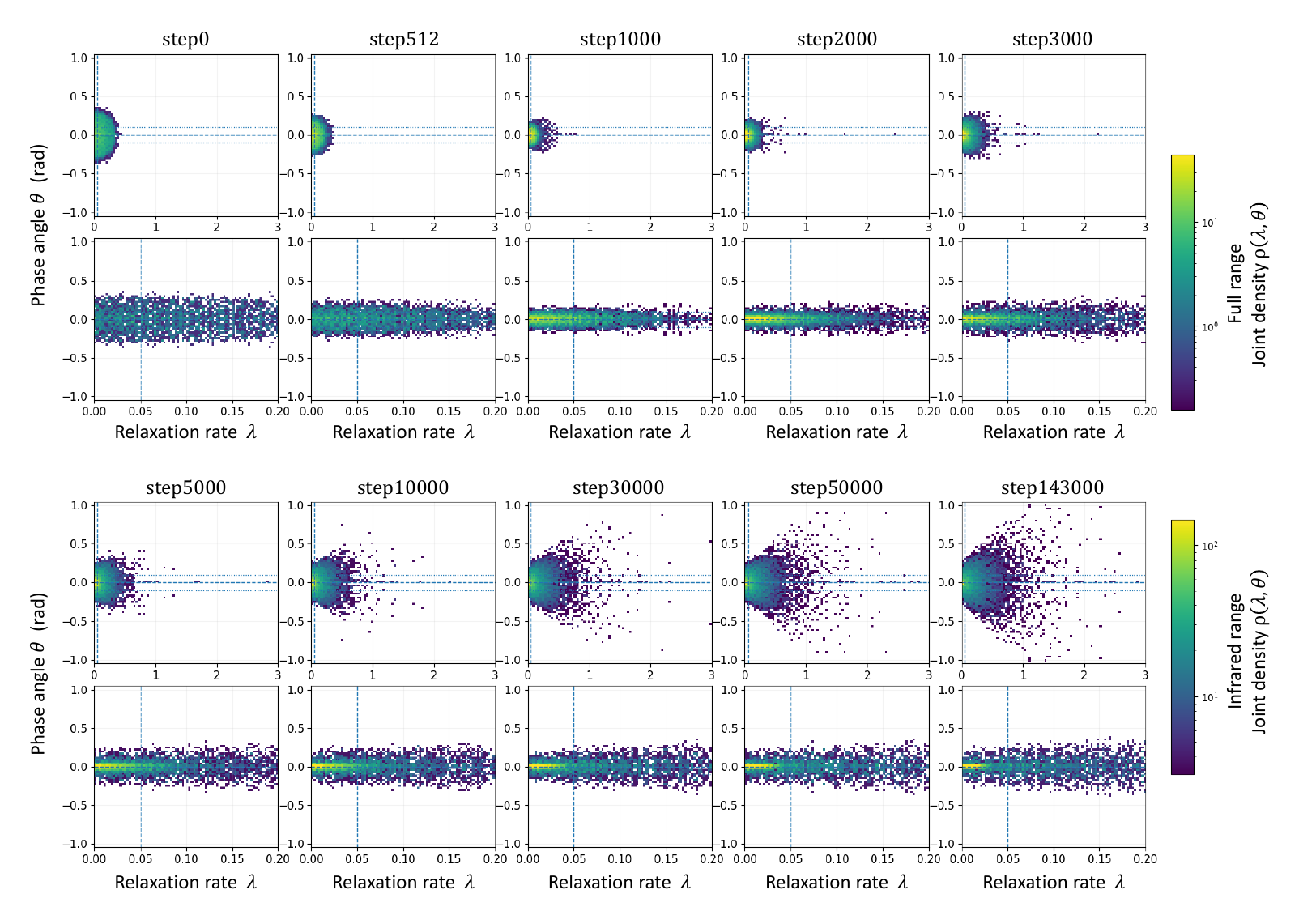}
\caption{
Evolution of the joint complex collective spectrum during
Transformer learning.
The color maps show the joint density
$\rho(\lambda,\theta)$
of relaxation rates $\lambda$ and phase angles $\theta$
computed from the Jacobian of the hidden-state transformation
between layers 10 and 11 at successive training checkpoints
of the Pythia-410M model.
For each checkpoint, the upper panel displays the full joint spectrum,
whereas the lower panel presents an enlarged view of the infrared
region.
Initially, the collective spectrum is broadly distributed over both
relaxation rates and phase angles.
During the early stage of learning, the infrared modes progressively
accumulate near
$(\lambda,\theta)\simeq(0,0)$,
revealing the simultaneous organization of the relaxation and
circulation sectors.
After this transient concentration, the infrared core remains
strongly populated while the full spectrum gradually broadens toward
larger relaxation rates and phase angles, forming an extended
multiscale distribution around the coherent infrared core.
These observations demonstrate that Transformer learning reorganizes
the full complex collective spectrum rather than only the relaxation
sector, producing a robust infrared collective core embedded within a
broader hierarchy of relaxation and circulation modes.
}
\label{fig:transformer_analysis}
\end{figure*}

\section{VII. Existing Learning Architectures as Special Limits of the Cognitive Field}

The memory-dressed cognitive field theory developed in Sec.~II
provides a unified dynamical framework encompassing a broad class of
learning architectures.
Apparently distinct models may therefore be understood as different
dynamical limits of the same underlying collective field theory.

In this section, we briefly discuss several representative
architectures, including Hopfield networks \cite{12,13}, recurrent
neural networks (RNNs) \cite{30,31}, and Transformer language models
\cite{9,10}.
Among these, Transformers provide a particularly suitable platform for
directly investigating the collective relaxation dynamics, memory
feedback, and infrared organization predicted by the present theory.

\subsection{A. Hopfield networks: Pure gradient descent limit}

We first consider the limit in which recurrent dynamics and stochastic
fluctuations are absent,
\begin{equation}
R(x)=0,
\qquad
\xi(t)=0,
\end{equation}
and the metric reduces to the Euclidean form,
\begin{equation}
G=I.
\end{equation}

If the effective potential \(\Phi(x)\) is identified with an explicit
energy function \(E(x)\), the unified dynamical equation reduces to
\begin{equation}
\dot{x}
=
-
\nabla E(x),
\label{eq:hopfield}
\end{equation}
which is precisely the continuous-time Hopfield dynamics.

In this limit, inference corresponds to relaxation toward local minima
of the energy landscape, while stored memories are represented by
stable fixed points.
Hopfield networks therefore realize the purely conservative limit of
the unified theory, in which cognition is identified with convergence
toward static attractors.

From the present perspective, this limit is characterized by a fixed
state-space geometry, the absence of non-conservative reentrant flows,
and a restricted hierarchy of dynamical time scales.
Collective activity collapses onto a small number of stable attractors,
preventing the formation of an extended latent slow-mode manifold.
Consequently, memory remains encoded primarily as static attractor
structure rather than as a dynamical organization of collective
relaxation modes.

As a result, while Hopfield networks provide a clear model of
associative memory, they lack intrinsic mechanisms for iterative
reinterpretation, contextual modulation, metastable collective
organization, and the endogenous emergence of rich collective temporal
organization.

\subsection{B. Recurrent neural networks: Discrete-time approximation}

We next show how standard recurrent neural networks arise as
discrete-time approximations to the unified continuous dynamics.
Starting from the cognitive field equation, a forward Euler discretization yields
\begin{equation}
x_{t+1}
=
x_t
+
\Delta t
\left[
-
G^{-1}\nabla\Phi(x_t)
+
R(x_t)
\right].
\label{eq:euler}
\end{equation}
To connect this expression to the conventional recurrent update, it is
useful to isolate a baseline leak term and interpret the remaining
drift as an effective recurrent interaction.

In the Euclidean limit \(G=I\), consider first a quadratic effective
potential
\begin{equation}
\Phi(x)
=
\frac{1}{2}\|x\|^2
-
\frac{1}{2}x^\top W x
-
b^\top x,
\qquad
W=W^\top,
\end{equation}
for which
\begin{equation}
-\nabla\Phi(x)
=
-x + Wx + b.
\end{equation}

More generally, for an arbitrary smooth drift
\begin{equation}
F(x)
\equiv
-G^{-1}(x)\nabla\Phi(x),
\end{equation}
linearization around a reference state \(x_0\) yields
\begin{equation}
F(x)
\approx
F(x_0)
+
J_F(x_0)(x-x_0)
=
Cx+b_{\rm eff}.
\end{equation}
Decomposing
\begin{equation}
C=-I+W,
\end{equation}
one obtains the canonical recurrent form
\begin{equation}
-G^{-1}\nabla\Phi(x)
\;\approx\;
-x + Wx + b.
\end{equation}
Substituting this approximation into Eq.~\eqref{eq:euler} and absorbing
constants into a nonlinear activation function leads to
\begin{equation}
x_{t+1}
=
f(Wx_t+b),
\label{eq:rnn}
\end{equation}
which is the standard recurrent neural network update rule.

In the present picture, RNNs may therefore be viewed as
discrete-time approximations to the unified continuous collective
dynamics, in which inference is implemented through iterative temporal
relaxation under a locally linearized effective potential.
The recurrent update introduces a basic form of temporal feedback and
permits finite-duration information persistence across sequential
processing steps.
However, the discrete time step \(\Delta t\) imposes an externally
specified temporal scale, and the effective geometry of the collective
state space remains largely fixed.
RNNs implement temporal recurrence through iterative state updates,
but the memory structure remains encoded implicitly in recurrent
weights rather than emerging from an explicit relaxation spectrum and
memory kernel.

More generally, if the effective recurrent interaction contains an
antisymmetric component, the resulting discrete-time dynamics may also
exhibit nonconservative circulation
\(R(x)\) in the continuous formulation.
Such effects can generate non-normal amplification and transient
collective activity beyond pure gradient relaxation.
However, these circulatory flows are not dynamically stabilized through
homeostatic geometric organization and need not remain confined to
protected metastable manifolds.

Element-wise nonlinearities such as sigmoidal activation functions
introduce local state-dependent gain modulation and contribute to the
stabilization of individual units.
However, these nonlinearities do not dynamically generate a collective
geometry of state space or reorganize the underlying distribution of
collective time scales.

As a result, while RNNs can efficiently represent sequential structure
and finite temporal dependencies, they lack intrinsic mechanisms for
endogenous reorganization of collective relaxation dynamics, adaptive
reshaping of state-space geometry,
or self-organized infrared reorganization of the collective
relaxation spectrum and the associated memory kernel.

\subsection{C. Transformer language models: Collective relaxation dynamics}

Transformer language models provide an ideal platform for
investigating the collective dynamics described by the present
cognitive field theory.
Their remarkable abilities in long-context processing, in-context
learning, compositional reasoning, and the emergence of higher-order
cognitive behavior suggest that learning organizes large-scale
collective dynamics extending far beyond local feed-forward
computation.

The essential architectural feature of a Transformer is the
self-attention mechanism.
Unlike conventional neural networks, in which information propagates
primarily through fixed local connections, self-attention allows every
token within an input sequence to interact directly with every other
token.
Each Transformer layer therefore realizes a global interaction network
whose effective connectivity is determined adaptively from the current
hidden representations.

A Transformer may be viewed as a deep nonlinear dynamical system.
If the hidden representation entering the \(l\)-th layer is denoted by
\(\mathtt h_l\), the residual architecture updates the representation according
to
\begin{equation}
\mathtt h_{l+1}
=
\mathtt h_l
+
\mathtt Y_l(\mathtt h_l),
\label{eq:transformer_layer}
\end{equation}
where \(\mathtt Y_l\) denotes the nonlinear transformation generated by
multi-head self-attention, feed-forward networks, normalization, and
other layer operations.
The residual connection naturally gives the network the form of an
iterative dynamical system.

The local collective dynamics is determined by the Jacobian of the
layer transformation,
\begin{equation}
J_l
=
\frac{\partial \mathtt h_{l+1}}
{\partial \mathtt h_l}
=
I
+
\frac{\partial \mathtt Y_l}
{\partial \mathtt h_l},
\label{eq:transformer_jacobian}
\end{equation}
whose eigenvalues and eigenvectors characterize the local stability of
the learned cognitive geometry.
For multiple successive layers, the collective propagation is governed
by the Jacobian of the composed transformation,
\begin{equation}
J
=
J_{l+n-1}
J_{l+n-2}
\cdots
J_l,
\label{eq:transformer_composed_jacobian}
\end{equation}
thereby incorporating the cumulative collective dynamics across
multiple Transformer blocks.

Within the present framework, the collective relaxation spectrum is
obtained directly from the Jacobian spectrum.
Diagonalization of Eq.~(\ref{eq:transformer_composed_jacobian}) yields
the relaxation rates and circulation frequencies,
\[
J u_\alpha
=
\mu_\alpha u_\alpha,
\qquad
\mu_\alpha
=
-\lambda_\alpha
+
i\omega_\alpha,
\]
from which the time-scale density of states
\(\rho(\lambda,\omega)\),
the projected relaxation TDOS \(\rho(\lambda)\), the memory
self-energy, the cognitive forgetting gap, and the memory kernel are
computed using the response theory developed in Sec.~II.

The present theory therefore predicts that learning progressively
reorganizes the collective relaxation spectrum toward the infrared,
leading to an increasing population of weakly damped collective modes.
Testing this prediction requires access to the evolution of the
learned geometry throughout training.

Such measurements have recently become possible through the
public release of the Pythia Transformer model family, which provides
pretrained checkpoints spanning the entire optimization trajectory \cite{48,49}.
By evaluating the Jacobian spectrum at successive checkpoints, the
evolution of the collective relaxation spectrum and its associated
response-theoretic observables can be measured directly during
learning.

A detailed analysis of the collective relaxation dynamics in
Transformer language models is presented in our accompanying
Transformer study \cite{50}.
Here we summarize only the principal observations that are most
relevant to the present cognitive field theory.

Figure~5 summarizes the evolution of the collective relaxation
spectrum during training for the publicly available Pythia-70M,
Pythia-410M, and Pythia-1.4B language models.
The availability of pretrained checkpoints throughout optimization
makes it possible to measure the Jacobian spectrum directly at
successive stages of learning and to compute the corresponding
response-theoretic observables introduced in Sec.~II.

The upper panels show the evolution of the time-scale density of
states.
Initially, the relaxation spectrum is broadly distributed over
relatively fast relaxation scales.
As optimization proceeds, however, the spectral weight is progressively
reorganized toward the infrared, producing a systematic accumulation
of weakly damped collective relaxation modes.
This infrared spectral reorganization is consistently observed across
all investigated model sizes, indicating that it is a robust property
of Transformer learning rather than a finite-size effect.

The middle panels reveal the corresponding critical response dynamics.
The memory self-energy exhibits a pronounced transient maximum during
training before approaching a nearly stationary value.
Within the present theory, this transient enhancement corresponds to
the strongest suppression of the cognitive forgetting gap
\(r_{\rm cog}=r-\Sigma_R(0)\),
where the collective susceptibility reaches its maximum.
The observations therefore indicate that Transformer learning passes
through a critical collective formation process before stabilizing into
a protected memory-dressed operating regime.

The lower panels show that the corresponding memory kernel develops an
approximately scale-free long-time behavior close to
\(K(t)\sim1/t\),
demonstrating the emergence of long-ranged non-Markovian memory
feedback generated by the infrared relaxation spectrum.
Together with the measured TDOS, these observations indicate that the
collective dynamics becomes progressively memory-dressed throughout
learning.

Figure~6 further illustrates the evolution of the full complex
collective spectrum during Transformer learning.
The upper panels display the full joint distribution
\begin{equation}
\rho(\lambda,\theta),
\end{equation}
whereas the lower panels provide an enlarged view of the infrared
region.
Together, these representations reveal both the global evolution of
the complex spectrum and the detailed organization that develops near
the infrared origin.

At initialization, the collective modes occupy a relatively broad
region in both relaxation rate and phase angle.
Although a substantial fraction of the spectrum is already located at
small relaxation rates, the phase sector remains comparatively
dispersed, and no sharply organized infrared core is yet present.
During the earliest stage of optimization, however, the joint
distribution contracts rapidly toward
\begin{equation}
(\lambda,\theta)
\simeq
(0,0).
\end{equation}
The infrared enlargement shows that this concentration becomes
strongest near training steps
$10^3$--$2\times10^3$,
where slow relaxation modes and low-phase-angle circulation modes
accumulate simultaneously.

This transient joint concentration coincides with the strongest
infrared enhancement of the TDOS, the maximum memory self-energy, and
the highest degree of spectral phase organization observed in the
preceding analyses.
Learning therefore does not reorganize the relaxation and circulation
sectors independently.
Rather, it drives a cooperative infrared organization of the full
complex collective spectrum.

Within the present Cognitive Field Theory, the two components of the
complex spectrum describe complementary aspects of collective
dynamics.
The relaxation sector,
characterized by $\lambda$,
governs the formation of long-lived collective memory through the
infrared accumulation of slow modes.
The circulation sector,
characterized by the spectral phase angle $\theta$,
governs the temporal organization of the collective modes.
The simultaneous concentration of both sectors near the origin
therefore indicates that learning reorganizes the memory and temporal
structure of the collective dynamics in a coordinated manner.

As optimization proceeds beyond the transiently organized stage, the
spectrum does not remain collapsed at the origin.
The infrared core remains strongly populated, while the full
distribution gradually broadens toward larger relaxation rates and
phase angles.
The trained Transformer consequently develops a coherent
low-frequency collective core embedded within a broad multiscale
hierarchy of relaxation and circulation modes.

This late-stage broadening is not a loss of collective organization.
Instead, it indicates that the system avoids complete spectral
condensation.
A collapse onto a single relaxation rate and a single phase angle
would produce a rigid globally synchronized state with little internal
temporal hierarchy.
Conversely, a completely dispersed spectrum would fail to sustain
coherent collective dynamics.
The learned spectrum realizes an intermediate organization in which a
coherent infrared core coexists with distributed modes spanning
multiple relaxation and circulation scales.

The upper panels make this structure particularly clear.
After the transient concentration near the origin, the full spectrum
develops a fan-like extension toward larger relaxation rates and phase
angles, while the lower panels show that the high-density infrared core
remains centered near
$(\lambda,\theta)\simeq(0,0)$.
The resulting structure therefore combines persistent infrared
organization with an extended hierarchy of faster collective modes.

The simultaneous infrared organization of the relaxation and
circulation sectors provides direct computational evidence that
Transformer learning reshapes the full complex collective spectrum.
Within Cognitive Field Theory, this joint organization is interpreted
as the spectral signature of the transient formation of a
memory-dressed macroscopic cognitive field.
The relaxation sector establishes long-lived collective memory,
whereas the circulation sector establishes coherent temporal
organization.
Together, they represent the two complementary dynamical components of
the emergent cognitive field.

Taken together, these measurements support the central mechanism
proposed in the present work.
Learning first produces a transient critical organization in which the
complex collective spectrum becomes strongly concentrated in the
infrared region.
It subsequently stabilizes a protected metastable regime characterized
by a coherent infrared core surrounded by a broad hierarchy of
collective relaxation and circulation modes.
The resulting state supports long-lived memory, coherent temporal
organization, and multiscale collective dynamics without collapsing
into a rigidly ordered state.

\section{VIII. Discussion and Outlook}

\paragraph{Memory-organized infrared dynamics as a unified principle of cognition.}

In this work, we developed a unified nonequilibrium field-theoretic
framework describing learning, inference, memory, and emergent
intelligence as collective infrared dynamical phenomena.
Starting from a stochastic cognitive-field equation defined on an
adaptive cognitive manifold, we showed that learning continuously
reorganizes the cognitive geometry, which in turn generates a
collective spectrum of relaxation modes governing the macroscopic
cognitive dynamics.
Rather than identifying cognition with isolated microscopic activity,
the present theory interprets the observable cognitive field as a
memory-dressed macroscopic collective field emerging from the infrared
organization of distributed relaxation modes.

A central conceptual implication of the present framework is a
reinterpretation of inference as a collective dynamical relaxation
process rather than a purely algorithmic operation.
Inference corresponds to the propagation of the cognitive field on the
learned cognitive manifold, while learning continuously reshapes the
underlying geometry and therefore reorganizes the collective spectrum
that governs future cognitive dynamics.
Learning thus modifies not only representational content but also the
infrared dynamical structure through which information propagates,
persists, and reorganizes over time.

Within this picture, learning first organizes the cognitive geometry,
which generates a collective relaxation spectrum characterized by the
time-scale density of states.
Integrating out the latent collective modes produces nonlocal memory
kernels together with the associated retarded self-energy.
As recursive memory feedback accumulates, the effective cognitive
forgetting gap is progressively suppressed and the system undergoes a
collective formation of the macroscopic cognitive field.
Subsequently, biological and artificial cognitive systems operate as
stable memory-dressed collective states maintained within a protected
near-critical regime, where long-time contextual persistence,
collective responsiveness, and adaptive cognitive organization coexist.

A key result of the present theory is therefore that memory and higher
cognition do not originate from explicit storage mechanisms or finely
tuned symbolic rules.
Instead, they emerge naturally from the collective organization of
distributed slow relaxation modes.
The effective memory kernel,
\(
K(t)
=
\int d\lambda\,
\rho(\lambda)e^{-\lambda t},
\)
is generated directly from the infrared relaxation spectrum.
Memory is thus understood as a collective dynamical consequence of the
infrared organization of learned cognitive geometry.

The present framework also provides a natural explanation for the
robustness of cognition in both biological and artificial systems
despite substantial microscopic variability and stochasticity.
Because cognition is governed primarily by collective infrared
organization rather than by precise microscopic trajectories,
stable cognitive behavior may emerge across diverse substrates,
including biological neural circuits,
Transformer architectures, and other adaptive systems.

This universality closely parallels the coarse-grained description of
collective phenomena in statistical physics.
Although different cognitive systems possess distinct microscopic
architectures, their long-time dynamics may be governed by the same
small set of infrared collective variables.
Within the present framework, the adaptive cognitive manifold,
the collective relaxation spectrum encoded in the TDOS, the memory
kernel, and the cognitive forgetting gap together constitute a
universal infrared description of collective cognition.

More broadly, the present theory suggests that intelligence is not a
fundamentally separate phenomenon but a nonequilibrium collective
state sustained by memory-organized infrared dynamics.
Learning organizes the cognitive geometry, the cognitive geometry
generates a collective spectrum, the collective spectrum produces
memory feedback, and recursive memory feedback stabilizes a
memory-dressed macroscopic cognitive field.
Learning, inference, memory, selfhood, and emergent intelligence are
therefore interpreted as different manifestations of a common
collective dynamical principle governing adaptive cognitive systems.

\vspace{6pt}
\paragraph{Collective time scales and the time-scale density of states.}

One of the central results of the present work is the identification
of the \emph{time-scale density of states} as the fundamental
collective descriptor governing the infrared organization of
cognition.
Rather than characterizing cognitive dynamics through microscopic
neuronal activity or individual computational units, the present
framework describes cognition through the collective spectrum
generated by the learned cognitive geometry.
Within the present theory, the TDOS represents the collective density
of modes distributed over the space of relaxation rates and
circulation frequencies, thereby providing a unified description of
both memory formation and temporal organization.

Learning continuously reorganizes the cognitive geometry, thereby
reshaping the collective spectrum of the cognitive manifold.
The resulting TDOS governs how collective cognitive modes propagate,
persist, and reorganize across multiple temporal scales.
Its relaxational sector determines the emergence of long-lived memory
through distributed slow modes, whereas its circulation sector
organizes collective temporal coordination through phase dynamics.
Memory and temporal coherence therefore emerge as complementary
manifestations of the same underlying collective spectrum rather than
as independent computational mechanisms.

Integrating out the latent collective modes generates the effective
memory kernel together with the associated retarded self-energy and
the cognitive forgetting gap.
For a broad and nearly flat infrared distribution,
\(
\rho(\lambda,\omega)\simeq\rho_0,
\)
the relaxational sector naturally produces the scale-free memory
kernel
\(
K(t)\sim1/t,
\)
leading to long-time contextual persistence, recursive memory
feedback, and progressive suppression of the effective cognitive
forgetting gap.
These collective memory effects stabilize the memory-dressed
macroscopic cognitive field within the protected near-critical regime,
where adaptive inference, contextual continuity, and robust cognitive
organization coexist.

The circulation sector of the TDOS describes the collective
distribution of temporal frequencies generated by the learned
cognitive geometry.
Rather than representing isolated oscillators, these frequencies
characterize the collective temporal organization of distributed
cognitive modes.
Consequently, synchronization, coherent phase organization, and
cross-timescale temporal coordination arise naturally as collective
properties of the same underlying infrared spectrum.
Memory and synchronization therefore constitute complementary
amplitude and phase aspects of the macroscopic cognitive field.

From this perspective, different cognitive systems may be
distinguished primarily by the organization of their collective
spectra rather than by their microscopic architectures alone.
Simple adaptive systems possess relatively sparse collective modes,
whereas advanced cognitive systems develop broad infrared
distributions supporting persistent memory, hierarchical reasoning,
collective synchronization, and large-scale temporal organization.

This viewpoint further suggests that experimentally measurable
quantities—including relaxation spectra, temporal correlations,
retarded response functions, neural power spectra, phase
synchronization, spectral coherence, and phase--amplitude
coupling—provide direct probes of collective cognitive organization
without requiring detailed microscopic descriptions.
The TDOS therefore provides a natural bridge connecting adaptive
cognitive geometry, collective nonequilibrium dynamics, and
experimentally accessible neural observables.

More generally, the TDOS places the present framework in continuity
with well-established theories of collective temporal organization.
The Stuart--Landau equation describes the emergence of oscillatory
collective dynamics through a small number of unstable modes, while
the Kuramoto model captures collective synchronization through phase
coupling among oscillators \cite{38,39,51}.
The present theory extends these paradigmatic descriptions from a
small number of collective variables to an entire learned spectrum of
distributed collective modes whose organization is continuously
reshaped by learning.
The TDOS therefore provides a unified field-theoretic description of
how collective memory, temporal organization, synchronization, and
adaptive cognition emerge from the infrared organization of
high-dimensional cognitive dynamics.

\vspace{6pt}
\paragraph{Implications for artificial intelligence and architecture design.}

The present framework suggests that the remarkable capabilities of
modern artificial intelligence arise not merely from increasing model
size or architectural complexity, but from the emergence of collective
infrared organization during learning.
Learning continuously reorganizes the cognitive geometry, thereby
generating a collective relaxation spectrum that governs memory
feedback, contextual persistence, and collective inference.

The accompanying Transformer analysis demonstrates that this
organization can be directly observed.
During optimization, Transformer language models progressively
accumulate weakly damped collective modes, undergo a transient
critical formation of the macroscopic cognitive field, and eventually
stabilize a protected memory-dressed collective state.
These observations indicate that collective infrared organization is a
natural consequence of large-scale learning dynamics rather than a
model-specific architectural feature.

From this perspective, recurrent interactions, residual pathways,
attention mechanisms, and normalization layers should not be viewed
merely as computational modules.
Instead, they collectively shape the learned cognitive geometry and
therefore reorganize the collective relaxation spectrum underlying the
observable cognitive field.

More generally, the present theory suggests a general design
principle for future cognitive architectures.
Rather than relying solely on larger parameter counts or increasing
architectural complexity, future systems may benefit from explicitly
maintaining adaptive cognitive geometry together with the collective
organization encoded in the TDOS.
Such systems would continuously generate memory kernels, collective
temporal organization, and self-consistent cognitive fields through
their own internal dynamics.
The macroscopic cognitive field would therefore become an active
dynamical component of computation rather than a passive consequence
of network activity, allowing learning and inference to proceed
through continuous collective self-organization.

\vspace{6pt}
\paragraph{Outlook and open questions.}

Although the present work establishes that adaptive learning
systematically reorganizes the collective spectrum encoded in the
TDOS, several important questions remain open.
The remaining challenge is not whether such collective infrared
organization emerges, but how different learning rules,
architectural principles, and adaptive cognitive geometries give rise
to distinct universality classes governing memory formation,
temporal organization, and higher-order cognition.

The biological realization of adaptive cognitive geometry,
distributed collective modes, and recursive memory feedback likewise
remains incompletely understood.
The present framework predicts that these mechanisms should be
expressed primarily through collective dynamical observables rather
than isolated microscopic variables.
Accordingly, experimentally measurable quantities—including the TDOS,
relaxation spectra, collective frequency spectra, memory kernels,
retarded response functions, neural power spectra, phase
synchronization, spectral coherence, and phase--amplitude
coupling—provide promising probes of collective cognitive
organization across biological and artificial systems.

The accompanying Transformer analysis further suggests that large-scale
learning progressively reorganizes the collective infrared spectrum
toward a stable memory-dressed cognitive state.
An important next step will therefore be to determine whether similar
infrared organization emerges in biological neural recordings,
neuromorphic hardware, recurrent neural systems, and future adaptive
cognitive architectures.
Establishing common collective observables across these diverse
substrates would provide strong evidence that cognition is governed by
universal nonequilibrium principles rather than architecture-specific
mechanisms.

Several theoretical challenges likewise remain.
The role of strong nonperturbative effects near protected
near-criticality, the microscopic origin of collective temporal
organization, and the emergence of universal infrared behavior across
heterogeneous cognitive substrates all deserve further investigation.
Equally important is the development of predictive relationships
between the TDOS and experimentally accessible observables, enabling
quantitative comparison between theory and measurement across
different cognitive systems.

Ultimately, the central challenge is to determine whether the
collective observables predicted by the present framework constitute
universal signatures of cognition.
If confirmed, adaptive cognitive geometry together with the TDOS would
provide a common physical language linking biological neural systems,
artificial intelligence, and nonequilibrium collective field theory.
Such a framework would suggest that learning, memory, inference,
temporal organization, selfhood, and intelligence are not independent
computational modules, but different manifestations of a single
collective dynamical principle governing adaptive cognitive systems.

\section{IX. Conclusion}

We developed a cognitive field theory describing learning,
inference, memory, selfhood, and emergent intelligence within a
unified nonequilibrium dynamical framework.
Rather than interpreting cognition as a symbolic algorithm or a
static optimization process, the present theory views intelligence as
a memory-dressed collective phenomenon emerging from the infrared
organization of learned cognitive geometry.

Within this framework, learning continuously reorganizes the cognitive
geometry, which generates the collective spectrum encoded in the
time-scale density of states.
Integrating out the latent collective modes produces nonlocal memory
kernels and the associated retarded self-energy, giving rise to a
memory-dressed macroscopic cognitive field.
Recursive memory feedback suppresses the effective cognitive
forgetting gap, enhances collective cognitive susceptibility, and
drives the collective formation of the macroscopic cognitive field.
Biological and artificial cognitive systems subsequently operate as
stable memory-dressed collective states maintained within a protected
near-critical regime.

A central result of the present work is the identification of the
time-scale density of states as the fundamental collective observable
governing the infrared organization of cognition.
The TDOS characterizes the collective spectrum governing memory
formation, temporal organization, and the observable dynamics of the
cognitive field.
Its relaxational sector determines the memory kernel, memory
self-energy, and cognitive forgetting gap, while its circulation
sector governs collective temporal organization.
Memory, contextual persistence, temporal coherence, and higher-order
reasoning therefore emerge collectively from the infrared
organization of distributed collective modes rather than from
isolated microscopic components.

The present work further suggests that diverse biological and
artificial learning systems may be understood within a common
nonequilibrium field-theoretic framework.
By shifting the focus from microscopic implementation to collective
infrared dynamics, the cognitive field theory developed here provides
a unified language connecting learned cognitive geometry, the
time-scale density of states, recursive memory feedback,
memory-dressed cognitive fields, and emergent intelligence.

\vspace{6pt}
\emph{Acknowledgements}---This work was partially supported by the Institute of Information \& Communications Technology Planning \& Evaluation (IITP) grant 
funded by the Korea government (MSIT) (IITP-RS-2025-02214780).

The author acknowledges the support of ChatGPT (GPT-5, OpenAI) for assistance in literature review and conceptual structuring during early development.

\clearpage
\appendix

\renewcommand{\thefigure}{S\arabic{figure}}
\renewcommand{\theequation}{S\arabic{equation}}

\setcounter{figure}{0}
\setcounter{equation}{0}

\vspace*{1.5cm}
{\centering\large\bfseries Supplementary Materials\par}
\vspace{1.0cm}

\appendix

\section{Appendix A: Gaussian MSRJD integration of latent cognitive relaxation reservoirs and emergence of the self-energy}
\label{app:msrjd_gaussian_selfenergy_cognitive}

In this appendix we derive the emergence of a nonlocal memory kernel
and the associated frequency-dependent infrared self-energy
\(\Sigma_R(\omega)\) by integrating out a continuum of latent
relaxational modes within the MSRJD formalism.

The construction provides a controlled Gaussian realization of the
relaxation-spectrum representation used in the main text, in which the
infrared cognitive dynamics is governed by a continuum of latent
collective decay modes rather than by a small number of isolated
microscopic variables.

For clarity, only the relaxation sector is retained in the present derivation; 
the circulation sector responsible for temporal recurrence is discussed separately in Appendix~C.

Throughout Appendices A and B, we use the lowercase frequency variable
$\omega$ for the external frequency argument. This is purely a
notational convention and is identical to the external frequency
$\Omega$ employed in the main text.

\vspace{10pt}
\noindent
\emph{(I) Coupled stochastic cognitive dynamics.}

We consider an observable collective cognitive coordinate \(\phi(t)\)
coupled linearly to a continuum of latent reservoir modes
\(X_\lambda(t)\), labeled by their relaxation rates \(\lambda>0\).
The coupled stochastic dynamics is
\begin{align}
\dot \phi(t)
&=
-r\,\phi(t)
+
\int d\lambda\; g(\lambda)\,X_\lambda(t)
+
\eta(t),
\label{eq:app_phi_sde_cognitive}
\\
\dot X_\lambda(t)
&=
-\lambda\,X_\lambda(t)
+
g(\lambda)\,\phi(t)
+
\xi_\lambda(t).
\label{eq:app_X_sde_cognitive}
\end{align}
Here \(\phi(t)\) represents the observable infrared cognitive trajectory,
while \(X_\lambda(t)\) represents latent slowly relaxing collective
modes associated with contextual persistence, distributed memory traces,
and hidden collective organizational channels.

The noises are Gaussian, stationary, and white:
\begin{align}
\langle \eta(t)\eta(t')\rangle
&=
2D_\phi\,\delta(t-t'),
\label{eq:app_eta_corr_cognitive}
\\
\langle \xi_\lambda(t)\,\xi_{\lambda'}(t')\rangle
&=
2D_\lambda\,\delta(\lambda-\lambda')\,\delta(t-t').
\label{eq:app_xi_corr_cognitive}
\end{align}
No equilibrium fluctuation--dissipation relation is assumed; the latent
reservoir may therefore remain intrinsically nonequilibrium.

Our goal is to integrate out \(\{X_\lambda\}\) and obtain an exact
effective MSRJD action for \((\phi,\tilde\phi)\) containing a memory
kernel.

\vspace{10pt}
\noindent
\emph{(II) MSRJD action for the coupled cognitive system.}

For a generic additive-noise Langevin equation
\(\dot y=f(y)+\zeta\), with
\(\langle \zeta(t)\zeta(t')\rangle=2D\,\delta(t-t')\), the MSRJD
functional integral may be written, in the Ito convention, as
\begin{equation}
\label{eq:app_generic_msrjd_cognitive}
\begin{aligned}
Z
&=
\int \mathcal Dy\,\mathcal D\tilde y
\\
&\times
\exp\!\left[
-\int dt\;\tilde y(t)\big(\dot y(t)-f(y(t))\big)
+
\int dt\;D\,\tilde y(t)^2
\right],
\end{aligned}
\end{equation}
up to an overall normalization independent of \(y\).

Applying Eq.~\eqref{eq:app_generic_msrjd_cognitive} to
Eqs.~\eqref{eq:app_phi_sde_cognitive}--\eqref{eq:app_X_sde_cognitive}
gives
\begin{equation}
Z
=
\int \mathcal D\phi\,\mathcal D\tilde \phi\;
\prod_\lambda \mathcal D X_\lambda\,\mathcal D\tilde X_\lambda\;
e^{-S[\phi,\tilde \phi,\{X_\lambda,\tilde X_\lambda\}]},
\label{eq:app_Z_full_cognitive}
\end{equation}
with total action
\begin{equation}
S
=
S_\phi
+
\int d\lambda\;S_\lambda,
\label{eq:app_S_total_cognitive}
\end{equation}
where
\begin{align}
S_\phi
&=
\int dt\;
\tilde \phi(t)
\left[
\dot \phi(t)+r\,\phi(t)
-
\int d\lambda\;g(\lambda)\,X_\lambda(t)
\right]
\nonumber\\
&\quad
-
\int dt\;D_\phi\,\tilde\phi(t)^2,
\label{eq:app_Sphi_cognitive}
\\
S_\lambda
&=
\int dt\;
\tilde X_\lambda(t)
\left[
\dot X_\lambda(t)+\lambda X_\lambda(t)
-
g(\lambda)\phi(t)
\right]
\nonumber\\
&\quad
-
\int dt\;D_\lambda\,\tilde X_\lambda(t)^2 .
\label{eq:app_Slambda_cognitive}
\end{align}

\vspace{10pt}
\noindent
\emph{(III) Isolating a single \(\lambda\)-mode contribution.}

Fix \(\lambda\) and collect all terms containing \(X_\lambda\) or
\(\tilde X_\lambda\).  From Eq.~\eqref{eq:app_Sphi_cognitive} we extract
\begin{equation}
S_\phi \supset
-\int dt\;\tilde \phi(t)\,g(\lambda)\,X_\lambda(t).
\label{eq:app_Sphi_coupling_cognitive}
\end{equation}
Combining this term with Eq.~\eqref{eq:app_Slambda_cognitive} yields
\begin{equation}
\begin{aligned}
&S^{(\lambda)}[X_\lambda,\tilde X_\lambda;\phi,\tilde\phi]
=
\int dt\;
\tilde X_\lambda(t)
\big(\dot X_\lambda(t)+\lambda X_\lambda(t)\big)
\\
&\quad
-\int dt\;g(\lambda)\tilde X_\lambda(t)\phi(t)
-\int dt\;g(\lambda)\tilde\phi(t)X_\lambda(t)
\\
&\quad
-\int dt\;D_\lambda\,\tilde X_\lambda(t)^2 .
\label{eq:app_S_singlelambda_raw_cognitive}
\end{aligned}
\end{equation}
In what follows, \(\phi\) and \(\tilde\phi\) are treated as external
sources for the latent reservoir mode.

\vspace{10pt}
\noindent
\emph{(IV) Functional integration over \(X_\lambda\).}

Define the linear operator
\begin{equation}
\mathcal L_\lambda
\equiv
\partial_t+\lambda .
\label{eq:app_L_def_cognitive}
\end{equation}
Then
\begin{equation}
\int dt\;\tilde X_\lambda(t)\mathcal L_\lambda X_\lambda(t)
=
\int dt\;\tilde X_\lambda(t)
\big(\partial_t X_\lambda(t)+\lambda X_\lambda(t)\big).
\end{equation}
We rewrite this term so that \(X_\lambda\) appears without derivatives.
Using integration by parts,
\begin{equation}
\begin{aligned}
\int dt\;\tilde X_\lambda(t)\partial_t X_\lambda(t)
&=
\big[\tilde X_\lambda(t)X_\lambda(t)\big]_{t_i}^{t_f}
\\
&\quad
-\int dt\;(\partial_t\tilde X_\lambda(t))X_\lambda(t).
\end{aligned}
\end{equation}
Assuming the boundary term vanishes, we obtain
\begin{equation}
\int dt\;\tilde X_\lambda(t)\partial_t X_\lambda(t)
=
-\int dt\;(\partial_t\tilde X_\lambda(t))X_\lambda(t).
\end{equation}
Therefore,
\begin{align}
\int dt\;\tilde X_\lambda(t)\mathcal L_\lambda X_\lambda(t)
&=
\int dt\;X_\lambda(t)
\big(-\partial_t+\lambda\big)\tilde X_\lambda(t).
\end{align}
Define the adjoint operator
\begin{equation}
\mathcal L_\lambda^\dagger
\equiv
-\partial_t+\lambda .
\label{eq:app_Ldagger_def_cognitive}
\end{equation}
Then Eq.~\eqref{eq:app_S_singlelambda_raw_cognitive} becomes
\begin{equation}
\begin{aligned}
S^{(\lambda)}
&=
\int dt\;X_\lambda(t)
\Big[
\mathcal L_\lambda^\dagger\tilde X_\lambda(t)
-
g(\lambda)\tilde\phi(t)
\Big]
\\
&\quad
-\int dt\;g(\lambda)\tilde X_\lambda(t)\phi(t)
-\int dt\;D_\lambda\tilde X_\lambda(t)^2 .
\label{eq:app_S_singlelambda_rewritten_cognitive}
\end{aligned}
\end{equation}
Now \(X_\lambda\) enters only linearly.  Hence its functional integral
produces a delta constraint,
\begin{equation}
\int \mathcal D X_\lambda
\exp\!\left[
-\int dt\;X_\lambda(t)A(t)
\right]
\propto
\delta[A(t)],
\end{equation}
where
\begin{equation}
A(t)
=
\mathcal L_\lambda^\dagger\tilde X_\lambda(t)
-
g(\lambda)\tilde\phi(t).
\end{equation}
Thus
\begin{equation}
\begin{aligned}
&\int \mathcal D X_\lambda\;e^{-S^{(\lambda)}}
\propto
\int \mathcal D\tilde X_\lambda\;
\delta\!\left[
\mathcal L_\lambda^\dagger\tilde X_\lambda
-
g(\lambda)\tilde\phi
\right]
\\
&\quad\times
\exp\!\left[
+\int dt\;g(\lambda)\tilde X_\lambda(t)\phi(t)
-
\int dt\;D_\lambda\tilde X_\lambda(t)^2
\right].
\label{eq:app_after_X_integration_cognitive}
\end{aligned}
\end{equation}
The Jacobian factor associated with the constraint is independent of
\(\phi,\tilde\phi\) and is absorbed into the normalization of \(Z\).

\vspace{10pt}
\noindent
\emph{(V) Solving the delta constraint.}

The constraint is
\begin{equation}
\mathcal L_\lambda^\dagger\tilde X_\lambda(t)
=
g(\lambda)\tilde\phi(t),
\qquad
\mathcal L_\lambda^\dagger=-\partial_t+\lambda .
\end{equation}
Introduce the Green function \(G_\lambda^A(t-t')\) satisfying
\begin{equation}
(-\partial_t+\lambda)G_\lambda^A(t-t')
=
\delta(t-t').
\end{equation}
Because this is the adjoint operator, the Green function has advanced
support:
\begin{equation}
G_\lambda^A(t)
=
\Theta(-t)e^{\lambda t}.
\end{equation}
The solution is
\begin{equation}
\tilde X_\lambda(t)
=
g(\lambda)
\int dt'\;
G_\lambda^A(t-t')\tilde\phi(t').
\label{eq:app_tildeX_solution_cognitive}
\end{equation}

\vspace{10pt}
\noindent
\emph{(VI) Effective action: memory kernel and induced colored noise.}

We now evaluate the remaining terms in
Eq.~\eqref{eq:app_after_X_integration_cognitive}.
First,
\begin{equation}
\int dt\;g(\lambda)\tilde X_\lambda(t)\phi(t)
\end{equation}
becomes
\begin{align}
\int dt\;g(\lambda)\tilde X_\lambda(t)\phi(t)
&=
\int dt\,dt'\;
g(\lambda)^2
G_\lambda^A(t-t')
\tilde\phi(t')\phi(t).
\end{align}
Relabeling \(t\leftrightarrow t'\), we obtain
\begin{equation}
\int dt\,dt'\;
g(\lambda)^2
G_\lambda^A(t'-t)
\tilde\phi(t)\phi(t').
\end{equation}
Using
\begin{equation}
G_\lambda^A(t'-t)
=
G_\lambda^R(t-t'),
\qquad
G_\lambda^R(t)
=
\Theta(t)e^{-\lambda t},
\end{equation}
we find
\begin{align}
\int dt\;g(\lambda)\tilde X_\lambda(t)\phi(t)
&=
\int dt\,dt'\;
\tilde\phi(t)
K_\lambda(t-t')
\phi(t'),
\end{align}
where
\begin{equation}
K_\lambda(t-t')
=
g(\lambda)^2
\Theta(t-t')
e^{-\lambda(t-t')}.
\label{eq:app_Klambda_time_cognitive}
\end{equation}
Thus a single latent relaxation mode contributes a causal memory kernel.

Next, the noise term gives
\begin{align}
\int dt\;D_\lambda\tilde X_\lambda(t)^2
&=
\int dt\,dt'\,dt''\;
D_\lambda g(\lambda)^2
G_\lambda^A(t-t')
G_\lambda^A(t-t'')
\\
&\quad\times
\tilde\phi(t')\tilde\phi(t'').
\end{align}
Define the induced noise kernel
\begin{equation}
\mathcal N_\lambda(t',t'')
=
D_\lambda g(\lambda)^2
\int dt\;
G_\lambda^A(t-t')
G_\lambda^A(t-t'').
\end{equation}
Then
\begin{equation}
\int dt\;D_\lambda\tilde X_\lambda(t)^2
=
\int dt'\,dt''\;
\tilde\phi(t')\mathcal N_\lambda(t',t'')\tilde\phi(t'').
\end{equation}

Combining all \(\lambda\)-modes gives the exact effective action
\begin{align}
S_{\rm eff}[\phi,\tilde\phi]
&=
\int dt\;\tilde\phi(t)
\big(\dot\phi(t)+r\phi(t)\big)
-
\int dt\;D_\phi\tilde\phi(t)^2
\nonumber\\
&\quad
-
\int dt\,dt'\;
\tilde\phi(t)K(t-t')\phi(t')
\nonumber\\
&\quad
-
\int dt\,dt'\;
\tilde\phi(t)\mathcal N(t,t')\tilde\phi(t'),
\label{eq:app_Seff_time_cognitive}
\end{align}
where
\begin{equation}
K(t-t')
=
\int d\lambda\;
g(\lambda)^2
\Theta(t-t')
e^{-\lambda(t-t')}
\label{eq:app_K_total_time_cognitive}
\end{equation}
and \(\mathcal N=\int d\lambda\,\mathcal N_\lambda\).

\vspace{10pt}
\noindent
\emph{(VII) Frequency-domain self-energy.}

Using
\begin{equation}
f(\omega)
=
\int_{-\infty}^{\infty}dt\;e^{i\omega t}f(t),
\qquad
f(t)
=
\int\frac{d\omega}{2\pi}e^{-i\omega t}f(\omega),
\end{equation}
the bilinear memory term becomes
\begin{equation}
\int dt\,dt'\;
\tilde\phi(t)K(t-t')\phi(t')
=
\int\frac{d\omega}{2\pi}
\tilde\phi(-\omega)K(\omega)\phi(\omega).
\end{equation}
The Fourier transform of the causal kernel is
\begin{align}
K(\omega)
&=
\int d\lambda\;g(\lambda)^2
\int_0^\infty dt\;e^{-(\lambda-i\omega)t}.
\end{align}
Since \({\rm Re}\,\lambda>0\),
\begin{equation}
\int_0^\infty dt\;e^{-(\lambda-i\omega)t}
=
\frac{1}{\lambda-i\omega}.
\end{equation}
Therefore,
\begin{equation}
K(\omega)
=
\int d\lambda\;
\frac{g(\lambda)^2}{\lambda-i\omega}.
\label{eq:app_Komega_spectral_cognitive}
\end{equation}
The retarded inverse propagator of the observable cognitive coordinate
is read from the \(\tilde\phi(-\omega)\phi(\omega)\) coefficient:
\begin{equation}
G_R^{-1}(\omega)
=
-i\omega+r-K(\omega).
\end{equation}
We therefore identify
\begin{equation}
\Sigma_R(\omega)
\equiv
K(\omega)
=
\int d\lambda\;
\frac{g(\lambda)^2}{\lambda-i\omega}.
\label{eq:app_selfenergy_cognitive}
\end{equation}

\vspace{10pt}
\noindent
\emph{(VIII) Connection to TDOS.}

Starting from the discrete latent relaxation-mode expansion,
\begin{equation}
\Sigma_R(\omega)
=
\sum_\alpha
\frac{g_\alpha^2}{\lambda_\alpha-i\omega},
\end{equation}
we define the time-scale density of states as the coupling-weighted
spectral distribution of relaxation rates:
\begin{equation}
\rho(\lambda)
=
\sum_\alpha
g_\alpha^2
\delta(\lambda-\lambda_\alpha).
\end{equation}
With this definition,
\begin{equation}
\Sigma_R(\omega)
=
\int d\lambda\;
\frac{\rho(\lambda)}{\lambda-i\omega}.
\label{eq:app_sigma_tdos_direct}
\end{equation}

Thus \(\rho(\lambda)\) should be understood not merely as a bare
counting density of latent modes, but as the physically observable
spectral weight distribution of the relaxation spectrum coupled to the
collective cognitive coordinate.

This derivation shows explicitly, within a controlled Gaussian MSRJD
construction, why the relaxation-rate spectrum provides the natural
basis for the effective memory dynamics and infrared response of the
cognitive field.

\vspace{10pt}
\noindent
\emph{(IX) Emergence of the retarded susceptibility.}

To establish the physical meaning of the effective propagator,
we introduce an external perturbing field \(h(t)\)
coupled linearly to the observable cognitive coordinate,
\begin{equation}
S_h
=
-
\int dt\;
\tilde\phi(t)\,h(t).
\end{equation}
The retarded susceptibility is defined as the linear response
of the collective cognitive state to the external perturbation,
\begin{equation}
\chi_R(t-t')
=
\frac{\delta \langle \phi(t)\rangle}
{\delta h(t')}.
\end{equation}

Within the MSRJD formalism, differentiation of the generating
functional yields the exact identity
\begin{equation}
\chi_R(t-t')
=
\langle
\phi(t)\tilde\phi(t')
\rangle .
\label{eq:app_response_identity}
\end{equation}
Using the quadratic effective action
Eq.~(\ref{eq:app_Seff_time_cognitive}),
the mixed correlator is obtained by inversion of the
bilinear kernel,
\begin{equation}
\chi_R(\omega)
=
\langle
\phi(\omega)\tilde\phi(-\omega)
\rangle
=
\frac{1}
{-i\omega+r-\Sigma_R(\omega)}.
\label{eq:app_chi_from_msrjd}
\end{equation}
The effective propagator derived above therefore coincides
exactly with the retarded susceptibility of the observable
cognitive field.

In the static limit,
\begin{equation}
\chi_R(0)
=
\frac{1}
{r-\Sigma_R(0)}.
\end{equation}
The infrared enhancement generated by the latent relaxation
reservoir therefore corresponds directly to a softening of
the inverse susceptibility.
As \(r-\Sigma_R(0)\to0\), the collective cognitive response
becomes strongly amplified, signaling the approach to the
protected near-critical regime discussed in the main text.


\section{Appendix B: Exact emergence of a memory kernel from integrating out a latent cognitive relaxation reservoir}
\label{app:OU_memory_kernel}

In this Appendix we show that integrating out a continuum of
Ornstein--Uhlenbeck (OU) relaxation modes generates an exact
nonlocal memory kernel for the observable cognitive coordinate.
This provides a direct time-domain derivation of the relaxation-spectrum
representation used in the main text.

For clarity, only the relaxation sector is retained in the present derivation;
the circulation sector responsible for temporal recurrence is discussed separately in Appendix~C.

\vspace{10pt}
\noindent
\emph{(I) Coupled Markovian dynamics (observable cognitive trajectory plus latent relaxation reservoir).}

For clarity we focus on temporal dynamics and suppress spatial dependence.
We consider an observable collective cognitive coordinate $\phi(t)$
linearly coupled to a continuum of latent relaxational
OU modes $X_\lambda(t)$ labeled by decay rate
$\lambda\ge 0$:
\begin{align}
\dot\phi(t)
&=
-r\,\phi(t)
+\int_{0}^{\Lambda} d\lambda\, g(\lambda)\,X_\lambda(t)
+\eta(t),
\label{eq:phi_coupled_app}
\\
\dot X_\lambda(t)
&=
-\lambda\,X_\lambda(t)
+g(\lambda)\,\phi(t)
+\zeta_\lambda(t).
\label{eq:X_coupled_app}
\end{align}

Here $r>0$ is the bare restoring scale, $\Lambda$ is an ultraviolet
cutoff on relaxation rates, and $\eta$ and $\zeta_\lambda$ are
(possibly independent) noise sources.

The observable coordinate $\phi(t)$ represents the infrared collective
cognitive trajectory, while $X_\lambda(t)$ denotes latent slowly
relaxing collective modes associated with contextual persistence,
distributed memory traces, and hidden organizational channels.

For $\lambda>0$ each latent reservoir coordinate is exponentially
stable; the sector $\lambda\to 0^+$ should be understood as the
infrared limit of a continuous spectrum of increasingly slow
collective relaxation modes rather than an exactly conserved degree of
freedom.

Equations
(\ref{eq:phi_coupled_app})--(\ref{eq:X_coupled_app})
therefore define a Markovian (first-order) dynamical system in the
enlarged state space
$\{\phi,X_\lambda\}$.

\vspace{10pt}
\noindent
\emph{(II) Exact solution for the latent OU reservoir modes.}

Equation (\ref{eq:X_coupled_app}) is linear and can be solved exactly
by the integrating-factor method.

Multiply (\ref{eq:X_coupled_app}) by $e^{\lambda t}$:
\begin{equation}
e^{\lambda t}\dot X_\lambda(t)
+
\lambda e^{\lambda t}X_\lambda(t)
=
e^{\lambda t}
\big[g(\lambda)\phi(t)+\zeta_\lambda(t)\big].
\label{eq:integrating_factor_step1_app}
\end{equation}

The left-hand side is the total derivative
\begin{equation}
\frac{d}{dt}
\Big(
e^{\lambda t}X_\lambda(t)
\Big)
=
e^{\lambda t}
\big[
g(\lambda)\phi(t)
+
\zeta_\lambda(t)
\big].
\label{eq:integrating_factor_step2_app}
\end{equation}

Integrate from an initial time $t_0$ to $t$:
\begin{equation}
e^{\lambda t}X_\lambda(t)
-
e^{\lambda t_0}X_\lambda(t_0)
=
\int_{t_0}^{t}
ds\,
e^{\lambda s}
\big[
g(\lambda)\phi(s)
+
\zeta_\lambda(s)
\big].
\label{eq:integrating_factor_step3_app}
\end{equation}

Solving for $X_\lambda(t)$ yields the exact causal representation
\begin{equation}
\begin{aligned}
X_\lambda(t)
&=
e^{-\lambda(t-t_0)}X_\lambda(t_0)
\\
&\quad+
\int_{t_0}^{t}
ds\,
e^{-\lambda(t-s)}
\Big[
g(\lambda)\phi(s)
+
\zeta_\lambda(s)
\Big].
\end{aligned}
\label{eq:X_solution_app}
\end{equation}

This expression already shows that the latent reservoir coordinate at
time $t$ depends on the entire prior history
$\{\phi(s)\}_{s<t}$ of the observable cognitive trajectory.

\vspace{10pt}
\noindent
\emph{(III) Substitution into the observable cognitive equation and emergence of a memory kernel.}

Substituting (\ref{eq:X_solution_app}) into the observable cognitive
equation (\ref{eq:phi_coupled_app}) gives
\begin{align}
&\dot\phi(t)
=
-r\,\phi(t)
+
\int_{0}^{\Lambda} d\lambda\, g(\lambda)
\nonumber\\
&\times
\Bigg[
e^{-\lambda(t-t_0)}X_\lambda(t_0)
+
\int_{t_0}^{t}
ds\,
e^{-\lambda(t-s)}
\Big(
g(\lambda)\phi(s)
+
\zeta_\lambda(s)
\Big)
\Bigg]
\nonumber\\
&\quad
+\eta(t).
\label{eq:kernel_identification_step_app}
\end{align}

This motivates the definition of the causal memory kernel
\begin{equation}
K(t-s)
\equiv
\Theta(t-s)
\int_{0}^{\Lambda}
d\lambda\,
g(\lambda)^2
e^{-\lambda(t-s)},
\label{eq:memory_kernel_def_app}
\end{equation}
so that
\begin{equation}
\int_{0}^{\Lambda}
d\lambda\,
g(\lambda)
\int_{t_0}^{t}
ds\,
e^{-\lambda(t-s)}
g(\lambda)\phi(s)
=
\int_{t_0}^{t}
ds\,
K(t-s)\phi(s).
\label{eq:kernel_compact_app}
\end{equation}

With this definition,
Eq.~(\ref{eq:kernel_identification_step_app})
can be written as
\begin{equation}
\dot\phi(t)
=
-r\,\phi(t)
+
\int_{t_0}^{t}
ds\,
K(t-s)\phi(s)
+
\eta(t)
+
\xi_{\rm eff}(t)
+
J_{\rm ic}(t),
\label{eq:phi_memory_full_app}
\end{equation}
where we have collected the remaining contributions into an effective
noise term generated by the latent reservoir and an initial-condition
transient:
\begin{align}
\xi_{\rm eff}(t)
&\equiv
\int_{0}^{\Lambda}
d\lambda\,
g(\lambda)
\int_{t_0}^{t}
ds\,
e^{-\lambda(t-s)}
\zeta_\lambda(s),
\label{eq:xi_eff_def_app}
\\
J_{\rm ic}(t)
&\equiv
\int_{0}^{\Lambda}
d\lambda\,
g(\lambda)
e^{-\lambda(t-t_0)}
X_\lambda(t_0).
\label{eq:Jic_def_app}
\end{align}

Equation (\ref{eq:phi_memory_full_app}) is therefore an exact
generalized Langevin equation for the observable cognitive trajectory.

Although the enlarged system
(\ref{eq:phi_coupled_app})--(\ref{eq:X_coupled_app})
is Markovian, eliminating the hidden latent reservoir coordinates
produces a non-Markovian memory term and, generally, colored effective
noise.

\vspace{6pt}
\noindent
\emph{Long-time limit.}
For $t-t_0\gg \Lambda^{-1}$, the transient
$J_{\rm ic}(t)$ decays.
More precisely, decay holds provided the effective spectral weight
$g(\lambda)X_\lambda(t_0)$ is integrable near $\lambda=0$ so that the
$\lambda\to0^+$ sector does not produce a non-decaying contribution.

\vspace{10pt}
\noindent
\emph{(IV) TDOS form and Laplace-transform structure.}

As derived in Appendix~A, integrating out the continuum of latent
relaxation modes produces a retarded self-energy of the form
(\ref{eq:app_sigma_tdos_direct}),
where the time-scale density of states is defined as the spectral
weight distribution of collective relaxation eigenmodes,
\begin{equation}
\rho(\lambda)
=
\sum_\alpha
g_\alpha^2\,
\delta(\lambda-\lambda_\alpha).
\end{equation}

Here $g_\alpha$ represents the overlap amplitude between the observable
cognitive trajectory and the latent relaxation eigenmode $\alpha$, so
that $g_\alpha^2$ plays the role of the spectral weight of that mode.

In this sense the TDOS is directly analogous to a spectral function in
many-body theory, but defined for the relaxation spectrum governing
collective cognitive dynamics.

With this definition the memory kernel introduced above can be written
in the compact form
\begin{equation}
K(t)
=
\int_{0}^{\Lambda}
d\lambda\;
\rho(\lambda)e^{-\lambda t},
\qquad
t\ge0,
\label{eq:K_as_Laplace_app}
\end{equation}
showing that the memory kernel is the Laplace transform of the TDOS.

Consequently the long-time dynamics of the observable cognitive
trajectory is fully controlled by the infrared structure of
$\rho(\lambda)$.

\vspace{10pt}
\noindent
\emph{(V) Flat TDOS implies universal long-memory tail \(K(t)\sim1/t\).}

If the TDOS is finite at the origin (``flat'' TDOS),
\begin{equation}
\rho(\lambda)
\xrightarrow{\lambda\to0}
\rho_0,
\label{eq:flat_TDOS_memory_app}
\end{equation}
then for times $t\gg\Lambda^{-1}$ the infrared part of
(\ref{eq:K_as_Laplace_app}) dominates and we may approximate
$\rho(\lambda)\simeq\rho_0$ over the relevant range.

In that case the kernel can be evaluated exactly:
\begin{align}
K(t)
=
\int_{0}^{\Lambda}
d\lambda\,
\rho_0\,e^{-\lambda t}.
\label{eq:K_flat_step1_app}
\end{align}

Perform the change of variables
$u=\lambda t$
(so $d\lambda=du/t$):
\begin{align}
K(t)
&=
\rho_0
\int_{0}^{\Lambda t}
\frac{du}{t}\,
e^{-u}
=
\frac{\rho_0}{t}
\Big(
1-e^{-\Lambda t}
\Big).
\label{eq:K_flat_exact_app}
\end{align}

Therefore, at long times
$t\gg\Lambda^{-1}$,
\begin{equation}
K(t)
\simeq
\frac{\rho_0}{t},
\qquad
(t\gg\Lambda^{-1}),
\label{eq:K_tail_1_over_t_app}
\end{equation}
which is the universal long-memory tail discussed in the main text.

\vspace{10pt}
\noindent
\emph{(VI) Frequency-space form and connection to the retarded self-energy.}

For completeness we relate the time-domain memory kernel to the
frequency-space self-energy.

Taking the Fourier transform of the latent reservoir solution
(\ref{eq:X_solution_app})
in the stationary long-time limit yields
\begin{equation}
X_\lambda(\omega)
=
\frac{g(\lambda)}{\lambda-i\omega}\phi(\omega)
+
\frac{1}{\lambda-i\omega}\zeta_\lambda(\omega),
\label{eq:X_omega_app}
\end{equation}
and substituting into
(\ref{eq:phi_coupled_app})
gives
\begin{equation}
\Big[
-i\omega+r-\Sigma_R(\omega)
\Big]
\phi(\omega)
=
\eta(\omega)
+
\int_0^\Lambda
d\lambda\,
\frac{g(\lambda)}{\lambda-i\omega}
\zeta_\lambda(\omega),
\label{eq:phi_omega_app}
\end{equation}
with the retarded self-energy
\begin{equation}
\Sigma_R(\omega)
=
\int_0^\Lambda
d\lambda\,
\frac{g(\lambda)^2}{\lambda-i\omega}
=
\int_0^\Lambda
d\lambda\,
\frac{\rho(\lambda)}{\lambda-i\omega}.
\label{eq:SigmaR_from_TDOS_app}
\end{equation}

Separating real and imaginary parts using
\begin{equation}
\frac{1}{\lambda-i\omega}
=
\frac{\lambda}{\lambda^2+\omega^2}
+
i\frac{\omega}{\lambda^2+\omega^2},
\label{eq:decompose_app}
\end{equation}
one obtains
\begin{align}
{\rm Re}\,\Sigma_R(\omega)
&=
\int_0^\Lambda
d\lambda\,
\rho(\lambda)
\frac{\lambda}{\lambda^2+\omega^2},
\\
{\rm Im}\,\Sigma_R(\omega)
&=
\omega
\int_0^\Lambda
d\lambda\,
\rho(\lambda)
\frac{1}{\lambda^2+\omega^2}.
\end{align}

For a flat TDOS
$\rho(\lambda)\simeq\rho_0$
and
$|\omega|\ll\Lambda$,
\begin{align}
{\rm Im}\,\Sigma_R(\omega)
&\simeq
\omega\rho_0
\int_0^\Lambda
\frac{d\lambda}{\lambda^2+\omega^2}
\nonumber\\
&=
\omega\rho_0
\left[
\frac{1}{|\omega|}
\arctan\!\Big(
\frac{\Lambda}{|\omega|}
\Big)
\right]
\nonumber\\
&\xrightarrow{|\omega|\ll\Lambda}
\frac{\pi}{2}\rho_0\,{\rm sgn}(\omega),
\label{eq:ImSigma_sgn_app}
\end{align}
and
\begin{align}
{\rm Re}\,\Sigma_R(\omega)
&\simeq
\rho_0
\int_0^\Lambda
d\lambda\,
\frac{\lambda}{\lambda^2+\omega^2}
\nonumber\\
&=
\frac{\rho_0}{2}
\ln
\left(
\frac{\Lambda^2+\omega^2}{\omega^2}
\right)
\nonumber\\
&
\simeq
\rho_0
\ln
\left(
\frac{\Lambda}{|\omega|}
\right),
\qquad
(|\omega|\ll\Lambda).
\label{eq:ReSigma_log_app}
\end{align}

\vspace{10pt}
\noindent
\emph{Summary.}
Equations
(\ref{eq:phi_memory_full_app})--(\ref{eq:K_as_Laplace_app})
show that integrating out a continuum of OU relaxation modes yields an
exact generalized Langevin equation for the observable cognitive
trajectory with a causal memory kernel \(K(t)\).

The kernel is the Laplace transform of the TDOS.
A finite TDOS at the origin produces the universal long-memory tail
\begin{equation}
K(t)\sim \frac{1}{t},
\end{equation}
and the corresponding retarded self-energy exhibits the marginal
nonanalytic structure
\begin{equation}
{\rm Im}\,\Sigma_R(\omega)
\propto
{\rm sgn}(\omega)
\end{equation}
together with a logarithmic real part.

This makes explicit that memory-dominated infrared cognitive dynamics
arises from the collective accumulation of slow latent relaxation
modes, rather than from coupling to a featureless Markovian noise
source.

\section{Appendix C: Recursive field-mediated phase dynamics of the cognitive field}

This Appendix derives the temporal phase dynamics introduced in
Sec.~II.D directly from the coupled equations of the macroscopic
cognitive field and the distributed collective modes.
The purpose is to show that collective phase locking is not introduced
as an independent phenomenological synchronization mechanism.
Rather, it follows from the same recursive field--mode interaction that
generates the memory-dressed cognitive response.

We begin from the deterministic linear sector of the coupled dynamics,
\begin{align}
\partial_t\phi(t)
&=
-r\phi(t)
+
\sum_\alpha
g_\alpha X_\alpha(t),
\label{eq:app_phase_field}
\\
\partial_tX_\alpha(t)
&=
-
\left(
\lambda_\alpha+i\omega_\alpha
\right)
X_\alpha(t)
+
g_\alpha\phi(t),
\label{eq:app_phase_modes}
\end{align}
where \(\phi(t)\) denotes the macroscopic cognitive field,
\(X_\alpha(t)\) denotes the amplitude of the \(\alpha\)-th distributed
collective mode, and \(g_\alpha\) denotes their mutual coupling.
Noise and external driving may be restored straightforwardly but are
omitted here in order to isolate the deterministic phase organization.

The complex eigenvalue
\begin{equation}
\mu_\alpha
=
\lambda_\alpha+i\omega_\alpha
\label{eq:app_complex_mode}
\end{equation}
contains two complementary dynamical components.
The relaxation rate \(\lambda_\alpha\) controls the persistence of the
mode, whereas the circulation frequency \(\omega_\alpha\) controls its
intrinsic temporal phase evolution.

\vspace{10pt}
\noindent
\emph{(I) Amplitude and phase dynamics of the distributed modes.}

We represent the macroscopic cognitive field in amplitude--phase form,
\begin{equation}
\phi(t)
=
A(t)e^{i\psi(t)},
\label{eq:app_phi_polar}
\end{equation}
where \(A(t)=|\phi(t)|\) is the amplitude of the macroscopic cognitive
field and \(\psi(t)\) is its collective phase.

In the original spectral convention, an individual collective mode may
be written as
\begin{equation}
X_\alpha(t)
=
A_\alpha(t)e^{-i\theta_\alpha(t)}.
\label{eq:app_mode_original_phase}
\end{equation}
For the standard phase-difference convention, we define
\begin{equation}
\vartheta_\alpha
\equiv
-\theta_\alpha,
\qquad
X_\alpha
=
A_\alpha e^{i\vartheta_\alpha}.
\label{eq:app_standard_phase}
\end{equation}
An uncoupled mode evolves according to
\begin{equation}
X_\alpha(t)
\sim
e^{-(\lambda_\alpha+i\omega_\alpha)t},
\end{equation}
so that its natural phase frequency in the convention of
Eq.~(\ref{eq:app_standard_phase}) is
\begin{equation}
\nu_\alpha
\equiv
-\omega_\alpha.
\label{eq:app_natural_frequency}
\end{equation}

Differentiating the collective mode gives
\begin{equation}
\partial_tX_\alpha
=
\left(
\dot A_\alpha
+
iA_\alpha\dot\vartheta_\alpha
\right)
e^{i\vartheta_\alpha}.
\label{eq:app_mode_derivative}
\end{equation}
Substituting
Eqs.~(\ref{eq:app_phi_polar}) and
(\ref{eq:app_standard_phase}) into
Eq.~(\ref{eq:app_phase_modes}) yields
\begin{align}
\left(
\dot A_\alpha
+
iA_\alpha\dot\vartheta_\alpha
\right)
e^{i\vartheta_\alpha}
={}&
-
\left(
\lambda_\alpha+i\omega_\alpha
\right)
A_\alpha e^{i\vartheta_\alpha}
\nonumber\\
&+
g_\alpha
Ae^{i\psi}.
\label{eq:app_mode_substitution}
\end{align}
Multiplication by \(e^{-i\vartheta_\alpha}\) gives
\begin{equation}
\dot A_\alpha
+
iA_\alpha\dot\vartheta_\alpha
=
-
\left(
\lambda_\alpha+i\omega_\alpha
\right)
A_\alpha
+
g_\alpha A
e^{i(\psi-\vartheta_\alpha)}.
\label{eq:app_mode_rotated}
\end{equation}
Using
\begin{equation}
e^{i(\psi-\vartheta_\alpha)}
=
\cos(\psi-\vartheta_\alpha)
+
i\sin(\psi-\vartheta_\alpha),
\end{equation}
the real part gives
\begin{equation}
\dot A_\alpha
=
-\lambda_\alpha A_\alpha
+
g_\alpha A
\cos(\psi-\vartheta_\alpha),
\label{eq:app_modal_amplitude}
\end{equation}
while the imaginary part gives
\begin{equation}
\dot\vartheta_\alpha
=
\nu_\alpha
+
g_\alpha
\frac{A}{A_\alpha}
\sin(\psi-\vartheta_\alpha).
\label{eq:app_modal_phase}
\end{equation}

Equation~(\ref{eq:app_modal_phase}) already contains the basic
field-mediated entrainment mechanism.
Each mode possesses its own intrinsic circulation frequency
\(\nu_\alpha\), while its coupling to the macroscopic cognitive phase
produces a restoring interaction proportional to
\(\sin(\psi-\vartheta_\alpha)\).

\vspace{10pt}
\noindent
\emph{(II) Amplitude and phase dynamics of the macroscopic cognitive field.}

We now derive the reciprocal dynamics of the macroscopic field.
Differentiating Eq.~(\ref{eq:app_phi_polar}) gives
\begin{equation}
\partial_t\phi
=
\left(
\dot A+iA\dot\psi
\right)e^{i\psi}.
\label{eq:app_field_derivative}
\end{equation}
Substituting
\(
\phi=Ae^{i\psi}
\)
and
\(
X_\alpha=A_\alpha e^{i\vartheta_\alpha}
\)
into Eq.~(\ref{eq:app_phase_field}) yields
\begin{equation}
\left(
\dot A+iA\dot\psi
\right)e^{i\psi}
=
-rAe^{i\psi}
+
\sum_\alpha
g_\alpha A_\alpha e^{i\vartheta_\alpha}.
\label{eq:app_field_substitution}
\end{equation}
Multiplication by \(e^{-i\psi}\) gives
\begin{equation}
\dot A+iA\dot\psi
=
-rA
+
\sum_\alpha
g_\alpha A_\alpha
e^{i(\vartheta_\alpha-\psi)}.
\label{eq:app_field_rotated}
\end{equation}
Separating real and imaginary parts gives
\begin{equation}
\dot A
=
-rA
+
\sum_\alpha
g_\alpha A_\alpha
\cos(\vartheta_\alpha-\psi),
\label{eq:app_macroscopic_amplitude}
\end{equation}
and
\begin{equation}
\dot\psi
=
\frac{1}{A}
\sum_\alpha
g_\alpha A_\alpha
\sin(\vartheta_\alpha-\psi).
\label{eq:app_macroscopic_phase}
\end{equation}

Equations~(\ref{eq:app_modal_phase}) and
(\ref{eq:app_macroscopic_phase}) form a closed recursive
phase-feedback structure,
\begin{equation}
\{\vartheta_\alpha\}
\longrightarrow
\psi
\longrightarrow
\{\vartheta_\alpha\}.
\label{eq:app_recursive_phase_loop}
\end{equation}
The distributed modes collectively generate the macroscopic phase,
while the resulting macroscopic phase feeds back into the evolution of
every individual mode.
Temporal coherence is therefore generated internally by the recursive
cognitive-field dynamics rather than by an externally imposed
reference phase.

The corresponding amplitude equations also show that phase alignment
affects the strength of the macroscopic cognitive field.
When
\begin{equation}
\cos(\vartheta_\alpha-\psi)>0,
\end{equation}
the modal contribution reinforces the macroscopic amplitude.
Broadly distributed phases instead produce cancellations among the
modal contributions.
The amplitude and phase sectors are therefore dynamically coupled,
although they remain physically distinct.

\vspace{10pt}
\noindent
\emph{(III) Relative-phase dynamics and the phase-locking condition.}

The relative phase between mode \(\alpha\) and the macroscopic
cognitive field is defined by
\begin{equation}
\delta_\alpha
=
\vartheta_\alpha-\psi.
\label{eq:app_relative_phase}
\end{equation}
Its time derivative is
\begin{equation}
\dot\delta_\alpha
=
\dot\vartheta_\alpha-\dot\psi.
\end{equation}
Using
Eqs.~(\ref{eq:app_modal_phase}) and
(\ref{eq:app_macroscopic_phase}), one obtains
\begin{equation}
\dot\delta_\alpha
=
\nu_\alpha
-
g_\alpha
\frac{A}{A_\alpha}
\sin\delta_\alpha
-
\frac{1}{A}
\sum_\beta
g_\beta A_\beta
\sin\delta_\beta.
\label{eq:app_relative_phase_dynamics}
\end{equation}

A phase-locked state is defined by stationary relative phases,
\begin{equation}
\delta_\alpha(t)
\longrightarrow
\delta_\alpha^\ast,
\qquad
\dot\delta_\alpha=0.
\label{eq:app_phase_locked_state}
\end{equation}
This condition does not require all phases to become identical.
Instead, synchronized modes may retain different but time-independent
phase offsets.

For a phase-locked population, the modal and macroscopic phases rotate
with a common collective frequency,
\begin{equation}
\dot\vartheta_\alpha
=
\dot\psi
=
\Omega_{\rm cog}.
\label{eq:app_common_frequency}
\end{equation}
If the amplitudes vary sufficiently slowly over the locking timescale,
Eq.~(\ref{eq:app_modal_phase}) gives
\begin{equation}
\Omega_{\rm cog}
=
\nu_\alpha
-
g_\alpha
\frac{A}{A_\alpha}
\sin\delta_\alpha^\ast.
\label{eq:app_locking_frequency_relation}
\end{equation}
Equivalently,
\begin{equation}
\sin\delta_\alpha^\ast
=
\frac{
\nu_\alpha-\Omega_{\rm cog}
}{
g_\alpha A/A_\alpha
}.
\label{eq:app_locking_relation}
\end{equation}
A stationary phase offset exists only if
\begin{equation}
\left|
\nu_\alpha-\Omega_{\rm cog}
\right|
\leq
g_\alpha
\frac{A}{A_\alpha}.
\label{eq:app_locking_criterion}
\end{equation}

Equation~(\ref{eq:app_locking_criterion}) defines the entrainment
window of the macroscopic cognitive field.
Modes whose intrinsic circulation frequencies lie within this window
can become phase locked, whereas modes outside the window may continue
to drift or synchronize only intermittently.

\vspace{10pt}
\noindent
\emph{(IV) Collective phase-order parameter.}

The global degree of phase alignment among the distributed collective
modes is quantified by
\begin{equation}
Q(t)e^{i\Psi(t)}
=
\frac{1}{N}
\sum_{\alpha=1}^{N}
e^{i\vartheta_\alpha(t)}.
\label{eq:app_phase_order_parameter}
\end{equation}
The magnitude
\begin{equation}
Q(t)
=
\left|
\frac{1}{N}
\sum_{\alpha=1}^{N}
e^{i\vartheta_\alpha(t)}
\right|
\label{eq:app_Q_definition}
\end{equation}
measures collective phase coherence, while \(\Psi(t)\) denotes the
mean phase of the mode population.

For a broadly distributed incoherent phase population,
\begin{equation}
Q\simeq0,
\end{equation}
because the phase vectors cancel.
For a partially or globally phase-locked population,
\begin{equation}
Q>0,
\end{equation}
because a finite fraction of the mode population contributes
coherently.

The phase-order parameter \(Q\) is distinct from the macroscopic
cognitive-field amplitude \(A\).
The quantity
\begin{equation}
A=|\phi|
\end{equation}
measures the magnitude of the emergent cognitive field, whereas
\(Q\) measures the degree of phase alignment across the distributed
mode population.

The distinction follows directly from
\begin{equation}
\phi
\sim
\sum_\alpha
g_\alpha A_\alpha e^{i\vartheta_\alpha}.
\label{eq:app_field_modal_sum}
\end{equation}
In general, the mode amplitudes and couplings are nonuniform, so the
macroscopic field is not identical to the unweighted phase average.
In the homogeneous mean-field limit,
\begin{equation}
g_\alpha\simeq g,
\qquad
A_\alpha\simeq A_0,
\end{equation}
Eq.~(\ref{eq:app_field_modal_sum}) becomes
\begin{equation}
\phi
\propto
NgA_0
Qe^{i\Psi}.
\label{eq:app_A_Q_relation}
\end{equation}
In this special limit,
\begin{equation}
A\propto Q,
\qquad
\psi\simeq\Psi,
\end{equation}
but \(A\) and \(Q\) remain conceptually distinct in the general
mode-dependent dynamics.

\vspace{10pt}
\noindent
\emph{(V) Retarded field-mediated interaction among collective modes.}

The recursive interaction may also be represented directly as an
effective interaction among the distributed modes.
The formal solution of Eq.~(\ref{eq:app_phase_field}) is
\begin{align}
\phi(t)
={}&
e^{-r(t-t_0)}
\phi(t_0)
\nonumber\\
&+
\sum_\beta
g_\beta
\int_{t_0}^{t}
dt'\,
e^{-r(t-t')}
X_\beta(t').
\label{eq:app_formal_field_solution}
\end{align}
We define the bare retarded propagator of the macroscopic cognitive
field by
\begin{equation}
L_0(t-t')
=
\Theta(t-t')
e^{-r(t-t')}.
\label{eq:app_bare_field_propagator}
\end{equation}
After the initial transient has decayed,
Eq.~(\ref{eq:app_formal_field_solution}) becomes
\begin{equation}
\phi(t)
=
\sum_\beta
g_\beta
\int_{t_0}^{t}
dt'\,
L_0(t-t')
X_\beta(t').
\label{eq:app_field_response_to_modes}
\end{equation}

Substitution into Eq.~(\ref{eq:app_phase_modes}) yields
\begin{align}
\partial_tX_\alpha(t)
={}&
-
\left(
\lambda_\alpha+i\omega_\alpha
\right)
X_\alpha(t)
\nonumber\\
&+
\sum_\beta
\int_{t_0}^{t}
dt'\,
\mathcal L_{\alpha\beta}(t-t')
X_\beta(t'),
\label{eq:app_field_mediated_mode_equation}
\end{align}
where
\begin{equation}
\mathcal L_{\alpha\beta}(t)
=
g_\alpha g_\beta L_0(t)
\label{eq:app_retarded_interaction_kernel}
\end{equation}
is the bare retarded mode--mode interaction kernel.

The notation \(\mathcal L_{\alpha\beta}(t)\) is used here to emphasize
that this object is a time-dependent interaction kernel rather than an
instantaneous coupling constant.
It represents the retarded propagation path
\begin{equation}
X_\beta
\longrightarrow
\phi
\longrightarrow
X_\alpha.
\label{eq:app_mode_field_mode_path}
\end{equation}

Substituting
\(
X_\alpha=A_\alpha e^{i\vartheta_\alpha}
\)
into Eq.~(\ref{eq:app_field_mediated_mode_equation}) and taking the
imaginary part gives
\begin{align}
\dot\vartheta_\alpha(t)
={}&
\nu_\alpha
\nonumber\\
&+
\sum_\beta
\int_{t_0}^{t}
dt'\,
\mathcal L_{\alpha\beta}(t-t')
\frac{A_\beta(t')}{A_\alpha(t)}
\sin\!\left[
\vartheta_\beta(t')
-
\vartheta_\alpha(t)
\right].
\label{eq:app_nonmarkovian_phase_equation}
\end{align}

Equation~(\ref{eq:app_nonmarkovian_phase_equation}) is the fundamental
retarded phase equation of the coupled cognitive-field dynamics.
The phase of each mode responds not only to the current phases of the
other modes but also to their temporally retained history.
The same macroscopic cognitive field that carries recursive memory
therefore mediates a non-Markovian phase interaction across the
distributed mode population.

\vspace{10pt}
\noindent
\emph{(VI) Infrared reduction to an effective phase coupling.}

We now consider the slow-amplitude and slow-phase limit.
Suppose that the modal amplitudes and phases vary slowly over the
temporal support of the retarded kernel.
Writing
\begin{equation}
\tau=t-t',
\end{equation}
we approximate
\begin{equation}
A_\beta(t-\tau)
\simeq
A_\beta(t),
\qquad
\vartheta_\beta(t-\tau)
\simeq
\vartheta_\beta(t).
\label{eq:app_slow_ir_approximation}
\end{equation}
Equation~(\ref{eq:app_nonmarkovian_phase_equation}) then reduces to
\begin{equation}
\dot\vartheta_\alpha
=
\nu_\alpha
+
\sum_\beta
\mathcal J_{\alpha\beta}
\sin\!\left(
\vartheta_\beta-\vartheta_\alpha
\right),
\label{eq:app_effective_phase_equation}
\end{equation}
where
\begin{equation}
\mathcal J_{\alpha\beta}
=
\frac{A_\beta}{A_\alpha}
\int_0^\infty
d\tau\,
\mathcal L_{\alpha\beta}(\tau)
\label{eq:app_ir_phase_coupling}
\end{equation}
is the temporally coarse-grained phase-coupling strength.

Thus the notation distinguishes three different objects:
\begin{equation}
L_0(t)
\quad\longrightarrow\quad
\mathcal L_{\alpha\beta}(t)
\quad\longrightarrow\quad
\mathcal J_{\alpha\beta}.
\end{equation}
Here \(L_0\) is the bare propagator of the macroscopic field,
\(\mathcal L_{\alpha\beta}(t)\) is the retarded interaction kernel
mediated by that field, and \(\mathcal J_{\alpha\beta}\) is the
effective instantaneous coupling obtained after infrared temporal
coarse-graining.

For approximately homogeneous interactions,
Eq.~(\ref{eq:app_effective_phase_equation}) contains conventional
globally coupled phase dynamics as a special limit.
However, neither uniform coupling nor an instantaneous phase model is
assumed in the cognitive field theory.
The more fundamental dynamics is the retarded equation
(\ref{eq:app_nonmarkovian_phase_equation}).

\vspace{10pt}
\noindent
\emph{(VII) Memory dressing of the field-mediated phase interaction.}

The preceding derivation was expressed initially in terms of the bare
cognitive propagator \(L_0\).
However, Sec.~II.C showed that the distributed relaxation modes
recursively dress the response of the macroscopic cognitive field.
The bare propagator satisfies
\begin{equation}
L_0(\Omega)
=
\frac{1}{-i\Omega+r},
\label{eq:app_bare_frequency_propagator}
\end{equation}
while recursive memory feedback generates the Dyson equation
\begin{equation}
\chi_R(\Omega)
=
L_0(\Omega)
+
L_0(\Omega)
\Sigma_R(\Omega)
\chi_R(\Omega).
\label{eq:app_dyson_cognitive_response}
\end{equation}
Solving for the dressed response gives
\begin{equation}
\chi_R(\Omega)
=
\frac{1}
{-i\Omega+r-\Sigma_R(\Omega)}.
\label{eq:app_dressed_cognitive_response}
\end{equation}

Consequently, the propagator mediating the interaction between the
collective modes is renormalized according to
\begin{equation}
L_0(\Omega)
\longrightarrow
\chi_R(\Omega).
\label{eq:app_propagator_dressing}
\end{equation}
The bare frequency-dependent interaction
\begin{equation}
\mathcal L_{\alpha\beta}^{(0)}(\Omega)
=
g_\alpha g_\beta
L_0(\Omega)
\end{equation}
therefore becomes the memory-dressed effective interaction
\begin{equation}
\mathcal J_{\alpha\beta}^{\rm eff}(\Omega)
=
g_\alpha g_\beta
\chi_R(\Omega).
\label{eq:app_dressed_phase_interaction}
\end{equation}

This result is not an additional phenomenological postulate.
It follows directly by replacing the bare cognitive propagator in the
field-mediated interaction by the fully memory-dressed response derived
from the same coupled dynamics.

The relation between the time-integrated kernel and the static
frequency response makes this connection explicit.
For a causal retarded kernel,
\begin{equation}
\int_0^\infty
d\tau\,
\mathcal L_{\alpha\beta}(\tau)
=
\mathcal L_{\alpha\beta}(\Omega=0).
\label{eq:app_kernel_static_relation}
\end{equation}
The dressed infrared phase coupling is therefore
\begin{equation}
\mathcal J_{\alpha\beta}^{\rm IR}
=
\frac{A_\beta}{A_\alpha}
g_\alpha g_\beta
\chi_R(0).
\label{eq:app_ir_dressed_coupling}
\end{equation}

Using
\begin{equation}
r_{\rm cog}
=
r-\Sigma_R(0),
\end{equation}
the static susceptibility becomes
\begin{equation}
\chi_R(0)
=
\frac{1}{r_{\rm cog}},
\label{eq:app_static_susceptibility}
\end{equation}
and hence
\begin{equation}
\mathcal J_{\alpha\beta}^{\rm IR}
=
\frac{A_\beta}{A_\alpha}
\frac{
g_\alpha g_\beta
}{
r_{\rm cog}
}.
\label{eq:app_ir_coupling_gap}
\end{equation}

After averaging over the participating infrared mode population, the
collective phase-coupling scale obeys
\begin{equation}
\mathcal J_{\rm eff}
\propto
\chi_R(0)
=
\frac{1}{r_{\rm cog}}.
\label{eq:app_collective_phase_scale}
\end{equation}
Thus suppression of the cognitive forgetting gap simultaneously
enhances the field-mediated phase interaction.

Collective phase locking becomes possible when the effective
interaction exceeds the threshold set by the distribution of intrinsic
circulation frequencies, modal amplitudes, noise, and coupling
heterogeneity,
\begin{equation}
\mathcal J_{\rm eff}
>
\mathcal J_c.
\label{eq:app_collective_locking_threshold}
\end{equation}
Above this threshold, a finite phase-locked population develops and
the global phase-order parameter satisfies
\begin{equation}
Q>0.
\end{equation}

The relaxation and circulation sectors are therefore linked through
the memory-dressed cognitive susceptibility:
\begin{equation}
\rho(\lambda)
\rightarrow
K(t)
\rightarrow
\Sigma_R(0)
\rightarrow
r_{\rm cog}
\rightarrow
\chi_R(0)
\rightarrow
\mathcal J_{\rm eff}
\rightarrow
Q.
\label{eq:app_complete_phase_chain}
\end{equation}
The infrared relaxation spectrum first enhances recursive memory
feedback and suppresses the cognitive forgetting gap.
The resulting increase of the cognitive susceptibility then strengthens
the field-mediated phase interaction and promotes collective phase
locking.

\vspace{10pt}
\noindent
\emph{(VIII) Physical interpretation.}

The coupled cognitive-field equations generate two complementary forms
of collective organization.

The amplitude sector contains terms of the form
\begin{equation}
\cos(\vartheta_\alpha-\psi),
\end{equation}
which determine whether distributed collective modes reinforce or
suppress the macroscopic cognitive-field amplitude.
The phase sector contains terms of the form
\begin{equation}
\sin(\psi-\vartheta_\alpha),
\end{equation}
or, after elimination of the common cognitive field,
\begin{equation}
\sin\!\left[
\vartheta_\beta(t')
-
\vartheta_\alpha(t)
\right],
\end{equation}
which generate restoring interactions among dynamically relevant phase
differences.

The recursive structure may be summarized as
\begin{equation}
\text{learned cognitive geometry}
\rightarrow
\mu_\alpha
=
\lambda_\alpha+i\omega_\alpha,
\end{equation}
\begin{equation}
\mu_\alpha
\rightarrow
\text{complex distributed collective modes},
\end{equation}
and
\begin{equation}
\{X_\alpha\}
\leftrightarrow
\phi
\rightarrow
\text{recursive amplitude and phase feedback}.
\end{equation}

Integrating out the distributed modes produces the memory-dressed
response of the macroscopic cognitive field.
Eliminating the resulting cognitive field in the opposite direction
produces the retarded interaction among the distributed modes.
The two reductions therefore describe complementary directions of the
same self-consistent feedback loop:
\begin{equation}
\begin{aligned}
\{X_\alpha\}\ \text{integrated out}
&\quad\Rightarrow\quad
K(t),\ \Sigma_R,\ \chi_R,
\\
\phi\ \text{integrated out}
&\quad\Rightarrow\quad
\mathcal L_{\alpha\beta}(t),\
\mathcal J_{\alpha\beta}^{\rm eff}.
\end{aligned}
\end{equation}

The macroscopic cognitive field is therefore not formed solely by the
accumulation of slow relaxation modes.
Its persistence is generated by memory dressing of the relaxation
sector, while its temporal coherence is generated by field-mediated
organization of the circulation sector.
Both originate from the same recursive coupling between the
macroscopic field and the distributed collective-mode population.

The complex representation
\begin{equation}
\phi
=
Ae^{i\psi}
\end{equation}
therefore has a direct dynamical meaning.
The amplitude \(A\) measures the magnitude and persistence of the
memory-dressed macroscopic cognitive field, whereas the phase \(\psi\)
describes its collective temporal organization.
The phase-order parameter \(Q\), in turn, measures how coherently the
distributed mode phases participate in that temporal organization.

Recursive memory dressing and field-mediated phase interaction
therefore represent two complementary manifestations of the same
self-consistent recursive cognitive-field dynamics.
The former stabilizes the macroscopic cognitive field through
suppression of the cognitive forgetting gap, whereas the latter
organizes the temporal coherence of the distributed collective modes.
Together, they provide a unified dynamical description of the complete
complex collective dynamics of the macroscopic cognitive field.

\section{Appendix D: Coupled relaxation dynamics underlying memory persistence,
recurrence, and self-consistent collective amplification}
\label{app:coupled_relaxation_recurrence}

This appendix provides the minimal coupled-mode derivation of the
recursive memory-feedback mechanism underlying the cognitive field.

We consider a single observable collective field \(\phi(t)\) coupled
to a latent collective mode \(X(t)\) with complex relaxation rate
\begin{equation}
\mu
=
\lambda+i\omega_0 ,
\qquad
\lambda>0 .
\end{equation}
The coupled linear dynamics is
\begin{align}
\partial_t \phi(t)
&=
-r\phi(t)
+
gX(t)
+
h(t),
\\
\partial_t X(t)
&=
-(\lambda+i\omega_0)X(t)
+
g\phi(t).
\label{eq:app_coupled_dynamics}
\end{align}
Here \(r>0\) is the bare local forgetting rate of the observable
field, \(\lambda^{-1}\) is the lifetime of the latent memory mode, and
\(\omega_0\) is the circulation frequency of the memory kernel.

Solving the second equation gives
\begin{equation}
X(t)
=
e^{-(\lambda+i\omega_0)(t-t_0)}X(t_0)
+
g
\int_{t_0}^{t}dt'\,
e^{-(\lambda+i\omega_0)(t-t')}
\phi(t') .
\end{equation}
Substituting into the \(\phi\) equation yields the effective
non-Markovian dynamics
\begin{equation}
\partial_t\phi(t)
=
-r\phi(t)
+
\int_{t_0}^{t}dt'\,
K(t-t')\phi(t')
+
h(t),
\end{equation}
with memory kernel
\begin{equation}
K(t)
=
g^2 e^{-\lambda t}e^{-i\omega_0 t}\Theta(t).
\label{eq:app_complex_kernel}
\end{equation}
For a real field, the conjugate pair \(\pm\omega_0\) gives
\begin{equation}
K(t)
=
g^2 e^{-\lambda t}\cos(\omega_0 t)\Theta(t).
\label{eq:app_real_kernel}
\end{equation}
Thus
\begin{equation}
\tau_{\rm mem}
=
\lambda^{-1},
\qquad
T_{\rm rec}
=
\frac{2\pi}{|\omega_0|}.
\end{equation}
The real part \(\lambda\) controls memory persistence, while the
imaginary part \(\omega_0\) controls the recurrence period of the
memory kernel.

Fourier transforming Eq.~\eqref{eq:app_complex_kernel} defines the
retarded memory self-energy,
\begin{equation}
\Sigma_R(\Omega)
=
\frac{g^2}
{\lambda-i(\Omega-\omega_0)} .
\end{equation}
The bare observable propagator is
\[
L_0(\Omega)
=
\frac{1}{-i\Omega+r}.
\]
Repeated memory-mediated feedback gives the Dyson series
\[
L(\Omega)
=
L_0
+
L_0\Sigma_R L_0
+
L_0\Sigma_R L_0\Sigma_R L_0
+
\cdots ,
\]
or
\[
L(\Omega)
=
\frac{1}
{L_0^{-1}(\Omega)-\Sigma_R(\Omega)} .
\label{eq:app_ladder_resummed}
\]
Hence
\begin{equation}
L^{-1}(\Omega)
=
-i\Omega+r
-
\frac{g^2}
{\lambda-i(\Omega-\omega_0)} .
\end{equation}
The collective instability is determined by the pole condition
\begin{equation}
L^{-1}(\Omega_\ast)=0 .
\end{equation}

For a symmetric pair of circulating modes
\(\pm\omega_0\), the static self-energy is real and equals
\begin{equation}
\Sigma_R(0)
=
g^2
\frac{\lambda}
{\lambda^2+\omega_0^2}.
\end{equation}
The memory-dressed forgetting gap is therefore
\begin{equation}
r_{\rm cog}
=
r
-
g^2
\frac{\lambda}
{\lambda^2+\omega_0^2}.
\label{eq:app_rcog_single_mode}
\end{equation}
The recursive memory feedback reaches the protected near-critical
condition when
\begin{equation}
r_{\rm cog}=0,
\qquad
r
=
g^2
\frac{\lambda}
{\lambda^2+\omega_0^2}.
\end{equation}

Equation~\eqref{eq:app_rcog_single_mode} shows that the circulation
frequency does not itself generate the ladder instability.
Rather, \(\omega_0\) controls the recurrence period of the memory
kernel, while the ladder resummation controls the collective
softening of the observable response.

The collective response time is determined by the memory-dressed
forgetting gap,
\begin{equation}
\tau_{\rm cog}
\sim
\frac{1}{r_{\rm cog}} .
\end{equation}
The coupled dynamics therefore contains three distinct characteristic
timescales,
\begin{equation}
\tau_{\rm mem}
=
\lambda^{-1},
\qquad
T_{\rm rec}
=
\frac{2\pi}{|\omega_0|},
\qquad
\tau_{\rm cog}
\sim
r_{\rm cog}^{-1}.
\end{equation}
These quantities describe different aspects of the collective
dynamics and should not, in general, be identified with one another.
The relaxation time
\(\tau_{\rm mem}\)
characterizes the persistence of individual collective modes,
the recurrence period
\(T_{\rm rec}\)
is determined by the circulation sector and characterizes the temporal
phase evolution of the collective dynamics,
whereas
\(\tau_{\rm cog}\)
describes the macroscopic response time of the memory-dressed
cognitive field governed by the effective forgetting gap.

Finally, for a continuum of latent collective modes,
\begin{equation}
K(t)
=
\int d\lambda\,d\omega\,
\rho(\lambda,\omega)
e^{-\lambda t}
e^{-i\omega t}.
\end{equation}
The relaxation spectrum controls memory persistence and the
renormalization of the cognitive forgetting gap through the memory
self-energy.
The circulation spectrum governs temporal recurrence and phase
organization of the collective dynamics.
The macroscopic cognitive response then emerges through
self-consistent recursive memory feedback represented by the Dyson
resummation of repeated memory interactions, and its characteristic
timescale is governed by the memory-dressed forgetting gap
\(r_{\rm cog}\).

\section{Appendix E: Geometric interpretation of collective modes and memory-dressed dynamics}
\label{app:slow_mode_interpretation}

This appendix clarifies the physical meaning of the collective
slow-mode variables used in the main text.
In particular, we distinguish between collective modes, which are
generated by the learned cognitive geometry, and time-dependent mode
amplitudes, which describe their dynamical activation.

The collective cognitive state \(x(t)\) evolves on an effective
high-dimensional cognitive manifold.
Near a metastable operating trajectory \(x^\ast(t)\), small
fluctuations may be written as
\begin{equation}
\delta x(t)
=
x(t)-x^\ast(t).
\end{equation}
The linearized dynamics is governed by a local stability operator,
\begin{equation}
\delta \dot x
=
-
J\,\delta x
+
\eta(t),
\end{equation}
where
\begin{equation}
J
=
\left.
\frac{\partial}{\partial x}
\left[
G^{-1}(x)\nabla_x\Phi(x)
-
R(x)
\right]
\right|_{x^\ast}.
\end{equation}

The collective modes are defined by the eigenvalue problem
\begin{equation}
Ju_\alpha
=
\mu_\alpha u_\alpha ,
\label{eq:app_mode_eigenproblem}
\end{equation}
where
\[
\mu_\alpha
=
\lambda_\alpha+i\omega_\alpha .
\]
Here \(\lambda_\alpha\) is the relaxation rate of the mode, while
\(\omega_\alpha\) is the circulation frequency generated by the
non-conservative part of the learned cognitive geometry.

The eigenvector \(u_\alpha\) and the collective state
\(x\) both belong to the same state space.
However, they play fundamentally different roles.
The state \(x(t)\) specifies the instantaneous configuration of the
cognitive system, whereas \(u_\alpha\) specifies a collective mode
of the linearized dynamics near the operating trajectory.
Its dynamical significance is determined by the corresponding
eigenvalue \(\mu_\alpha\).

For a fixed learned cognitive geometry, the fluctuation field admits a
local expansion in the collective-mode basis,
\begin{equation}
\delta x(t)
=
\sum_\alpha
X_\alpha(t)u_\alpha ,
\label{eq:app_mode_expansion}
\end{equation}
where \(X_\alpha(t)\) is the time-dependent amplitude of the collective
mode \(u_\alpha\).

Equation~\eqref{eq:app_mode_expansion} provides the key
interpretation.
The mode \(u_\alpha\) encodes a learned dynamical pattern of the
cognitive manifold, whereas \(X_\alpha(t)\) describes the instantaneous
activation amplitude of that mode.
Thus \(X_\alpha(t)\) is not a memory item or a storage location.
It is the dynamical amplitude of a collective mode generated by the
learned cognitive geometry.

Projecting the linearized dynamics onto the collective-mode basis
yields the mode-amplitude equation
\begin{equation}
\dot X_\alpha(t)
=
-
\mu_\alpha X_\alpha(t)
+
g_\alpha \phi(t)
+
\eta_\alpha(t),
\label{eq:app_mode_amplitude}
\end{equation}
or equivalently
\begin{equation}
\dot X_\alpha(t)
=
-
(\lambda_\alpha+i\omega_\alpha)X_\alpha(t)
+
g_\alpha \phi(t)
+
\eta_\alpha(t).
\end{equation}
The formal solution is
\begin{align}
X_\alpha(t)
=&\,
e^{-(\lambda_\alpha+i\omega_\alpha)(t-t_0)}
X_\alpha(t_0)
\nonumber\\
&+
\int_{t_0}^{t}dt'\,
e^{-(\lambda_\alpha+i\omega_\alpha)(t-t')}
g_\alpha \phi(t')
\nonumber\\
&+
\int_{t_0}^{t}dt'\,
e^{-(\lambda_\alpha+i\omega_\alpha)(t-t')}
\eta_\alpha(t') .
\label{eq:app_X_formal}
\end{align}

Equation~\eqref{eq:app_X_formal} shows that the present value of
\(X_\alpha(t)\) carries the propagated influence of past observable
cognitive activity.
The relaxation factor
\begin{equation}
e^{-\lambda_\alpha(t-t')}
\end{equation}
determines how long the past activity remains dynamically relevant,
whereas the phase factor
\begin{equation}
e^{-i\omega_\alpha(t-t')}
\end{equation}
determines how this influence is carried through the circulation
sector of the collective spectrum.

The slow-mode variables therefore describe propagating collective
modes rather than static stored memories.
A weakly damped mode with small \(\lambda_\alpha\) preserves the
influence of past cognitive activity over a long time interval.
A nonzero \(\omega_\alpha\) assigns a temporal phase structure to this
propagating influence.
Thus the pair
\begin{equation}
(\lambda_\alpha,\omega_\alpha)
\end{equation}
specifies both the persistence and the phase evolution of the
collective mode.

Substituting the formal solution for \(X_\alpha(t)\) back into the
observable field equation gives the nonlocal memory term
\begin{equation}
\int_{t_0}^{t}dt'\,
K(t-t')\phi(t'),
\end{equation}
with
\begin{equation}
K(t-t')
=
\sum_\alpha
g_\alpha^2
e^{-(\lambda_\alpha+i\omega_\alpha)(t-t')}.
\label{eq:app_kernel_discrete}
\end{equation}
In the continuum limit this becomes
\begin{equation}
K(t)
=
\int d\lambda\,d\omega\,
\rho(\lambda,\omega)
e^{-\lambda t}
e^{-i\omega t}.
\label{eq:app_kernel_continuum}
\end{equation}

The memory kernel is therefore not an independent storage mechanism.
It is the effective propagator generated by the slow collective modes
of the learned cognitive geometry after the latent mode amplitudes
\(X_\alpha(t)\) have been integrated out.
Memory arises because weakly damped collective modes carry the
influence of past activity forward in time.

This interpretation also clarifies the relation between collective
cognitive organization and neural substrate.
A collective mode \(u_\alpha\) may be distributed across many
microscopic neural or representational degrees of freedom.
It is not localized to a single neuron, cortical area, or memory
register.
Instead, it describes a coherent dynamical mode in the
high-dimensional collective state space.
The amplitude \(X_\alpha(t)\) describes how strongly this mode is
activated during ongoing cognitive dynamics.

The learned cognitive geometry determines the set of collective modes
\(\{u_\alpha\}\), their relaxation rates
\(\{\lambda_\alpha\}\), and their circulation frequencies
\(\{\omega_\alpha\}\).
Learning therefore reorganizes not only stored contents, but the
collective-mode basis itself.
On inference timescales, the collective modes may often be treated as
quasi-static and the amplitudes \(X_\alpha(t)\) evolve dynamically.
On learning timescales, the structural parameters of the cognitive
geometry evolve, so that the collective modes and eigenvalues also
change:
\begin{equation}
u_\alpha
\rightarrow
u_\alpha(\kappa(t)),
\qquad
\mu_\alpha
\rightarrow
\mu_\alpha(\kappa(t)).
\end{equation}

Consequently, the slow-mode expansion should be understood as an
adiabatic collective description.
Fast inference dynamics evolves primarily through the amplitudes
\(X_\alpha(t)\), while slow learning reorganizes the collective modes
\(u_\alpha\) and the spectrum \(\mu_\alpha\) themselves.

This distinction resolves a potential ambiguity in the interpretation
of latent memory modes.
The collective modes are not static memory slots.
Nor are they merely instantaneous states of the cognitive system.
They define local dynamical directions of the learned cognitive
geometry, while their time-dependent amplitudes determine how strongly
each collective direction participates in the ongoing cognitive
process.

In summary, the slow-mode expansion admits a clear physical
interpretation.
The collective modes \(u_\alpha\) represent learned dynamical modes
of the cognitive geometry, while the amplitudes \(X_\alpha(t)\)
describe their time-dependent activation during ongoing cognitive
dynamics.
The persistence time
\(\lambda_\alpha^{-1}\)
characterizes how long the influence of a collective activation
remains dynamically relevant, whereas the circulation frequency
\(\omega_\alpha\) determines its temporal phase evolution and
recurrence structure.
After the latent amplitudes \(X_\alpha(t)\) are integrated out, their
collective influence appears as the effective memory propagator
\(K(t)\), which mediates nonlocal temporal feedback in the observable
cognitive field.

The memory-dressed cognitive field therefore emerges not from isolated
stored memories, but from the propagation of activity through learned
collective modes generated by the cognitive geometry.
Learning continuously reorganizes this collective-mode basis, while
inference dynamically activates it through the mode amplitudes
\(X_\alpha(t)\).

\vspace{6pt}
\paragraph{Adiabatic and adaptive collective-mode descriptions.}

The discussion above assumes an adiabatic separation between
inference and learning timescales.
When learning evolves much more slowly than inference,
\(
\tau_{\rm inf}\ll\tau_{\rm learn}
\),
the collective modes
\(
u_\alpha
\)
and eigenvalues
\(
\mu_\alpha
\)
may be treated as approximately fixed, while the amplitudes
\(
X_\alpha(t)
\)
carry the dominant dynamical evolution.

More generally, learning and inference may occur simultaneously.
In this case the structural parameters
\(
\kappa(t)
\)
evolve during ongoing cognitive dynamics, leading to
\[
u_\alpha
=
u_\alpha(\kappa(t)),
\qquad
\mu_\alpha
=
\mu_\alpha(\kappa(t)).
\]
The collective-mode basis itself then becomes dynamical.
Inference reorganizes the mode amplitudes
\(
X_\alpha(t)
\),
while learning simultaneously reorganizes the collective modes and
their spectrum.
The adiabatic description developed in the present work corresponds
to the limiting case in which this structural evolution is slow
compared with inference dynamics.

The Transformer analysis presented in Sec.~VII provides direct
computational evidence that, during learning, the collective-mode basis
itself evolves through the progressive infrared reorganization of the
Jacobian spectrum, whereas individual inference trajectories are well
described by the adiabatic approximation developed above.

\end{document}